%% file: main.tex
\pdfoutput=1
\documentclass[11pt]{article}
\usepackage[utf8]{inputenc}%
\usepackage[tracking=true, letterspace=100, expansion=false]{microtype}
\usepackage[T1]{fontenc}
\usepackage{amsmath}%
\usepackage{amsfonts}%
\usepackage{amsthm}%
\usepackage{tikz} %
\usepackage[capposition=bottom]{floatrow}
\usepackage{booktabs}
\usepackage{tabularx}
\usepackage{siunitx}
\usepackage{pdflscape}%
\usepackage[normal, online, flushleft]{threeparttable}%
\usepackage[labelformat=simple]{subcaption}
\usepackage{multirow}
\usepackage{graphicx}
\newtheorem{assumption}{Assumption}

\newtheorem{proposition}{Proposition}
\newtheorem{corollary}{Corollary}
\theoremstyle{definition}
\newtheorem{example}{Example}
\theoremstyle{remark}
\newtheorem{remark}{Remark}
\usepackage[margin=1.2in]{geometry}
\usepackage[onehalfspacing]{setspace}
\usepackage{mathtools}
\DeclarePairedDelimiter\indicatorfence{\{}{\}}

\DeclarePairedDelimiter\ip{\langle}{\rangle}
\usepackage{bbm} %
\newcommand\1{\operatorname{\mathbbm{1}}\indicatorfence}

\DeclareMathOperator{\var}{var}
\DeclareMathOperator{\cov}{cov}
\DeclareMathOperator*{\argmin}{argmin}
\DeclareMathOperator{\diag}{diag}
\usepackage[title]{appendix}
\usepackage{etoolbox} %
\AtBeginEnvironment{appendices}{%
  \crefalias{section}{appendix}
  \counterwithin{figure}{section}
  \counterwithin{table}{section}
}

\usepackage[
bibencoding=utf8,%
maxbibnames=3, %
minbibnames=1, %
backend=biber,%
sortlocale=en_US,%
style=apa,%
uniquename=false,%
url=false,%
sortcites=false,
doi=false,%
eprint=false%
]{biblatex}
\usepackage{changepage}

\usepackage{xcolor}%
\definecolor{webbrown}{rgb}{.6,0,0}%
\usepackage{hyperref} %
\hypersetup{%
  breaklinks = true,%
  colorlinks = true,%
  anchorcolor = webbrown,%
  citecolor = webbrown,%
  filecolor = webbrown,%
  linkcolor = webbrown,%
  menucolor = webbrown,%
  urlcolor= webbrown,%
  citebordercolor= 1 0 0,%
  menubordercolor=1 0 0,%
  urlbordercolor=1 0 0,%
  runbordercolor=1 0 0,%
  pdftitle={Contamination Bias in Linear Regressions},
  pdfauthor={Goldsmith-Pinkham, Hull, and Kolesár}
}

\usepackage{cleveref}
\crefname{appsec}{appendix}{appendices}
\crefname{appsubsec}{appendix}{appendices}
\crefname{assumption}{assumption}{assumptions}

\usepackage[nolist]{acronym}
\begin{acronym}
  \acro{CI}{confidence interval}%
  \acro{OLS}{ordinary least squares}%
  \acro{CLT}{central limit theorem}%
  \acro{IV}{instrumental variables}%
  \acro{ATE}{average treatment effect}%
  \acro{RCT}{randomized control trial}%
  \acro{VAM}{value-added model}%
  \acro{LAN}{locally asymptotically normal}%
  \acro{DiD}{difference-in-differences}%
  \acro{OVB}{omitted variables bias}
  \acro{FWL}{Frisch-Waugh-Lovell}
  \acro{TWFE}{two-way fixed effects}
  \acro{EW}{easiest-to-estimate weighting}
  \acro{CW}{easiest-to-estimate common weighting}
\end{acronym}

\newcommand{\dd}{\mathcal{D}}
\newcommand{\kk}{\mathcal{K}}
\newcommand{\lcw}{\lambda^{\textnormal{CW}}}
\newcommand{\hlcw}{\hat{\lambda}^{\textnormal{CW}}}
\usepackage{accents}
\newcommand{\dbtilde}[1]{\accentset{\approx}{#1}}

\newcommand{\overbar}[1]{\mkern 1.5mu\overline{\mkern-1.5mu#1\mkern-1.5mu}\mkern 1.5mu}

\newcolumntype{Y}{>{\centering\arraybackslash}X}

\begin{document}

\title{Contamination Bias in Linear Regressions\thanks{\noindent Contact:
    \href{mailto:paul.goldsmith-pinkham@yale.edu}{paul.goldsmith-pinkham@yale.edu},
    \href{mailto:peter_hull@brown.edu}{peter\_hull@brown.edu}, and
    \href{mailto:mkolesar@princeton.edu}{mkolesar@princeton.edu}. We thank Alberto Abadie, Jason Abaluck, Isaiah Andrews, Josh Angrist, Tim Armstrong, Kirill Borusyak, Kyle Butts, Clément de Chaisemartin, Peng Ding, Len Goff, Jin Hahn, Xavier D'Haultfœuille, Simon Lee, Bernard Salanié, Pedro Sant'Anna, Tymon Słoczyński, Isaac Sorkin, Jonathan Roth, Jacob Wallace, Stefan Wager, and numerous seminar participants for helpful comments. Hull acknowledges support from National Science Foundation Grant SES-2049250.
    Kolesár acknowledges support by the Sloan Research Fellowship and by the
    National Science Foundation Grant SES-22049356. Mauricio C\'{a}ceres Bravo, Jerray Chang, William Cox, and Dwaipayan Saha provided expert research assistance. An earlier draft of this paper circulated under the title ``On Estimating Multiple Treatment Effects with Regression.''}}%
\author{Paul Goldsmith-Pinkham\\Yale University \and Peter Hull\\Brown University \and Michal Kolesár\\Princeton University}%
\date{\today}

  \pagenumbering{Alph} %
\begin{titlepage}
  \maketitle
  \thispagestyle{empty}
\begin{adjustwidth*}{0.2cm}{0.2cm}
\begin{abstract} %
  \noindent We study regressions with multiple treatments and a set of controls
  that is flexible enough to purge omitted variable bias. We show that these
  regressions generally fail to estimate convex averages of heterogeneous
  treatment effects---instead, estimates of each treatment's effect are
  contaminated by non-convex averages of the effects of other treatments. We
  discuss three estimation approaches that avoid such contamination bias,
  including the targeting of easiest-to-estimate weighted average effects. A
  re-analysis of nine empirical applications finds economically and
  statistically meaningful contamination bias in observational studies; contamination bias in experimental studies is more limited due to smaller variability in propensity scores.
\end{abstract}
\end{adjustwidth*}
\end{titlepage}
\pagenumbering{arabic}
\setstretch{1.25}

\section{Introduction}

Consider a linear regression of an outcome $Y_i$ on a vector of treatments $X_i$ and a vector of flexible controls $W_i$. The treatments are assumed to be as good as randomly assigned conditional on the controls. For example, $X_i$ may indicate the assignment of individuals $i$ to different interventions in a stratified \ac{RCT}, with the randomization protocol varying across some experimental strata indicators in $W_i$. Or, in an education \ac{VAM}, $X_i$ might indicate the matching of students $i$ to different teachers or schools with $W_i$ including measures of student demographics and lagged achievement which yield a credible selection-on-observables assumption. The regression might also be the first stage of an \ac{IV} regression leveraging the assignment of multiple decision-makers (e.g.\ bail judges) indicated in $X_i$, which is as-good-as-random conditional on some controls $W_i$. These sorts of regressions are widely used across many fields in economics.\footnote{Prominent \acp{RCT} where randomization probabilities vary across strata include Project STAR \parencite{krueger1999experimental} and the RAND Health Insurance Experiment \parencite{manning1987health}. Prominent \ac{VAM} examples include studies of teachers \parencite{NBERw14607,chetty2014measuring}, schools \parencite{angrist2017leveraging,ahpw20,mountjoy2020returns}, and healthcare institutions \parencite{abaluck2020mortality,geruso2020all}. Prominent ``judge \ac{IV}'' examples include \textcite{kling2006incarceration,maestas2013does,dobbie2015debt}.}

This paper shows that such multiple-treatment regressions generally fail to estimate convex weighted averages of heterogeneous causal effects, and discusses solutions to this problem. The problem may be surprising given an influential result in \textcite{angrist98}, showing that regressions on a single binary treatment $D_i$ and flexible controls $W_i$ estimate a convex average of treatment effects whenever $D_i$ is conditionally as good as randomly assigned. We show that this result does not generalize to multiple treatments: regression estimates of each treatment's effect are generally contaminated by a non-convex average of the effects of other treatments. Thus, the regression coefficient for a given treatment arm incorporates the effects of \emph{all} arms.

We first derive a general characterization of such \emph{contamination bias} in multiple-treatment regressions.\footnote{\label{fn:contamination_terminology}Our use of the term ``contamination'' follows \textcite{sun2021estimating}, and differs from its use in some analyses of clinical trials \parencite[e.g.][]{keoghbrown07} to describe settings where members of one treatment group receive the treatment of another group---what economists typically call ``non-compliance''. Our ``bias'' terminology refers to an implication of our result: if a given treatment has constant effects, but the other treatment effects are heterogeneous, the regression estimand is inconsistent for the given treatment effect.} We show the core problem by
focusing on the special case of a set of mutually exclusive treatment indicators, though our characterization applies even when the treatments are not restricted to be binary or mutually exclusive. To separate the problem from the typical challenge of \ac{OVB}, we assume a best-case scenario where the covariate parametrization is flexible enough to include the treatment propensity scores (e.g., with a linear covariate adjustment, we assume that the propensity scores are linear in the covariates). This condition holds trivially if the only covariates are strata indicators. Under these conditions, we show that the regression coefficient on each treatment identifies a convex weighted average of its causal effects plus a contamination bias term given by a linear combination of the causal effects of other treatments, with weights that sum to zero. Thus, each treatment effect estimate will generally incorporate the effects of other treatments, unless the effects are uncorrelated with the contamination weights. Since these weights sum to zero some are necessarily negative---further complicating the interpretation of the coefficients.

Contamination bias arises because regression adjustment for the confounders in $W_i$ is generally insufficient for making the other treatments ignorable when estimating a given treatment's effect, even when this adjustment is flexible enough to avoid \ac{OVB}. To see this intuition clearly, suppose the only controls are strata indicators. %
\ac{OVB} is avoided when the treatments are as good as randomly assigned within strata. But because the treatments enter the regression linearly, the \textcite{angrist98} result implies that the causal interpretation of a \emph{given} treatment's coefficient is only guaranteed when its assignment depends linearly on both the strata indicators \emph{and} the other treatment indicators. With mutually exclusive treatments, this condition fails because the dependence is inherently nonlinear---the probability of assignment to a given treatment is zero if an individual is assigned to one of the other treatments, regardless of their stratum, but strata indicators affect the treatment probability otherwise. Such dependence generates contamination bias.

Contamination bias also arises under an alternative ``model-based'' identifying
assumption that---rather than making assumptions on the treatment's ``design''
(i.e.\ propensity scores)---posits that the covariate specification spans the
conditional mean of the potential outcome under no treatment, $Y_i(0)$. In a
linear model with unit and time fixed effects, this reduces to the
parallel trends restriction often used in \ac{DiD} and event study regressions. It is
common for $X_i$ to include multiple indicators in such settings---for example,
the leads and lags relative to a treatment adoption date used to support the
parallel trends assumption or estimate treatment effect
dynamics.\footnote{Alternatively $X_i$ may indicate multiple contemporaneous
  treatments, as in certain ``mover'' regressions.} We show that replacing the
restriction on propensity scores in our characterization with an assumption on $Y_i(0)$ generates an
additional issue: the own-treatment weights are negative whenever the implicit
propensity score model used by the regression to partial out the covariates and
the other treatments fits probabilities greater than one. This result shows that
the negative weighting and contamination bias issues documented previously in
the context of two-way fixed effects regressions
\parencite[e.g.,][]{goodman2021difference, sun2021estimating, de2020aer, de2020two, callaway2021difference,
  borusyak2021revisiting,wooldridge2021mundlak,hull2018movers} are more
general---and conceptually distinct---problems.\footnote{Our analysis also
  relates to issues with interpreting multiple-treatment \ac{IV} estimates
  \parencite{behaghel2013robustness, kirkeboen2016field,    kline2016headstart,hull2018isolateing,leeMultivalued,bhuller_sigstad}.}
Negative weighting arises because regressions leveraging model-based
restrictions on $Y_i(0)$ may fit treatment probabilities exceeding one.
Contamination bias arises because additive covariate adjustments don't account
for the non-linear dependence of a given treatment on the other treatments and
covariates. This generates a different form of propensity score misspecification: a
non-zero fitted probability of a given treatment, even when one of the other
treatments is known to be non-zero.\footnote{While our results are framed in the context of a causal model,
we show how analogous results apply to descriptive regressions which seek to estimate averages of conditional group contrasts without assuming a causal framework---as in studies of outcome disparities across multiple racial or ethnic groups, studies of regional variation in healthcare utilization or outcomes, or studies of industry wage gaps.}

We then discuss three solutions to the contamination bias problem, and their
trade-offs. These solutions apply when the propensity scores are
non-degenerate, such as in an \ac{RCT} or other ``design-based'' regression specification.\footnote{\label{fn:did_solutions}Solving the contamination bias problem under model-based identification approaches requires either targeting subpopulations of the
  treated or applying substantive restrictions on the conditional
  means of potential outcomes under treatment. We do not
  explore this case as it has already been studied extensively in the \ac{DiD} context
  \parencite[e.g.][]{de2020two, sun2021estimating, callaway2021difference,borusyak2021revisiting,wooldridge2021mundlak}.} First, a conceptually principled solution is
to adapt approaches to estimating the \ac{ATE} of a conditionally ignorable
binary treatment to the multiple
treatment case \parencite[e.g.][]{cattaneo10, chernozhukov2018double,ChNeSi21,ReZu20,
  GrPi22}. For example, one could run a regression that includes
interactions between the treatments and demeaned controls, or combine such regression with inverse propensity score weighting for doubly-robust estimation. Such \ac{ATE} estimators work well under strong overlap of the covariate
distribution for units in each treatment arm. But they may be imprecise under limited overlap or be outright infeasible with
overlap failures---common scenarios in observational studies \parencite{crump2009dealing}.

This practical consideration motivates an alternative approach: estimating a
weighted average of treatment effects, as regression does in the binary
treatment case, while avoiding the contamination bias with multiple-treatments.
We derive the weights that are easiest to estimate, in the sense of minimizing a
semiparametric efficiency bound under homoskedasticity. This \ac{EW} scheme is
always convex; it corresponds to weighting schemes previously proposed in
\textcite{crump2006dealing}, \textcite{liBalancing}, and \textcite{lili19}. The
weights also coincide with the implicit linear regression weights when the
treatment is binary (i.e.\ the \textcite{angrist98} case).
In the multiple treatment case, the \ac{EW} scheme that allows the weights to be
treatment specific can be implemented by a simple second solution: a linear
regression which restricts estimation to the individuals who are either in the
control group or the treatment group of interest. Since the weights are
treatment-specific, these one-treatment-at-a-time regressions preclude direct
comparisons across treatment arms. The third solution is to impose common
weights across treatments in the \ac{EW} scheme; these weights can be
implemented using a weighted regression approach. We show how researchers can
gauge the extent of contamination bias in practice and implement these tools
with a new R and Stata package, \texttt{multe}.\footnote{The package is available at
  CRAN (R) and \url{https://github.com/gphk-metrics/stata-multe} (Stata).}

We study the empirical relevance of contamination bias in nine applications: six
\acp{RCT} with stratified randomization and three observational studies of
racial disparities. We find economically and statistically significant
bias in two of the three observational studies with no evidence for bias in any of the experimental studies. In a detailed analysis of one experiment---the Project STAR trial---we show that the lack of contamination bias is driven by small variation in the contamination weights, rather than limited effect
heterogeneity. This analysis highlights the importance of conducting
contamination bias diagnostics, particularly in observational studies where covariates are expected to generate high variability in propensity scores, and thus likely in contamination weights.

We structure the rest of the paper as follows. \Cref{sec:example} illustrates
contamination bias in a simple stylized setting. \Cref{sec:problem} characterizes the general
problem, and
discusses connections to previous analyses. \Cref{sec:solutions} provides three solutions, and gives guidance for
measuring and avoiding contamination bias in practice. \Cref{sec:applications}
illustrates these tools in nine applications. \Cref{sec:conclusion}
concludes. \Cref{apdx:proofs} collects all proofs and extensions. \Cref{apdx:did} discusses the connection between our contamination bias characterization and that in the \ac{DiD} literature. Details on the applications and additional exhibits are given in \Cref{apx:empirical_apps,apx:additional_fig}.

\section{Motivating Example}\label{sec:example}

We build intuition for the contamination bias problem in two simple examples. We
first review how regressions on a single randomized binary treatment and binary
controls identify a convex average of heterogeneous treatment effects. We then
show how this result fails to generalize when we introduce an additional
treatment arm. We base these examples on a stylized version of the Project STAR
experiment, which we return to as an application in \Cref{sec:star}. The
simple structure of these examples helps isolate the core mechanisms of
contamination bias. Later sections consider non-experimental settings with
richer control specifications, both theoretically and empirically.

\subsection{Convex Weights with One Randomized Treatment}
Consider the regression of an outcome $Y_i$ on a single treatment indicator
$D_{i}\in\{0,1\}$, a single binary control $W_i\in\{0,1\}$, and an intercept:
\begin{equation}\label{eq:simple_ci}
    Y_i=\alpha + \beta D_i+\gamma W_i +U_i.
\end{equation}
By definition, $U_i$ is a mean-zero regression residual that is uncorrelated with $D_i$ and $W_i$. %
For example, analysing the Project STAR trial, \textcite{krueger1999experimental} primarily studied the effect of
small class size $D_i$ on the test scores $Y_i$ of kindergartners
indexed by $i$. Project STAR randomized students to classes within schools, with the fraction of students assigned to small classes varying by school due to the varying number of total students in each school. To account for this, \textcite{krueger1999experimental} included school fixed effects as controls. Such specifications are often found in stratified \acp{RCT} with varying treatment assignment rates across a set of pre-treatment strata. If we imagine two such strata, demarcated by a binary indicator $W_{i}$, then \cref{eq:simple_ci} corresponds to a stylized two-school version of a Project STAR regression.

We wish to interpret the coefficient $\beta$ in terms of the causal effects of $D_i$ on $Y_i$. For this we use potential outcome notation, letting $Y_i(d)$ denote the test score of student $i$ when $D_i=d$. Individual $i$'s treatment effect is then given by $\tau_{1i}=Y_i(1)-Y_i(0)$, and we can write realized achievement as $Y_i=Y_i(0)+\tau_{1i}D_i$. Since treatment assignment is random within schools, $D_i$ is conditionally independent of potential outcomes given $W_i$: $\left(Y_i(0), Y_i(1)\right)\perp D_i\mid W_i$.

\textcite{angrist98} showed that regression coefficients like $\beta$ identify a convexly-weighted average of within-strata \acp{ATE}. In our Project STAR example, this result shows that:
\begin{equation}\label{eq:simple_angrist}
    \beta=\phi\tau_{1}(0)+(1-\phi)\tau_{1}(1), \quad\text{where}\quad
  \phi= \frac{\var(D_i\mid W_i=0)\Pr(W_i=0)}{\sum_{w=0}^{1}
  \var(D_i\mid W_i=w)\Pr(W_i=w)}\in [0,1]
\end{equation}
gives a convex weighting scheme, and $\tau_{1}(w)=E[Y_i(1)-Y_i(0)\mid W_i=w]$ is the \ac{ATE} in
school $w\in\{0,1\}$. Thus, in our example the coefficient $\beta$ identifies a weighted average of school-specific small classroom effects $\tau_{1}(w)$ across the two schools.

\Cref{eq:simple_angrist} can be derived by applying the \ac{FWL} Theorem. The multivariate regression coefficient $\beta$ can be written as a univariate regression coefficient from regressing $Y_i$ onto the population residual $\tilde{D}_i$ from regressing $D_i$ onto $W_i$ and a constant:
\begin{equation}\label{eq:fwl_simple}
    \beta=\frac{E[\tilde{D}_{i}Y_i]}{E[\tilde{D}_i^2]}=
    \frac{E[\tilde{D}_{i}Y_i(0)]}{E[\tilde{D}_i^2]}+\frac{E[\tilde{D}_{i}D_{i}\tau_{1i}]}{E[\tilde{D}_i^2]},
\end{equation}
where we substitute the potential outcome model for $Y_i$ in the second equality. Since $W_i$ is binary, the propensity score $E[D_i\mid W_i]$ is linear and the residual $\tilde{D_i}$ is mean independent of $W_i$ (not just uncorrelated with it): $E[\tilde{D_i}\mid W_i]=0$. Therefore,
\begin{equation}\label{eq:simple_meanzero}
    E[\tilde{D}_i Y_i(0)]= E[E[\tilde{D}_i Y_i(0)\mid W_i]]=E[E[\tilde{D}_i\mid W_i]E[Y_i(0)\mid W_i]]=0.
\end{equation}
The first equality in \cref{eq:simple_meanzero} follows from the law of iterated expectations, the second equality follows by the conditional random assignment of $D_i$ and the third equality uses $E[\tilde{D_i}\mid W_i]=0$. Hence, the first summand in \cref{eq:fwl_simple} is zero. Analogous arguments show that
\begin{equation*}
    E[\tilde{D}_{i}D_{i}\tau_{1i}]
  = E[E[\tilde{D}_i D_i\tau_{1i}\mid W_i]]=E[E[\tilde{D}_{i} D_i\mid W_i]E[\tau_{1i}\mid W_i]]
  =E[\var(D_i\mid W_i)\tau_{1}(W_i)],
\end{equation*}
where $\var(D_i\mid W_i)=E[\tilde{D}_i^2\mid W_i]$ gives the conditional variance of the small-class treatment within schools. Since $E[\var(D_i\mid W_i)]=E[E[\tilde{D}_i^2\mid W_i]]=E[\tilde{D}_i^2]$, it follows that we can write the second summand in \cref{eq:fwl_simple} as
\begin{equation*}
  \beta=\frac{E[\var(D_{i}\mid W_{i})\tau_{1}(W_i)]}{E[\var(D_{i}\mid W_{i})]}=\phi\tau_{1}(0)+(1-\phi)\tau_{1}(1),
\end{equation*}
proving the representation of $\beta$ in \cref{eq:simple_angrist}.

The key fact underlying this derivation is that the residual $\tilde{D}_i$ from the auxiliary regression of the treatment $D_i$ on the other regressors $W_i$ is mean-independent of $W_{i}$. By the \ac{FWL} theorem, treatment coefficients like $\beta$ can always be represented as in \cref{eq:fwl_simple} even without this property. We next show, however, that the remaining steps in the derivation of \cref{eq:simple_angrist} fail when an additional treatment arm is included. This failure can be attributed to the fact that the auxiliary \ac{FWL} regression delivers a treatment residual that is uncorrelated with---but not mean-independent of---the other regressors. The lack of mean independence leads to an additional term in the expression for the regression coefficient.

\subsection{Contamination Bias with Two Randomized Treatments}
In reality, Project STAR randomized students to three mutually exclusive conditions within schools: a control group with a regular class ($D_i=0$), a treatment that reduced class size ($D_i=1$), and a treatment that introduced full-time teaching aides ($D_i=2$). We incorporate this extension of our stylized example by considering a regression of student achievement $Y_i$ on a vector of two treatment indicators, $X_{i} = (X_{i1}, X_{i2})'$, where $X_{ik}=\1{D_i=k}$ indicates assignment to treatment $k=1,2$.  We continue to include a constant and the school indicator $W_i$ as controls, yielding the regression
\begin{equation}\label{eq:simple_bias_regression}
  Y_i=\alpha+  \beta_{1}X_{i1}+\beta_{2}X_{i2}+\gamma W_i+U_i.
\end{equation}
The observed outcome is now given by $Y_i=Y_i(0)+\tau_{i1}X_{i1}+\tau_{i2}X_{i2}$, with $\tau_{i1}=Y_i(1)-Y_i(0)$ and $\tau_{i2}=Y_i(2)-Y_i(0)$ denoting the potentially heterogeneous effects of a class size reduction and introduction of a teaching aide, respectively. As before, we analyze this regression by assuming ${X}_i$ is conditionally independent of the potential achievement outcomes $Y_i(d)$ given the school indicator $W_i$: $\left(Y_i(0), Y_i(1), Y_i(2)\right)\perp {X}_i \mid W_i$.

To analyze the coefficient on $X_{i1}$, we again use the \ac{FWL} theorem to write
\begin{equation}\label{eq:three_fwl}
  \beta_1=\frac{E[\dbtilde{X}_{i1}Y_i]}{E[\dbtilde{X}_{i1}^2]}=\frac{E[\dbtilde{X}_{i1}Y_{i}(0)]}{E[\dbtilde{X}_{i1}^2]}+\frac{E[\dbtilde{X}_{i1}X_{i1}\tau_{i1}]}{E[\dbtilde{X}_{i1}^2]}+\frac{E[\dbtilde{X}_{i1}X_{i2}\tau_{i2}]}{E[\dbtilde{X}_{i1}^2]},
\end{equation}
where $\dbtilde{X}_{i1}$ again denotes a population residual, but now from regressing $X_{i1}$ on $W_i$, a constant, \emph{and} $X_{i2}$.
Unlike before, this residual is uncorrelated with but \emph{not} mean-independent of the remaining regressors $(W_i, X_{i2})$ because the dependence between $X_{i1}$ and $X_{i2}$ is non-linear. When $X_{i2}=1$, $X_{i1}$ must be zero regardless of the value of $W_i$ (because they are mutually exclusive) while if $X_{i2}=0$ the mean of $X_{i1}$ does depend on $W_i$ unless the treatment assignment is completely random. Thus, in general, $\dbtilde{X}_{i1}\neq X_{i1}-E[X_{i1}\mid W_i, X_{i2}]$.

Because $\dbtilde{X}_{i1}$ does not coincide with a conditionally de-meaned $X_{i1}$, we can not generally reduce \cref{eq:three_fwl} to an expression involving only the effects of the first treatment arm, $\tau_{i1}$. It turns out that we nevertheless still have $E[\dbtilde{X}_{i1}Y_{i}(0)]=0$, as in \cref{eq:simple_meanzero}, since the auxilliary regression residuals are still uncorrelated with any individual characteristic like $Y_i(0)$.\footnote{To see this, note that in the auxiliary regression
  $X_{i1} = \mu_{0} + \mu_{1}X_{i2} + \mu_{2}W_{i} + \dbtilde{X}_{i1}$ we can
  partial out $W_i$ and the constant from both sides to write
  $\tilde{X}_{i1} = \mu_{1}\tilde{X}_{i2} + \dbtilde{X}_{i1}$. Thus,
  $\dbtilde{X}_{i1} = \tilde{X}_{i1} - \mu_{1}\tilde{X}_{i2}$ is a linear
  combination of residuals which, per \cref{eq:simple_meanzero}, are
  both uncorrelated with $Y_i(0)$. It follows that
  $E[\dbtilde{X}_{i1}Y_i(0)]=0$.} The regression thus does not suffer
from \ac{OVB}. However, we do not generally have
$E[\dbtilde{X}_{i1}X_{i2}\tau_{i2}]=0$. Instead, simplifying \cref{eq:three_fwl} by the same steps as before leads to the expression
\begin{equation}\label{eq:simple_bias}
    \beta_{1}=E[\lambda_{11}(W_i)\tau_{1}(W_i)] +
    E[\lambda_{12}(W_i)\tau_{2}(W_i)]
 \end{equation}
as a generalization of  \cref{eq:simple_angrist}. Here $\lambda_{11}(W_i)=E[\dbtilde{X}_{i1}X_{i1}\mid W_i]/E[\dbtilde{X}_{i1}^2]$ can be shown to be non-negative and to average to one, similar to the $\phi$ weight in \cref{eq:simple_angrist}. Thus, if not for the second term in \cref{eq:simple_bias}, $\beta_1$ would similarly identify a convex average of the conditional \acp{ATE} $\tau_{1}(W_i)=E[Y_i(1)-Y_i(0)\mid W_i]$. But precisely because $\dbtilde{X}_{i1}\neq X_{i1}-E[X_{i1}\mid W_i, X_{i2}]$, this second term is generally present:  $\lambda_{12}(W_i)=E[\dbtilde{X}_{i1}X_{i2}\mid W_i]/E[\dbtilde{X}_{i1}^2]$ is generally non-zero, complicating the interpretation of $\beta_1$ by including the conditional effects of the other treatment $\tau_{2}(W_i)=E[Y_i(2)-Y_i(0)\mid W_i]$.

The second \emph{contamination bias} term in \cref{eq:simple_bias} arises because the residualized small class treatment $\dbtilde{X}_{i1}$ is not conditionally
independent of the second full-time aide treatment $X_{i2}$ within schools, despite being uncorrelated with $X_{i2}$ by construction. This can be seen by viewing $\dbtilde{X}_{i1}$ as the result of an equivalent two-step residualization. First, both $X_{i1}$ and $X_{i2}$ are de-meaned within schools: $\tilde{X}_{i1} = X_{i1} - E[X_{i1}\mid W_{i}] = X_{i1} - p_{1}(W_{i})$ and $\tilde{X}_{i2} = X_{i2} - E[X_{i2}\mid W_{i}] = X_{i2} - p_{2}(W_{i})$ where $p_j(W_i)=E[X_{ij}\mid W_i]$ gives the propensity score for treatment $j$. Second, a bivariate regression of $\tilde{X}_{i1}$ on $\tilde{X}_{i2}$ is used to generate the residuals $\dbtilde{X}_{i1}$.  When the propensity scores vary across the schools (i.e.\ $p_j(0)\neq p_j(1)$), the relationship between these residuals varies by school, and the line of best fit between  $\tilde{X}_{i1}$ and $\tilde{X}_{i2}$ averages across this relationship. As a result, the line of best fit does not isolate the conditional (i.e.\ within-school) variation in $X_{i1}$: the remaining variation in $\dbtilde{X}_{i1}$ will tend to predict $X_{i2}$ within schools, making the \emph{contamination weight} $\lambda_{12}(W_i)$  non-zero.

\subsection{Illustration and Intuition}\label{sec:illustr-intu}

A simple numerical example helps make the contamination bias problem concrete. Suppose in the previous setting that school $0$ (indicated by $W_i=0$) assigned only 5 percent of the students to the small classroom treatment, with 45 percent of the students assigned to the full-time aide treatment and the rest assigned to the control group. In school $1$ (indicated by $W_i=1$), there was a substantially larger push for students to be placed into treatment groups with 45 percent of students assigned to a small classroom, 45 percent assigned to a classroom with a full-time aide, and only 10 percent assigned to the control group. Therefore, $p_1(0) = 0.05$ and $p_2(0) = 0.45$ while $p_1(1) = p_2(1) = 0.45$. Suppose that the schools have the same number of students, so that $\Pr(W_{i} = 1) = 0.5$. It then follows from the above formulas that $\lambda_{12}(0) = 99/106$ and $\lambda_{12}(1) =-99/106$.

As reasoned above, the contamination weights are non-zero here because the
within-school correlation between the residualized treatments, $\tilde{X}_{i1}$
and $\tilde{X}_{i12}$, is heterogeneous: in school $0$ it is about $-0.2$, so
that the value of the demeaned class aide treatment is only weakly predictive of
the small classroom treatment, while in school $1$ it is highly predictive with
correlation $-0.8$.
\Cref{fig:example_misspecification} in \Cref{apx:additional_fig} illustrates
this graphically, showing that because the overall regression of
$\tilde{X}_{i1}$ on $\tilde{X}_{i2}$ averages over these two correlations, the regression residuals are predictive of the value of the class aide treatment.

To illustrate the potential magnitude of bias in this example, suppose that classroom reductions have no effect on student achievement (so $\tau_1(0)=\tau_1(1)=0$), but that the effect of a teaching aide varies across schools. In school $1$ the aide is highly effective, $\tau_2(1)=1$, (which may be the reason for the higher push in this school to place students into treatment groups) but in school $0$, the aide has no effect, $\tau_2(0)=0$. By \cref{eq:simple_bias}, the regression coefficient on the first treatment identifies
\begin{equation*}
  \beta_1=E[\lambda_{11}(W_i)\cdot 0] +
           E[\lambda_{12}(W_i)\tau_{2}(W_{i})] = 0 + (-99/106 \times 1 +99/106 \times 0)/2 \approx -0.47.
\end{equation*}
Thus, in this example, a researcher would conclude that small classrooms have a sizable negative effect on student achievement---equal in magnitude to around half of the true teaching aide effect in school $1$---despite the true small-classroom effect being zero for all students. This treatment effect coefficient can be engineered to match an arbitrary magnitude and sign by varying the heterogeneity of the teaching aide effects across schools.

To build further intuition for \cref{eq:simple_bias}, it is useful to consider two cases where the contamination bias term is zero. First, note that since regression residuals are by construction uncorrelated with the included regressors, $E[\lambda_{12}(W_i)]=E[\dbtilde{X}_{i1}X_{i2}]/E[\dbtilde{X}_{i1}^2]=0$. Therefore, $E[\lambda_{12}(W_i)\tau_2(W_i)]=E[\lambda_{12}(W_i)\tau_2(W_i)]-E[\lambda_{12}(W_i)]E[\tau_2(W_i)]=\cov(\lambda_{12}(W_i), \tau_2(W_i))$. If the average effects of the teaching aide treatment are constant across the two schools, $\tau_2(1)=\tau_2(0)$, then $\tau_2(W_i)$ is constant, and this covariance is zero such that contamination bias disappears. More generally, when the average teaching aide treatment effects across schools $\tau_2(W_i)$ exhibit idiosyncratic variation, in the sense that they have a weak covariance with the contamination weights across schools, the contamination bias term will be small.

Second, consider the case where $X_{i1}$ and $X_{i2}$ are independent conditional on $W_i$---such as when the small classroom and teacher aid interventions are independently assigned within schools, in contrast to the previously assumed mutual exclusivity of these treatments. In this case the conditional expectation $E[X_{i1}\mid W_i, X_{i2}]=E[X_{i1}\mid W_i]$ will be linear, since $X_{i1}$ and $X_{i2}$ are unrelated given $W_i$, and will thus be identified by the auxiliary regression of $X_{i1}$ on $W_i$,  $X_{i2}$, and a constant. Consequently, the $\dbtilde{X}_{i1}$ residuals will coincide with $X_{i1}-E[X_{i1}\mid W_i]$. The coefficient on $X_{i1}$ in \cref{eq:simple_bias_regression} can therefore be shown to be equivalent to the previous \cref{eq:simple_angrist}, identifying the same convex average of $\tau_1(w)$. This case highlights that dependence across treatments is necessary for the contamination bias to arise.

\section{General Problem}\label{sec:problem}

We now derive a general characterization of the contamination bias problem, in regressions of an outcome $Y_i$ on a $K$-dimensional treatment vector $X_i$ and flexible transformations of a control vector $W_i$. We focus on the case of mutually exclusive indicators $X_{ik}=\1{D_i=k}$ for values of an underlying treatment $D_i\in\{0,\dotsc, K\}$ (with the $\1{D_i=0}$ indicator omitted). We extend the characterization to a general (i.e.\ potentially non-binary) $X_i$ in \Cref{apdx:general}.

We suppose the effects of $X_i$ on $Y_i$ are estimated by a partially linear model:
\begin{equation}\label{eq:partially_linear}
  Y_i=X_i^\prime\beta+g(W_i)+U_i,
\end{equation}
where $\beta$ and $g$ are defined as the minimizers of expected squared residuals $E[U_i^2]$:
\begin{equation}\label{eq:ssr_minimizer}
  (\beta, g)=  \argmin_{\tilde{\beta}\in\mathbb{R}^{K}, \tilde{g}\in\mathcal{G}}
  E[(Y_i-{X}_i^\prime\tilde{\beta}-\tilde{g}(W_i))^{2}]
\end{equation}
for some linear space of functions $\mathcal{G}$. This setup nests linear covariate
adjustment by setting
$\mathcal{G}=\{\alpha+w^\prime\gamma\colon[\alpha, \gamma^\prime]^\prime\in\mathbb{R}^{1+\dim(W_i)}\}$,
in which case \cref{eq:partially_linear} gives a linear regression of $Y_i$ on
${X}_i$, $W_i$, and a constant. The setup also
allows for more flexible covariate adjustments---such as by specifying
$\mathcal{G}$ to be a large class of ``nonparametric'' functions \parencite[e.g.][]{robinson88}.

Two examples highlight the generality of this setup:
\begin{example}[\emph{Multi-Armed \ac{RCT}}]\label{ex1}
 $W_i$ is a vector of mutually-exclusive indicators for experimental strata, within which $X_i$ is randomly assigned to individuals $i$. $g$ is linear.
\end{example}
\begin{example}[\emph{Two-Way Fixed Effects}]\label{ex2}
 $i=(j, t)$ indexes panel data, with a fixed set of units $j=1,\dotsc, n$
 observed over periods $t=1,\dotsc, T$. $W_{i} = (J_i, T_{i})$ where $J_i=j$ and
 $T_i=t$ denote the underlying unit and period, and  $g(W_i)=\alpha+(\1{J_i=2},
 \dotsc, \1{J_i=n}, \1{T_i=2}, \dotsc, \1{T_i=T})^\prime\gamma$ includes unit and period indicators.
 $X_i$ contains indicators for leads and lags relative to a deterministic treatment adoption date, $A(j)\in\{1, \dotsc, T, \infty\}$ (with at least one lead excluded to prevent collinearity).
\end{example}
\noindent \Cref{ex1} nests the motivating \ac{RCT} example in \Cref{sec:example}, allowing for an arbitrary number of experimental strata in $W_i$ and multiple treatment arms in $X_i$. \Cref{ex2} shows that our setup can also nest the kind of regressions considered in a recent literature on  \ac{DiD}  and related regression specifications \parencite[e.g.][]{goodman2021difference,hull2018movers, sun2021estimating, de2020aer,de2020two,callaway2021difference, borusyak2021revisiting,wooldridge2021mundlak}.  We elaborate on the connections to this literature in \Cref{apdx:did} by considering general \ac{TWFE} specifications with non-random treatments. These include specifications with multiple static treatment indicators, as in ``mover regressions'' that leverage over-time transitions, as well as dynamic event study specifications.\footnote{\label{fn_dd}Some papers in this \ac{DiD} literature study issues we do not consider, such as when researchers fail to include indicators for all relevant treatment states, which will generally add bias terms to our decomposition of $\beta$, below. Similarly, we do not consider multicollinearity issues like in \textcite{borusyak2021revisiting} by assuming a unique solution to \cref{eq:ssr_minimizer}. For event studies this means we assume some units are never treated, with $A(j)=\infty$.}

As a first step towards characterizing the treatment coefficient vector $\beta$, we solve the minimization problem in \cref{eq:ssr_minimizer}. Let $\tilde{X}_i$ denote the residuals from projecting $X_i$ onto the control specification, with elements $\tilde{X}_{ik}=X_{ik}-\argmin_{\tilde{g}\in\mathcal{G}} E[(X_{ik}-\tilde{g}(W_i))^{2}]$. It follows from the projection theorem \parencite[e.g.][Theorem 11.1]{vdv98} that
\begin{equation}\label{eq:beta_general}
  \beta=E[\tilde{{X}}_i\tilde{{X}}_i^\prime]^{-1}E[\tilde{X}_{i} Y_i].
\end{equation}
Applying the \ac{FWL} theorem, each treatment coefficient can be written $\beta_k=E[\dbtilde{X}_{ik}Y_i]/E[\dbtilde{X}_{ik}^2]$
where $\dbtilde{X}_{ik}$ is the residual from regressing $\tilde{X}_{ik}$ on
$\tilde{X}_{i,-k}=(\tilde{X}_{i1}, \dotsc, \tilde{X}_{i,k-1}, \tilde{X}_{i,
  k+1}, \dotsc, \tilde{X}_{iK})'$. Letting
$E^{*}[X_{ik}\mid X_{i,-k}, W_i]$ denote the projection of $X_{ik}$ onto the space
$\{X_{i,-k}'\tilde{\delta}+\tilde{g}(W_{i})
\colon \tilde{\delta}\in\mathbb{R}^{K-1}, \tilde{g}\in\mathcal{G}\}$, we may
write these residuals as
$\dbtilde{X}_{ik}=X_{ik}-E^{*}[X_{ik}\mid X_{i,-k}, W_{i}]$.

\subsection{Causal Interpretation}\label{sec:caus-interpr}

We now consider the interpretation of each treatment coefficient $\beta_k$ in terms of causal effects. Let $Y_i(k)$ denote the potential outcome of unit $i$ when $D_i=k$. Observed outcomes are given by $Y_i=Y_i(D_i)=Y_i(0)+X_i^\prime\tau_i$ where $\tau_i$ is a vector of treatment effects with elements $\tau_{ik}=Y_i(k)-Y_i(0)$. We denote the conditional expectation of the vector of treatment effects given the controls by $\tau(W_i)=E[\tau_i\mid W_i]$, so that $\tau_k(W_i)$ is the conditional \ac{ATE} for the $k$th treatment. We let $p(W_i)=E[X_i\mid W_i]$ denote the vector of propensity scores, so that $p_{k}(W_i) = \Pr(D_i=k\mid W_i)$. Our characterization of contamination bias doesn't require the propensity scores to be bounded away from $0$ and $1$ and in fact allows them to be degenerate, i.e.\  $p_k(w)\in\{0,1\}$ for all $w$. This is the case in \Cref{ex2}, since $X_i$ is a non-random function of $W_i$. We return to practical questions of propensity score support in \Cref{sec:solutions}.

We make two assumptions to interpret $\beta_k$ in terms of the effects $\tau_i$. First, we assume mean-independence of the potential outcomes and treatment, conditional on the controls:
\begin{assumption}\label{ass:mean_indep}
$E[Y_i(k)\mid D_i, W_i]=E[Y_i(k)\mid W_i]$ for all $k$.
\end{assumption}
\noindent A sufficient condition for this assumption is that the treatment is randomly assigned conditional on the controls, making it conditionally independent of the potential outcomes:
\begin{equation}\label{eq:full_ci}
    \left(Y_i(0), \dotsc, Y_i(K)\right)\perp {D}_i \mid W_i.
\end{equation}
Such conditional random assignment appears in \Cref{ex1}. In \Cref{ex2}, where treatment is a non-random function of the unit and time indices in $W_i$, \Cref{ass:mean_indep} holds trivially.

Second, we assume $\mathcal{G}$ is specified such that that one of two conditions holds:
\begin{assumption}\label{ass:model_or_design}
Let $\mu_0(w)=E[Y_i(0)\mid W_i=w]$ and recall $p_{k}(w)= E[X_{ik}\mid W_i=w]$. Either
\begin{equation}\label{eq:pscore_in_G}
p_{k}\in\mathcal{G}
\end{equation}
for all $k$, or
  \begin{equation}\label{eq:model_based}
    \mu_0\in\mathcal{G}.
  \end{equation}
\end{assumption}
\noindent The first condition requires the covariate adjustment to be flexible enough to capture each treatment's propensity score. For example, with a linear specification for $g$,
\cref{eq:pscore_in_G} requires the propensity scores to be linear in $W_i$
\parencite[cf.\ eq. (30) in][]{AnKr99handbook}. This condition holds trivially in \Cref{ex1}, since $W_i$ is a vector of indicators for groups within which $X_i$ is randomly assigned. When this condition holds, the projection of the treatment onto the covariates coincides with the vector of propensity scores, and the projection residuals coincide with the conditionally demeaned treatment vector $\tilde{X}_i=X_i-p(W_i)$.

In \Cref{ex2}, with $X_i$ being a deterministic function of unit and time indices and $g(W_i)$ including unit and time fixed effects, \cref{eq:pscore_in_G} fails because the propensity scores are binary---they cannot be captured by a linear combination of the \acp{TWFE}. However, \cref{eq:model_based} is satisfied by a parallel trends assumption: that the average untreated potential outcomes $Y_i(0)$ are linear in the unit and time effects. We elaborate on this setup in \Cref{apdx:did}.\footnote{Identification based on \cref{eq:pscore_in_G} can be seen as ``design-based'' in that it only restricts the treatment assignment process. Identification based on \cref{eq:model_based} can be seen as ``model-based'' in that it makes no assumptions on the treatment assignment process but specifies a model for the unobserved untreated potential outcomes.}

Under either condition in \Cref{ass:model_or_design}, the specification of controls is flexible enough to avoid \ac{OVB}. To see this formally, suppose all treatment effects are constant: $\tau_{ik}=\tau_k$ for all $k$. This restriction lets us write $Y_i=Y_i(0)+X_i^\prime\tau$, where $\tau$ is a vector collecting the constant effects. The only source of bias when regressing $Y_i$ on $X_i$ and controls is then the unobserved variation in the untreated potential outcomes $Y_i(0)$. But it follows from the expression for $\beta$ in \cref{eq:beta_general} that there is no such \ac{OVB} when \Cref{ass:model_or_design} holds:
\begin{equation*}
  \beta
  =E[\tilde{{X}}_i\tilde{{X}}_i^\prime]^{-1}
  ({E[\tilde{{X}}_{i} Y_i(0)]}+E[\tilde{{X}}_{i} \tilde{X}_i^\prime]\tau)\\
  =E[\tilde{{X}}_i\tilde{{X}}_i^\prime]^{-1}
  \underbrace{E[\tilde{X}_{i} E[Y_i(0)\mid W_i]]}_{=0}+\tau=\tau.
\end{equation*}
Here the first equality uses the fact that $E[\tilde{{X}}_{i} X_i^\prime]=E[\tilde{X}_i\tilde{X}_i^\prime]$ because $\tilde{X}_i$ is a vector of projection residuals, and the second equality uses the law of iterated expectations and  \Cref{ass:mean_indep}. Under \cref{eq:pscore_in_G}, $E[\tilde{X}_{i} \mid W_i]=0$, so that the term in braces is zero by another application of the law of iterated expectations: $E[\tilde{X}_{i} E[Y_i(0)\mid W_i]]=E[E[\tilde{X}_{i} \mid W_i] E[Y_i(0)\mid W_i]]=0$.
It is likewise zero under \cref{eq:model_based} since $\tilde{X}_{i}$ is by definition of projection orthogonal to any function in $\mathcal{G}$ such that
$E[\tilde{X}_{i} E[Y_i(0)\mid W_i]]=E[\tilde{X}_{i}\mu_{0}(W_{i})]=0$.
Hence, \ac{OVB} is avoided in the constant-effects case so long as either the propensity scores or the untreated potential outcomes are spanned by the control specification. Versions of this double robustness property have been previously observed in, for instance, \textcite{RoMaNe92}.

When treatment effects are heterogeneous but ${X}_i$ contains a \emph{single} treatment indicator, $\beta$ identifies a weighted average of the conditional effects $\tau(W_i)$. Specifically, since by the previous argument we still have $E[\tilde{X}_{i}Y_i(0)]=0$, it follows from \cref{eq:beta_general} that
\begin{equation}\label{eq:efficient_weigthing_binary}
\beta=\frac{E[\tilde{X}_{i} X_i\tau_i]}{E[\tilde{X}_i^2]}=
E[\lambda_{11}(W_i)\tau(W_i)],
\quad\text{with}\quad\lambda_{11}(W_i)=\frac{E[\tilde{X}_{i} X_i\mid W_i]}{E[\tilde{X}_i X_i]},
\end{equation}
where the second equality uses iterated expectations and the identity $E[\tilde{{X}}_{i}^2]=E[\tilde{X}_i{X}_i]$. Under \cref{eq:pscore_in_G}, $E[\tilde{X}_i X_i\mid W_i]=E[\tilde{X}_i^2\mid W_i]=\var(X_i\mid W_i)$, so the weights further simplify to $\lambda_{11}(W_i)=\frac{\var(X_i\mid W_i)}{E[\var(X_i\mid W_i)]}\ge 0$. This extends the \textcite{angrist98} result to a general control specification; versions of this extension appear in, for instance, \textcite{AnKr99handbook}, \textcite[Chapter 3.3]{AnPi09}, and \textcite{ArSa16}.

This result provides a robustness rationale for estimating the effect of a single as-good-as-randomly assigned treatment with a partially linear model~\eqref{eq:partially_linear}: so long as the specification of $\mathcal{G}$ is rich enough to make \cref{eq:pscore_in_G} hold, $\beta$ will identify a convex average of heterogeneous treatment effects. In \Cref{sec:solutions} we will derive another rationale for targeting $\beta$ in this model, showing that the weights $\lambda_{11}(W_i)$ minimize the semiparametric efficiency bound (conditional on the controls) for estimating some weighted-average treatment effect.

Our first proposition shows that with multiple treatments, the interpretation of
$\beta$ becomes more complicated because of contamination bias:
\begin{proposition}\label{theorem:main-result}\allowdisplaybreaks
  Under \Cref{ass:mean_indep,ass:model_or_design}, the treatment coefficients in~\eqref{eq:partially_linear} identify
  \begin{equation}\label{eq:betak_decomposition}
    \beta_{k}=E[\lambda_{kk}(W_{i})\tau_{k}(W_{i})]+
    \sum_{\ell\neq k}E[\lambda_{k\ell}(W_{i})\tau_{\ell}(W_{i})],
  \end{equation}
  where, recalling that $E^{*}[X_{ik}\mid X_{i,-k}, W_i]$ gives the projection of $X_{ik}$ onto the space
$\{X_{i,-k}'\tilde{\delta}+\tilde{g}(W_{i})
\colon \tilde{\delta}\in\mathbb{R}^{K-1}, \tilde{g}\in\mathcal{G}\}$,
  \begin{align*}
    \lambda_{kk}(W_{i})&=\frac{E[\dbtilde{X}_{ik}X_{ik}\mid W_{i}]}{E[\dbtilde{X}_{ik}^{2}]}
                         =\frac{p_{k}(W_{i})(1-E^{*}[X_{ik}\mid X_{i,-k}=0,W_{i}])}{E[\dbtilde{X}_{ik}^{2}]},
                         \qquad\text{and}\\
    \lambda_{k\ell}(W_{i})&=\frac{E[\dbtilde{X}_{ik}X_{i\ell}\mid W_{i}]}{E[\dbtilde{X}_{ik}^{2}]}=
                            -\frac{p_{\ell}(W_{i})E^{*}[X_{ik}\mid
                            X_{i\ell}=1,W_{i}]}{E[\dbtilde{X}_{ik}^{2}]}
  \end{align*}
  with $E[\lambda_{kk}(W_{i})]=1$ and
  $E[\lambda_{k\ell}(W_{i})]=0$. Furthermore, if \cref{eq:pscore_in_G} holds,
  $\lambda_{kk}(W_{i})\geq 0$.
\end{proposition}

\noindent \Cref{theorem:main-result} shows that the coefficient on $X_{ik}$ in
\cref{eq:partially_linear} is a sum of two terms. The first term is a weighted
average of conditional \acp{ATE} $\tau_{k}(W_i)$, with \emph{own treatment weights}
$\lambda_{kk}(W_i)$ that average to one---generalizing the characterization of
the single-treatment case, \cref{eq:efficient_weigthing_binary}. The expression
for $\lambda_{kk}$ implies that these weights are convex if the implicit linear
probability model used to compute $\dbtilde{X}_{ik}$ fits
probabilities that lie below one,
 $E^{*}[X_{ik}\mid X_{i,-k}=0,W_{i}]\leq 1$.
The second term is a weighted average of treatment effects for \emph{other}
treatments $\tau_{\ell}(W_i)$, with \emph{contamination weights} $\lambda_{k\ell}(W_i)$ that average
to zero. Because the contamination weights are zero on average, they must be
negative for some values of the controls unless they are all identically
zero.\footnote{\Cref{theorem:main-result} complements an algebraic result in
  \textcite[][, Section 7.1]{chatto_zubiz_2021}, which shows that the regression
  estimator of $\beta_k$ can be written in terms of weighted sample averages of
  outcomes among units in different treatment arms (regardless of whether
  \Cref{ass:mean_indep,ass:model_or_design} hold). In contrast, our analysis
  interprets regression \emph{estimands} in terms of weighted averages of
  conditional \acp{ATE} under a broad class of identifying assumptions. In a finite-population setting, \textcite{aaiw20} show that $\beta$ identifies matrix-weighted averages of individual treatment effect vectors $\tau_i$; however, they do not discuss the interpretation of the estimand.} This is
the case when the implicit linear probability model correctly predicts that
$X_{ik}=0$ if $X_{i\ell}=1$.

Hence, if the linear probability model is correctly specified, i.e.
$E[X_{ik}\mid {X}_{i,-k}, W_i]=X_{i, -k}'\alpha+g_{k}(W_i)$ for some vector
$\alpha$ and $g_{k}\in\mathcal{G}$, the contamination weights
$\lambda_{k\ell}(W_{i})$ are zero and the own treatment weights
$\lambda_{kk}(W_{i})$ are positive. This is the analog of
condition~\eqref{eq:pscore_in_G} if we interpret $X_{ik}$ as a binary treatment
of interest and $X_{i, -k}'\alpha+g_{k}(W_i)$ as a specification for the
controls. In other words, the assignment of treatment $k$ must be additively
separable between $X_{i,-k}$ and $W_{i}$. However, with mutually exclusive
treatments, this won't be the case unless treatment assignment is
unconditionally random. In particular, since $X_{ik}$ must equal zero if the
unit is assigned to one of the other treatments regardless of the value of
$W_i$, under correct specification it must be the case that $\alpha_\ell=-g_k(W_i)$ for all elements $\alpha_{\ell}$ of
$\alpha$. This in turn implies that the assignment of treatment $k$ doesn't depend on
$W_i$, which is impossible unless the propensity score $p_k(W_i)$ is
constant.

Thus, misspecification in the linear probability model will generally yield
nonsensical fitted probabilities $E^{*}[X_{ik}\mid X_{i\ell}=1,W_{i}]\neq 0$ that
generate non-zero contamination weights $\lambda_{k\ell}(W_{i})$. Furthermore,
if the misspecification also yields fitted probabilities
$E^{*}[X_{ik}\mid X_{i,-k}=0,W_{i}]> 1$, we will have negative own treatment
weights. The last part of \Cref{theorem:main-result} shows that such
nonsensible predictions are ruled out if \cref{eq:pscore_in_G} holds.

We make four further remarks on our general characterization of contamination bias:

\begin{remark}\label{remark:bias_magnitude}
  Since the contamination weights are mean zero, we may write the contamination
  bias term as
  $E[\lambda_{k\ell}(W_i)\tau_{\ell}(W_i)]=\cov(\lambda_{k\ell}(W_i), \tau_\ell(W_i))$.
  Thus, the treatment coefficient $\beta_{k}$ does not suffer from contamination
  bias if the contamination weights $\lambda_{k\ell}(W_i)$ are uncorrelated with
  the conditional \acp{ATE} $\tau_\ell(W_i)$. This is trivially true if the
  other treatments are homogeneous, i.e.\ when $\tau_{\ell}(W_i)=\tau_{\ell}$.
  More generally, contamination bias will be small if the contamination weight
  exhibits weak covariance with the conditional \acp{ATE}. Since
  $\cov(\lambda_{k\ell}(W_i), \tau_\ell(W_i))
  =\operatorname{cor}(\lambda_{k\ell}(W_i), \tau_\ell(W_i))\operatorname{sd}(\lambda_{k\ell}(W_i))\operatorname{sd}(\tau_\ell(W_i))$,
  this is the case when (i) the factors influencing treatment effect
  heterogeneity are largely unrelated to the factors influencing the treatment
  assignment process in the sense that
  $\operatorname{cor}(\lambda_{k\ell}(W_i), \tau_\ell(W_i))$ is close to zero,
  (ii) the contamination weights display limited variability, and/or (iii)
  treatment effect heterogeneity in the other treatments $\ell\neq k$ is
  limited.
\end{remark}

\begin{remark}
Since the weights in \cref{eq:betak_decomposition} are functions of the variances $E[\dbtilde{X}_{ik}^{2}]$ and covariances $E[\dbtilde{X}_{ik}X_{i\ell}]$ and $E[\dbtilde{X}_{ik}X_{ik}]$, they are identified and can be used to further characterize each $\beta_k$ coefficient. For example, the contamination bias term can be bounded by the identified contamination weights $\lambda_{k\ell}(W_i)$ and bounds on the heterogeneity in conditional \acp{ATE} $\tau_\ell(W_i)$.
\end{remark}

\begin{remark}
The results in \Cref{theorem:main-result} are stated for the case when $X_{i}$ are mutually exclusive treatment indicators. In \Cref{apdx:general} we relax this assumption to allow for combinations of non-mutually exclusive treatments (either discrete or continuous). In this case, the own-treatment weights $\lambda_{kk}(W_{i})$ may be negative even if \cref{eq:pscore_in_G} holds.
\end{remark}

\begin{remark}\label{remark:descriptive_bias}
While we derived \Cref{theorem:main-result} in the context of a causal model, an analogous result follows for descriptive regressions that do not assume potential outcomes or impose \Cref{ass:mean_indep}. Consider, specifically, the goal of estimating an average of conditional group contrasts $E[Y_i\mid D_i=k, W_i=w]-E[Y_i\mid D_i=0,W_i=w]$ with a partially linear model~\cref{eq:partially_linear} and replace condition~\eqref{eq:model_based} with an assumption that  $E[Y_i\mid D_i=0,W_i=w]\in\mathcal{G}$. The steps that lead to \Cref{theorem:main-result} then show that such regressions also generally suffer from contamination bias: the coefficient on a given group indicator averages the conditional contrasts across all other groups, with non-convex weights. Furthermore, the weights on own-group conditional contrasts are not necessarily positive. These sorts of conditional contrast comparisons are therefore not generally robust to misspecification of the conditional mean, $E[Y_i\mid D_i, W_i]$.
\end{remark}

\subsection{Implications}\label{sec:discussion_thm}

\Cref{theorem:main-result} shows that treatment effect heterogeneity can induce two conceptually distinct issues in flexible regression estimates of treatment effects. First, with either single or multiple treatments, there is a negative weighting of a treatment's \emph{own} effects when projecting the treatment indicator onto other treatment indicators and covariates yields fitted values exceeding one, i.e.\ when $E^{*}[X_{ik}\mid X_{i,-k}=0,W_{i}]>1$. This issue is relevant in various \ac{DiD} regressions and related approaches which rely on a model of untreated potential outcomes that ensures \cref{eq:model_based} holds (e.g.\ parallel trends assumptions) but which potentially misspecify the assignment model in \cref{eq:pscore_in_G}. Although the recent \ac{DiD} literature focuses on \ac{TWFE} regressions, \Cref{theorem:main-result} shows such negative weighing can arise more generally---such as when researchers allow for linear trends, interacted fixed effects, or other extensions of the basic parallel trends model. None of these alternative specifications for $g$ are in general flexible enough to capture the degenerate propensity scores and hence ensure that $E^{*}[X_{ik}\mid X_{i,-k}=0,W_{i}]\leq 1$.

Second, in the multiple treatment case, there is a potential for contamination bias from \emph{other} treatment effects---regardless of which condition in \Cref{ass:model_or_design} holds. This form of bias is relevant whenever one uses an additive covariate adjustment, no matter how flexibly the covariates are specified. Versions of this problem have been noted in, for example, the \textcite{sun2021estimating} analysis of \ac{DiD} regressions with treatment leads and lags or the \textcite{hull2018movers} analysis of mover regressions (see \Cref{apdx:did}).\footnote{The negative weights issue raised in \textcite{de2020aer} (when $K=1$), and the related issue that own-treatment weights may be negative in \textcite{sun2021estimating} and \textcite{de2020two} (when $K>1$), arise because the treatment probability is not linear in the unit and time effects. If \cref{eq:pscore_in_G} holds with $K=1$, \Cref{theorem:main-result} shows $\beta$ estimates a convex combination of treatment effects. This covers the setting considered in Theorem 1(iv) in \textcite{athey2022design}.  In their Comment 2, \textcite{athey2022design} say that ``the sum of the weights [used in Theorem 1(iv)] is one, although some of the weights may be negative''. \Cref{theorem:main-result} shows these weights are, in fact, non-negative.} \Cref{theorem:main-result} shows such contamination bias arises much more broadly, however.

The characterization in \Cref{theorem:main-result} also relates to concerns in interpreting multiple-treatment \ac{IV} estimates with heterogeneous effects \parencite{behaghel2013robustness,kirkeboen2016field,kline2016headstart,hull2018isolateing,leeMultivalued,bhuller_sigstad}. This connection comes from viewing \cref{eq:partially_linear} as the second stage of an \ac{IV} model estimated by a control function approach; in the linear \ac{IV} case, for example, $g(W_i)$ can be interpreted as giving the residuals from a first-stage regression of $X_i$ on a vector of valid instruments $Z_i$. In the single-treatment case, the resulting $\beta$ coefficient has an interpretation of a weighted average of conditional local average treatment effects under the appropriate first-stage monotonicity condition \parencite{ia94}. But as in \Cref{theorem:main-result} this interpretation fails to generalize when $X_i$ includes multiple mutually-exclusive treatment indicators: each $\beta_k$ combines the local effects of treatment $k$ with a non-convex average of the effects of other treatments.

Finally, \Cref{theorem:main-result} has implications for single-treatment \ac{IV} estimation with multiple instruments and flexible controls if the first stage has the form of \cref{eq:partially_linear}, where now $Y_i$ is interpreted as the treatment and $X_i$ gives the vector of instruments. \Cref{theorem:main-result} shows that the first-stage coefficients on the instruments $\beta_k$ will not generally be convex weighted average of the true first-stage effects $\tau_{ik}$. Because of this non-convexity, the regression specification may fail to satisfy the effective monotonicity condition even when $\tau_{ik}$ is always positive: the cross-instrument contamination of causal effects may cause monotonicity violations, even when specifications with individual instruments do not. This issue is distinct from previous concerns over monotonicity failures in multiple-instrument designs \parencite{mueller2015criminal, frandsen2019judging, norris2019examiner, mogstad2021identification}, which are generally also present in such just-identified specifications. It is also distinct from concerns about insufficient flexibility in the control specification when monotonicity holds unconditionally \parencite{blandhol2022tsls}.

This new monotonicity concern may be especially important in ``examiner'' \ac{IV} designs, which exploit the conditional random assignment to multiple decision-makers. Many studies leverage such variation by computing average examiner decision rates, often with a leave-one-out correction, and use this ``leniency'' measure as a single instrument with linear controls. These \ac{IV} estimators can be thought of as implementing versions of a jackknife \ac{IV} estimator \parencite{aik99}, based on a first stage that uses examiner indicators as instruments, similar to \cref{eq:partially_linear}. \Cref{theorem:main-result} thus raises a new concern with these \ac{IV} analyses when controls (such as time fixed effects) are needed to ensure ignorable treatment assignment.

\section{Solutions}\label{sec:solutions}

We now discuss three solutions to the contamination bias problem raised by
\Cref{theorem:main-result}, each targeting a distinct causal parameter. First,
in \Cref{sec:estim-aver-treatm}, we discuss estimation of unweighted \acp{ATE}.
The other two solutions target weighted averages of individual treatment effects
using an \acf{EW} scheme in that the weights minimize the semiparametric
efficiency bound for estimating weighted \acp{ATE} under homoskedasticity. In
the second solution, the weights are allowed to vary across treatments, while in
the third, they are constrained to be common across treatments. In
\Cref{sec:effic-weight} we characterize these estimation targets, while in
\Cref{sec:guidance} we discuss how to estimate them; we also outline our proposed
guidance to researchers in measuring contamination bias.

Implementing the first solution requires strong overlap (i.e.\ that treatment
propensity scores are bounded away from zero and one) while the other two
solutions require nonempty overlap, ruling out fully degenerate propensity
scores. Solutions allowing for degenerate propensity scores require either
targeting subpopulations of the treated or adding substantive restrictions on
conditional means of treated potential outcomes (beyond \cref{eq:model_based},
which only restricts untreated potential outcomes). We refer readers to
\textcite{de2020two, sun2021estimating,
  callaway2021difference,borusyak2021revisiting,wooldridge2021mundlak} for such
solutions in the context of \ac{DiD} regressions.

\subsection{Estimating Average Treatment Effects}\label{sec:estim-aver-treatm}

Many estimators exist for the \ac{ATE} of binary treatments---see
\textcite{imbens2009recent} and \textcite{abadie2018econometric} for reviews. Several of these approaches extend naturally to multiple treatments: including
matching on covariates or the propensity score, inverse propensity score
weighting, balancing weights, interacted regression, or doubly-robust
methods (see, among others, \textcite{cattaneo10}, \textcite{ReZu20}, \textcite{ChNeSi21}, and
\textcite{GrPi22}). Here we summarize the last two approaches.

For the interacted regression solution, we adapt the implementation for the binary treatment case
discussed in \textcite[Section 5.3]{imbens2009recent} to multiple treatments.
Specifically, consider the specification:
\begin{equation}\label{eq:partial_linear_wint}
  Y_{i} = X_i'\beta +q_{0}(W_i)+\sum_{k=1}^{K}X_{ik}\left(q_{k}(W_i)-E[q_{k}(W_i)]\right)+ \dot{U}_{i},
\end{equation}
where $q_{k}\in\mathcal{G}$, $k=0,\dotsc, K$ and we continue to define $\beta$ and the functions $q_k$ as minimizers of $E[\dot{U}_i^2]$. When $\mathcal{G}$ consists of linear functions, \cref{eq:partial_linear_wint} specifies a linear regression of $Y_i$ on $X_i$, $W_i$, a constant, and the interactions between each treatment indicator $X_{ik}$ and the demeaned control vector $W_i-E[W_i]$. Define $\mu_{k}(w)=E[Y_{i}(k)\mid W_{i}=w]$ for $k=0,\dotsc, K$, so that $\tau_{k}(w)=\mu_{k}(w)-\mu_{0}(w)$. If \Cref{ass:mean_indep} holds and $\mathcal{G}$ is furthermore rich enough to ensure $\mu_{k}\in\mathcal{G}$ for $k=0,\dotsc, K$ then $\beta=\tau$. Moreover, $q_{k}(w)=\tau_{k}(w)$ for $k=1,\dotsc, K$, such that the regression identifies both the unconditional and conditional \acp{ATE}.

The added interactions in \cref{eq:partial_linear_wint} ensure that each treatment coefficient $\beta_{k}$ is determined only by the outcomes in treatment arms with $D_{i}=0$ and $D_{i}=k$, avoiding the contamination bias in \Cref{theorem:main-result}. Demeaning the $q_k(W_i)$ in the interactions ensures they are appropriately centered to interpret the coefficients on the uninteracted $X_{ik}$ as \acp{ATE}.

Estimation of \cref{eq:partial_linear_wint} is conceptually straightforward for
parametric $q_{k}$. In particular, if $\mathcal{G}$ consists of linear
functions, one simply estimates
\begin{equation}\label{eq:ate_linear}
Y_{i}=\alpha_{0}+\sum_{k=1}^{K}X_{ik}\tau_{k}+W_{i}^\prime\alpha_{W, 0}+\sum_{k=1}^{K}X_{ik}(W_{i}-\overbar{W})^\prime\gamma_{W, k}+\dot{U}_i.
\end{equation}
by \ac{OLS}, where  $\overbar{W}=\frac{1}{N}\sum_i W_i$ is the sample average of the covariate
vector. More generally, to increase the plausibility of the key assumption that
$\mu_{k}\in\mathcal{G}$, one may constrain $\mathcal{G}$ only by nonparametric
smoothness assumptions. Given a sequence of basis functions
$\{b_{j}(W_{i})\}_{j=1}^{\infty}$, such as polynomials or splines, one then
approximates $q_{k}$ with a linear combination of the first $J$ terms, with $J$
increasing with the sample size, thus tailoring the model complexity to data
availability. Given a choice of $J$, estimation and inference can proceed
as in the parametric case; the only difference is that the baseline
covariates $W_{i}$ in \cref{eq:ate_linear} are replaced by the basis vector
$(b_{1}(W_{i}), \dotsc, b_{J}(W_{i}))'$ and $\overbar{W}$ is replaced by the
sample average of this expansion. This estimator
has been studied in the binary treatment case by \textcite{ChHoTa08} and \textcite{ImNeRi07}, with the latter
providing a detailed analysis of how to choose $J$ and the former showing that this
sieve estimator achieves the semiparametric efficiency bound under strong
overlap: it
is impossible to construct another regular estimator of the \ac{ATE} with
smaller asymptotic variance.

An attractive alternative approach combines the interacted regression with
inverse propensity score weighting. Instead of using \ac{OLS} to
estimate \cref{eq:partial_linear_wint} one uses weighted least squares, weighting observations by the inverse of some estimate $\hat{p}_{D_{i}}(W_{i})$ of the propensity score (see,
e.g., \textcite{robins1994estimation, wooldridge2007inverse,
  sloczynski2018general}). An advantage of this approach is that it is doubly-robust: the estimator is consistent so long as either the propensity score
estimator is consistent or the outcome model is correct (i.e.
$\mu_{k}\in\mathcal{G}$). A recent literature shows how the double robustness
property, when combined with cross-fitting, reduces the sensitivity of the ATE
estimate to overfitting or regularization bias in estimating the nuisance
functions $p_{k}$ and $\mu_{k}$. Cross-fitting also allows for using more flexible methods to
approximate $p_{k}$ and $\mu_{k}$, including modern machine learning methods
\parencite[see, e.g.][]{chernozhukov2018double,ceinr22,ChNeSi21}.

Either approach should work reliably in stratified \acp{RCT} and other settings
with strong overlap. But under weak overlap, when propensity scores are not
bounded away from zero and one, all of these \ac{ATE} estimators may be
imprecise and have poor finite-sample behavior. This is not a shortcoming of the
specific estimator; indeed, \textcite{KhTa10} show that under weak overlap,
$\sqrt{N}$-estimation of the \ac{ATE} is not possible. Furthermore, if some
propensity scores attain values of zero or one, the \ac{ATE} is not even
point-identified. These results formalize the intuition that it is difficult or
impossible to estimate the counterfactual outcomes for units with extreme
propensity scores.\footnote{One approach to limited overlap is trimming: i.e.,
  dropping observations with extreme propensity scores
  \parencite{crump2006dealing,crump2009dealing, yang2016propensity}. As with the
  estimators we derive next, trimming estimators shift the estimand from
  \ac{ATE} to easier-to-estimate weighted averages of conditional \acp{ATE}.}
Such extreme propensity scores are common in observational settings. The
solutions we discuss next downweight these difficult-to-estimate counterfactuals
to address this practical challenge.

\subsection{Easiest-to-Estimate Averages of Treatment Effects}\label{sec:effic-weight}

Suppose in a sample of observations $i=1, \dots, N$ we wish to estimate a
weighted average of conditional potential outcome contrasts
$\sum_{i=1}^{N}\lambda(W_{i})
\sum_{k=0}^{K}c_{k}\mu_{k}(W_{i})/\sum_{i=1}^{N}\lambda(W_{i})$, where
$\mu_{k}(W_{i})=E[Y_{i}(k)\mid W_{i}]$, $c$ is a $(K+1)$-dimensional contrast
vector with elements $c_{k}$, and $\lambda(W_i)$ is some weighting
scheme.\footnote{In a slight abuse of notation relative to \Cref{sec:problem},
  the weights $\lambda$ here are not required to average to one. Instead, we
  scale the estimand by the sum of the weights, $\sum_{i=1}^{N}\lambda(W_{i})$.}
We focus on two specifications for the contrast vector, leading to two
alternatives the \ac{ATE} target. First, for separately estimating the effect of
each treatment $k$, we set $c_{k}=1$, $c_{0}=-1$ and set the remaining entries
of $c$ to $0$. The contrast of interest then becomes
$\sum_{i=1}^{N}\lambda(W_{i})\tau_{k}(W_{i})/\sum_{i=1}^{N}\lambda(W_{i})$, the
weighted \ac{ATE} of treatment $k$. Second, we specify
$c$ so as to allow us to simultaneously contrast the effects of all $K$
treatments---we discuss this further below. For each contract vector $c$, we
characterize in this section the \acf{EW} scheme $\lambda(W_i)$ that leads to
the smallest possible standard errors under homoskedasticity. We discuss
estimation of the corresponding estimands in \Cref{sec:guidance}.

This optimization problem has four motivations. First, there is a robustness motivation:
a researcher would like to estimate a given contrast as precisely as possible,
at least under the benchmark of constant treatment effects, while being robust
to the possibility that the effects are heterogeneous. While the optimization
problem does not impose convexity, it turns out that the \ac{EW} scheme is
convex. Hence, the resulting estimand identifies a convex average of conditional
contrasts under heterogeneous treatment effects, and avoids any contamination
bias. Such a robustness property presumably underlies the popularity of
regression as a tool for estimating the effect of a binary treatment: the
regression estimator is efficient under homoskedasticity and constant treatment
effects while, by the \textcite{angrist98} result, retaining a causal
interpretation under heterogeneous
effects.\footnote{\label{fn:convex_motivation}There are several motivations for the interest in convex weights. First, $\lambda(W_i)\ge 0$ ensures
  the estimand captures average effects for \emph{some} well-defined (and
  characterizable) subpopulation. Second, it prevents what \textcite{strlb17}
  call a sign-reversal: if $\tau_k(w)$ has the same sign for all $w$ ($+,0$
  or $-$), then the estimand will also have this sign. %
  \textcite{blandhol2022tsls} call such estimands ``weakly causal.'' Finally,
  the estimand satisfies a population version of what \textcite{rslr07} call
  boundedness: the estimand lies in the support of $\tau_k(w)$.}

Second, the \ac{EW} scheme gives a bound on the information available in the data: if the scheme yields overly large standard errors, inference on other treatment effects (such as the unweighted \ac{ATE}) must be at least as uninformative. Computing the \ac{EW} standard errors thus reveals whether informative conclusions for \emph{any} treatment effect estimand are only possible under additional assumptions or with the aid of additional data.
In fact, we show below that in the binary treatment case the \ac{EW} scheme is
exactly the same as that used by regression. Recall that in the binary treatment
case, the regression treatment weights are proportional to the conditional
variance of treatment, $\var(D_i\mid W_i)=p_{1}(W_i)(1-p_{1}(W_i))$. Because
these weights tend to zero as $p_{1}(W_{i})$ tends to zero or one, regression
downweights observations with extreme propensity scores where the estimation of
counterfactual outcomes is difficult, avoiding the poor finite-sample behavior
of \ac{ATE} estimators under weak overlap and allowing for informative inference
even when one cannot precisely estimate the unweighted \ac{ATE}.

Third, the \ac{EW} scheme can be viewed as offering an intermediate point along
a particular robustness-precision ``possibility frontier.'' The \ac{ATE}
estimator based on the interacted specification in \cref{eq:partial_linear_wint}
lies on one end of this frontier, being the most robust to treatment effect
heterogeneity (i.e.\ retaining a clear interpretation regardless of the form of
$\tau(w)$ or how it relates to the propensity scores). But this robustness comes
at the cost of imprecision and non-standard inference under weak overlap. The
regression estimator based on \cref{eq:partially_linear} lies on the other end
of the frontier: it is likely to be precise even when overlap is weak (and is
efficient under homoskedasticity if the partly linear model in
\cref{eq:partially_linear} is correct, such that treatment effects are
constant). But this precision comes at the cost of contamination bias under
heterogeneous treatment effects. The \ac{EW} scheme lies in between these
extremes, purging contamination bias and retaining good performance under weak
overlap by giving up explicit control over the treatment effect weighting,
letting it be data-determined.\footnote{There are other approaches to resolving
  the robustness-precision tradeoff, such as seeking precise estimates subject
  to the weights $\lambda$ remaining ``close'' to one, or placing some
  restrictions on the form of effect heterogeneity, in contrast to leaving it
  completely unrestricted as we do here (see \textcite{mst18} for an example of
  this approach in an \ac{IV} setting). We leave these alternatives to future
  research.}

Finally, while the derivation of the \ac{EW} scheme is motivated by statistical precision concerns, the resulting estimand can be seen as identifying the impact of a policy that manipulates the treatment via a particular incremental propensity score intervention. We discuss this interpretation in \Cref{remark:policy_relevance} below.

We derive the \ac{EW} scheme in two steps. First, we establish a precision
benchmark---a semiparametric efficiency bound---for estimation of a given
weighted average of treatment effects under the idealized scenario that the
propensity score is known. Second, we determine which weights $\lambda$ minimize
the bound.

The following \namecref{theorem:seb} establishes the first step of our derivation:
\begin{proposition}\label{theorem:seb}
  Suppose \cref{eq:full_ci} holds in an i.i.d.\ sample of size $N$, with known non-degenerate propensity scores
  $p_{k}(W_{i})$. Let $\sigma^{2}_{k}(W_{i})=\var(Y_{i}(k)\mid W_{i})$. Consider the
  problem of estimating the weighted average of contrasts
  \begin{equation*}
    \theta_{\lambda, c}=\frac{1}{\sum_{i=1}^{N}\lambda(W_{i})}
    \sum_{i=1}^{N}\lambda(W_{i})\sum_{k=0}^{K}c_{k}\mu_{k}(W_{i}),
  \end{equation*}
  where the weighting function $\lambda$ and contrast vector $c$ are both known. Suppose the weighting function satisfies $E[\lambda(W_i)]\neq 0$, and that the second moments of $\lambda(W_i)$ and $\mu(W_i)$ are bounded. Then, conditional on the controls $W_1,\dots, W_N$, the semiparametric efficiency bound is almost-surely given by
  \begin{equation}\label{eq:seb}
    \mathcal{V}_{\lambda, c}=   \frac{1}{E[\lambda(W_{i})]^{2}}E\left[\sum_{k=0}^{K}\frac{\lambda(W_{i})^{2}c_{k}^{2}
        \sigma_{k}^{2}(W_{i})}{p_{k}(W_{i})}\right].
  \end{equation}
\end{proposition}

\noindent As formalized in the \Cref{sec:proof_p2} proof, $\mathcal{V}_{\lambda,
  c}$ establishes the lower bound on the asymptotic variance of any regular
estimator of $\theta_{\lambda, c}$ under the idealized case of known propensity
scores.\footnote{The efficiency bound for the population analog
  $\theta^*_{\lambda,
    c}=E[\lambda(W_i)\sum_{k=0}^{K}c_{k}\mu_{k}(W_{i})]/E[\lambda(W_i)]$ has an
  additional term,
  $E[\lambda(W_{i})^{2}(\sum_{k=0}^{K}c_{k}\mu_{k}(W_{i})-\theta_{\lambda,
    c}^{*})^{2}]/E[\lambda(W_{i})]^{2}$, reflecting the variability of the
  conditional average contrast. The variance-minimizing weights for
  $\theta^*_{\lambda, c}$ thus depend on the nature of treatment effect
  heterogeneity. By focusing on $\theta_{\lambda, c}$, we avoid this term, which
  allows us give the characterization in \cref{eq:lambdastar} without any assumptions about heterogeneity in treatment effects.}

To establish the second step, we minimize~\cref{eq:seb} over $\lambda$.
Simple algebra shows that the \ac{EW} scheme is (up to an arbitrary constant) given by
\begin{equation}\label{eq:lambdastar}
  \lambda^{*}_{c}(W_{i})= \left(\sum_{k=0}^{K}\frac{c_{k}^{2}\sigma^{2}_{k}(W_{i})}{p_{k}(W_{i})}\right)^{-1}.
\end{equation}
Observe that this scheme delivers convex weights, $\lambda^*_{c}\geq 0$, even
though convexity was not imposed in the optimization. Hence, there is no cost in
precision if we restrict attention to convex weighted averages of conditional
\acp{ATE}.

When the contrast vector is selected to estimate the weighted average effect of a particular treatment $k$, a corollary to \Cref{theorem:seb} is that regression weights are the easiest-to-estimate:
\begin{corollary}\label{cor:angrist}
For some $k\ge 1$, let $c^{k}$ be a vector with elements $c^{k}_{j}=1$ if $j=k$, $c^{k}_{j}=-1$ if $j=0$, and $c^{k}_{j}=0$ otherwise. Suppose that the conditional variance of relevant potential outcomes is homoskedastic: $\sigma^{2}_{k}(W_{i})=\sigma^{2}_{0}(W_{i})=\sigma^{2}$. Then the variance-minimizing weighting scheme is given by $\lambda^{*}_{c^k}=\lambda^{k}$, where
\begin{equation}\label{eq:single_contrast1}
  \lambda^{k}(W_i)=\frac{p_{0}(W_i)p_{k}(W_i)}{p_{0}(W_i)+p_{k}(W_i)}.
\end{equation}
\end{corollary}
\noindent Per \cref{eq:efficient_weigthing_binary}, the weighting $\lambda^k$
coincides with the weighting of conditional \acp{ATE} from the
partially linear model~\eqref{eq:partially_linear} when it is fit only on
observations with $D_{i}\in\{0,k\}$, provided
$p_{k}/(p_{k}+p_{0})\in\mathcal{G}$.\footnote{This follows since the propensity
  score in the subsample is given by
  $\Pr(D_{i}=k\mid W_{i},
  D_{i}\in\{0,k\})=\frac{p_{k}(W_{i})}{p_{0}(W_{i})+p_{k}(W_{i})}$, so that
  $\lambda^k(W_i)$ in \cref{eq:single_contrast1} equals the conditional variance
  of the treatment indicator times the probability of being in the subsample.}
\Cref{cor:angrist} thus gives a precision justification for estimating the
effect of any given treatment $k$ by a partially linear regression in the
subsample with $D_{i}\in\{0,k\}$ under a homoskedasticity benchmark,
complementing the robustness motivation discussed
earlier.\footnote{\label{fn:homoskedasticity_bench}As usual, homoskedasticity is
  a tractable baseline: the arguments in favor of \ac{OLS} following
  \Cref{cor:angrist} can be extended to favor a (feasible) weighted least
  squares regression when $\sigma^2(W_i)$ is consistently estimable.} To
estimate the effects of all treatments one can run $K$ such
one-treatment-at-a-time regressions, one for each treatment arm. Plugging
\cref{eq:single_contrast1} into \cref{eq:seb} reveals that the asymptotic
variance is bounded so long as the overlap between the covariate distribution in
each treatment arm is nonempty, i.e.
$P(p_{k}(W_{i})>\varepsilon\cap p_{0}(W_{i})>\varepsilon)>\varepsilon$ for some
$\varepsilon$.

For binary treatments, \textcite[Corollary 5.2]{crump2006dealing}
and \textcite[Corollary
1]{liBalancing} show that the weighting $p_1(W_i)(1-p_{1}(W_i))$ minimizes the
asymptotic variance of a particular class of inverse propensity score weighted
estimators. Our \Cref{cor:angrist} extends the property to all regular
estimators, and to multiple treatments.

\begin{remark}\label{remark:interacted_regression}
  The one-treatment-at-a-time regression can also be motivated as a direct
  solution to contamination bias in the partially linear regression
  in~\cref{eq:partially_linear}. In particular, as discussed in
  \Cref{sec:caus-interpr}, contamination bias arises because the implicit linear
  probability model $E^{*}[X_{ik}\mid X_{i,-k}, W_{i}]$ incorrectly imposes
  additive separability between $X_{i,-k}$ and $W_{i}$. To solve this issue, one
  can include interactions between the controls and $X_{i,-k}$. This is similar
  to the interacted regression in \cref{eq:partial_linear_wint}, except we
  exclude the interaction $X_{ik}(q_{k}(W_{i})-E[q_{k}(W_{i})])$. Simple algebra
  shows that this regression is equivalent to the one-treatment-at-a-time
  regression.
\end{remark}
\begin{remark}\label{remark:policy_relevance}
  The population analog of the estimand implied by the weighting in
  \Cref{cor:angrist}, $E[\lambda_k(W_i)\tau_k(W_i)]/E[\lambda_k(W_i)]$, also
  identifies the effect of a particular marginal policy intervention. Consider the
  effects of a class of policies indexed by a scalar $\delta$ that restrict
  treatments to $\{0,k\}$ by increasing the propensity score of treatment $k$ to
  $p_k^{\delta}(W_i)$ and setting
  $p_0^{\delta}(W_i)=1-p_k^{\delta}(W_i)$.\footnote{\label{fn:two_contrasts}With
    multiple treatments, policy relevance of any contrast only involving two
    treatments will generally require the policy to restrict the number of
    treatments to preclude flows in and out of multiple treatment states. For
    instance, the ATE gives the effect of comparing two policies: one makes only
    treatment $k$ available, while the other makes only treatment $0$
    available.} Then the marginal effect of the increasing the policy intensity
  $\delta$ per unit treated at $\delta=0$ is given by
  $E[\partial p_k^{\delta}(W_i)/\partial \delta\cdot \tau(W_i)]/E[\partial
  p_k^\delta(W_i)/\partial \delta]$ \parencite[see][for derivation and
  discussion]{ZhOp22}. Thus, the weights
  $\lambda_k(W_i)=\frac{p_{0}(W_i)p_{k}(W_i)}{p_{0}(W_i)+p_{k}(W_i)}$ identify
  the marginal policy effect when they correspond to the derivative
  $\partial p_k^\delta(W_i)/\partial \delta$.
  For example, \textcite{ZhOp22} show this holds for
  policies that increase the log odds of a single binary treatment by a constant
  $\delta$---such as by increasing the intercept in a logit model for treatment.
\end{remark}

A shortcoming of the \ac{EW} scheme in \Cref{cor:angrist} is that it is
treatment-specific, precluding comparisons of the
weighted-average effects across treatments.\footnote{Formally, for treatments
  $1$ and $2$, we estimate the weighted averages
  $\sum_{i}\lambda^{1}(W_i)\tau_{1}(W_i)/\sum_{i}\lambda^{1}(W_i)$ and
  $\sum_{i}\lambda^{2}(W_i)\tau_{2}(W_i)/\sum_{i}\lambda^{2}(W_i)$. Because the
  weights $\lambda^1$ and $\lambda^2$ differ, the difference between these
  estimands cannot generally be written as a convex combination of conditional
  treatment effects $\tau_{1}(W_i)-\tau_{2}(W_i)$. This critique also applies to the own-treatment weights in \Cref{theorem:main-result}. Thus even without contamination bias one may find the implicit multiple-treatment regression weighting deficient.}
This issue is especially salient when the control group is arbitrarily chosen, such as in teacher \ac{VAM} regressions which omit an arbitrary teacher from estimation and seek causal comparisons across all teachers.

We thus turn to the question of how \Cref{theorem:seb} can be used to select a
weighting scheme which allows for simultaneous comparisons across all treatment
arms. Suppose that the contrast of interest is drawn at random from a given
marginal treatment distribution $\Pr(D_i=k)=\pi_k$, so that $c_{j}=1$ with
probability $\pi_{j}(1-\pi_{j})/(1-\sum_{k=0}^{K}\pi_{k}^{2})$ and $c_{j}=-1$
with the same probability.\footnote{Formally, we draw two treatments at random
  from the given marginal distribution, discarding the draw if the two
  treatments are equal.} Let $F_\pi$ denote this distribution over the (now
random) contrasts. If the researcher wishes to report an accurate contrast
estimate but needs to commit to a weighting scheme before knowing the contrast
of interest, it is optimal to minimize the expected variance
\begin{equation*}
  \int \mathcal{V}_{\lambda, c} dF_\pi(c)
= \frac{1}{E[\lambda(W_{i})]^{2}(1-\sum_{k=0}^{K}\pi_{k}^{2})}
  \sum_{k=0}^{K}E\left[\frac{\lambda(W_{i})^{2}2\pi_{k}(1-\pi_{k})\sigma_{k}^{2}(W_{i})}{p_{k}(W_{i})}\right].
\end{equation*}
Minimizing this expression over $\lambda$ is equivalent to minimizing
\cref{eq:seb} with $c_{k}^{2}=2\pi_k(1-\pi_k)$, which yields
\cref{eq:lambdastar} with this contrast specification as the optimal weighting.
Thus, the optimal weights are proportional to
$\left(\sum_{k=0}^{K}\frac{\pi_{k}(1-\pi_{k})\sigma^{2}_{k}(W_{i})}{p_{k}(W_{i})}\right)^{-1}$.
Specializing to the homoskedastic case leads to the following result:
\begin{corollary}\label{theorem:F-weighting}
  Let $F_\pi$ denote the distribution over possible contrast vectors such that $P_{F_\pi}(c_k=1)=P_{F_\pi}(c_k=-1)=\pi_{j}(1-\pi_{j})/(1-\sum_{k=0}^{K}\pi_{k}^{2})$. Suppose that $\sigma^{2}_{k}(W_{i})=\sigma^{2}$ for all $k$. Then the
  weighting scheme minimizing the average variance bound
  $\int \mathcal{V}_{\lambda, c} dF_\pi(c)$ is given by:
  \begin{equation*}
    \lcw(W_{i})=\left(\sum_{k=0}^{K}\frac{\pi_{k}(1-\pi_{k})}{p_{k}(W_{i})}\right)^{-1}.
  \end{equation*}
\end{corollary}
\noindent The \ac{CW} scheme $\lcw$ generalizes the intuition behind the single
binary treatment (\Cref{cor:angrist}), placing lower weight on strata with
extreme propensity scores. When the treatment is binary, $K=1$, the $\pi_{k}$'s
do not matter and the \ac{CW} scheme reduces to that in~\Cref{cor:angrist}:
$\lcw(W_{i})=\lambda^{1}(W_{i})=\lambda^{0}(W_{i})=p_{1}(W_{i})p_{0}(W_{i})$.
With multiple treatments, however, the weights $\lcw$ remain the same for every
treatment---allowing for simultaneous comparisons across all treatment pairs
$(k, \ell)$.

There are two natural choices for the marginal treatment probabilities $\pi$.
First, when equally interested in all contrasts, one can set
$\pi_k=1/(K+1)$. This weighting scheme was previously proposed by
\textcite{lili19}; our characterization of it in terms of optimizing a semiparametric
efficiency bound is, to our knowledge, novel. Second, if more common treatments
are of greater interest, we may set $\pi_k$ to the empirical treatment
probabilities $N^{-1}\sum_i X_{ik}$. This weighting targets precise estimation
of contrasts involving more common treatments at the expense of contrasts
involving less common treatments. We use this choice in our empirical
applications in \Cref{sec:applications}. For either choice of weights, the
resulting asymptotic variance in \cref{eq:seb} remains bounded so long as the
overlap between covariate distributions in each treatment arm is not empty:
$P(\cap_{k=0}^{K}p_{k}(W_{i})>\varepsilon)>\varepsilon$ for some $\varepsilon$.
Non-empty overlap is a substantially weaker assumption than strong overlap, needed for
$\sqrt{N}$-estimation of the unweighted \ac{ATE}, which requires this
probability to equal one. For instance, in the nine empirical applications below, non-empty overlap always holds, but strong overlap fails in six.

\subsection{Practical Guidance in Measuring and Avoiding Contamination Bias}\label{sec:guidance}

A researcher interested in estimating the effects of multiple mutually exclusive
treatments with regression can use \Cref{theorem:main-result} to measure the
extent of contamination bias in their estimates. When the propensity score is
not fully degenerate, they can further estimate one of the alternative
estimation targets discussed in the previous subsections. Here we provide
practical guidance on both procedures, which we illustrate empirically in the
next section.

For simplicity, we focus on the case where $g$ is linear and \cref{eq:partially_linear} is estimated by \ac{OLS}. We suppose \Cref{ass:mean_indep} and both conditions in \Cref{ass:model_or_design} hold, such that all propensity scores $p_k$ and potential outcome conditional expectation functions $\mu_k$ are linearly spanned by the controls $W_i$. These conditions hold, for example, when $W_i$ contains a set of mutually exclusive group indicators. When $\mathcal{G}$ is unrestricted, the recommendations in this section would require non-parametric approximations for $g$ analogous to those discussed in \Cref{sec:estim-aver-treatm}.

Under this setup, we can decompose the \ac{OLS} estimator $\hat{\beta}$ from the uninteracted regression
\begin{equation}\label{eq:uninteracted_linear}
    Y_i=\alpha + \sum_{k=1}^{K} X_{ik}\beta_k + W_i'\gamma +U_i,
\end{equation}
to obtain a sample analog of the decomposition in \Cref{theorem:main-result}. To
this end, note that the own-treatment and contamination bias weights in
\Cref{theorem:main-result} are identified by the linear regression of $X_i$ on
the residuals $\tilde{X}_i$. Specifically, $\lambda_{k\ell}(W_{i})$ is given by
the $(k, \ell)$th element of the $K\times K$ matrix $\Lambda(W_{i})=  E[\tilde{X}_{i}\tilde{X}_{i}']^{-1}E[\tilde{X}_{i}X_{i}'\mid W_{i}]$,
which can be estimated by its sample analog
  $\hat{\Lambda}_{i}=(\dot{X}'\dot{X})^{-1}\dot{X}_{i}X_{i}',$
where $\dot{X}_{i}$ is the sample residual from an \ac{OLS} regression of $X_{i}$ on $W_{i}$ and a constant and $\dot{X}$ is a matrix collecting these sample residuals. The $(k, \ell)$th element of $\hat\Lambda_i$ estimates the weight that observation $i$ puts on the $\ell$th treatment effect in the $k$th treatment coefficient. For $k=\ell$ this is an estimate of the own-treatment weight in \Cref{theorem:main-result}; for $k\neq\ell$ this is an estimate of a contamination weight.

Under linearity, the $k$th conditional \ac{ATE} may be written as $\tau_k(W_{i})=\gamma_{0,k}+W_{i}^\prime\gamma_{W, k}$, where $\gamma_{0,k}$ and $\gamma_{W, k}$ are coefficients in the interacted regression specification
\begin{equation}\label{eq:linear_gamma}
    Y_i=\alpha_{0}+\sum_{k=1}^{K}X_{ik}\gamma_{0, k}+W_{i}^\prime\alpha_{W, 0}+\sum_{k=1}^{K}X_{ik}W_{i}^\prime\gamma_{W, k}+\dot{U}_i.
\end{equation}
Estimating \cref{eq:linear_gamma} by \ac{OLS} yields estimates $\hat{\tau}_k(W_i)=\hat\gamma_{0,k}+W_{i}'\hat\gamma_{W, k}$. For each observation $i$, we stack the set of conditional \ac{ATE} estimates in a $K\times 1$ vector $\hat\tau (W_i)$.

Using the \ac{OLS} normal equations, we then obtain a sample analog of the population decomposition in \Cref{theorem:main-result}:
\begin{equation}\label{eq:beta_hat}
    \hat{\beta}= \sum_{i=1}^{N}\diag(\hat{\Lambda}_{i})\hat{\tau}(W_{i}) +
    \sum_{i=1}^{N}[\hat{\Lambda}_{i}-\diag(\hat{\Lambda}_{i})] \hat{\tau}(W_{i}).
\end{equation}
The first term estimates the own-treatment effect components, $E[\lambda_{kk}(W_{i})\tau_{k}(W_{i})]$, while the second term estimates the contamination bias components, $\sum_{\ell\neq k}E[\lambda_{k\ell}(W_{i})\tau_{\ell}(W_{i})]$. If the contamination bias term is large for some $\hat\beta_k$, it suggests the estimate of the $k$th treatment effect is substantially impacted by the effects of other treatments. Researchers can also compare the first term of \cref{eq:beta_hat} to other weighted averages of own-treatment effects, including the ones discussed next, to gauge the impact of the regression weighting $\diag(\hat{\Lambda}_{i})$.\footnote{When the covariates are not saturated, it is possible that the estimated weighting function $\hat{\Lambda}(w)=\frac{1}{N}\sum_{i=1}^{N}\1{W_{i}=w}\hat{\Lambda}_{i}$ is not positive-definite for some or all $w$. In particular, the diagonal elements of $\hat{\Lambda}(w)$ need not all be positive. However, it is guaranteed that the diagonal of $\hat{\Lambda}(w)$ sums to one and the non-diagonal weights sum to zero, since $\sum_{i=1}^{N}\hat{\Lambda}_{i}=I_{k}$.}

Further analysis of the estimated weights $\hat\lambda_{k\ell}(w)=\frac{\sum_{i=1}^{N}\1{W_i=w}\hat{\Lambda}_{i, k\ell}}{\sum_{i=1}^{N}\1{W_i=w}}$ can shed more light on the regression estimates in $\hat\beta$. For example, the contamination weights for $\ell\neq k$ can be plotted against the treatment effect estimates $\hat\tau_\ell(W_i)$ to visually assess the sources of contamination bias. Low bias may arise from limited treatment effect heterogeneity, small contamination weights, or a low correlation between the two.

Estimation of the  unweighted \ac{ATE} and the \ac{EW} and \ac{CW} schemes is also straightforward under the linearity assumptions. First, estimating \cref{eq:ate_linear} by \ac{OLS} yields estimates of the unweighted \acp{ATE} $\tau_k=E[\tau_k(W_i)]$. The estimates are numerically equivalent to $\hat{\tau}_k=\hat\gamma_{0,k}+\overbar{W}^\prime\hat\gamma_{W, k}$, where $\hat\gamma_{0,k}$ and $\hat\gamma_{W, k}$ are \ac{OLS} estimates of \cref{eq:linear_gamma}.

Second, the \ac{EW} scheme from
\Cref{cor:angrist} can be estimated using the uninteracted one-treatment-at-a-time regression
\begin{equation}\label{eq:1-at-a-time}
      Y_{i}=\ddot{\alpha}_{k}+X_{ik}\ddot{\beta}_{k}+W_{i}^\prime\ddot{\gamma}_{k}+
      \ddot{U}_{ik},
\end{equation}
where we only use observations assigned either to treatment $k$ or the control group.

The third solution is to estimate the \ac{CW} scheme $\lcw$ from
\Cref{theorem:F-weighting}. We use inverse propensity
score weighting in our applications below: we regress $Y_{i}$ onto $X_{i}$ and a constant,
weighting each observation by $\hlcw(W_i)/\hat{p}_{D_i}(W_i)$ where
$\hat{p}_{k}(W_{i})$ denotes estimated propensity scores from a multinomial
logit model and
\begin{equation}\label{eq:common_weight_estimate}
  \hlcw(W_{i})=
  \left(\sum_{k=0}^{K}\frac{\pi_{k}(1-\pi_{k})}{\hat{p}_{k}(W_{i})}\right)^{-1}
\end{equation}
is an estimate of $\lcw$. When the weights $\pi$ are uniform, this estimator
reduces to the estimator studied in \textcite{lili19}. The resulting estimator
can be written as
\begin{equation}\label{eq:weighted_regression}
  \hat{\beta}_{\hlcw, k}=
  \frac{1}{\sum_{i=1}^{N}\frac{\hlcw(W_{i})}{\hat{p}_{k}(W_{i})}X_{ik}}
  \sum_{i=1}^{N}\frac{\hlcw(W_{i})}{\hat{p}_{k}(W_{i})}X_{ik}Y_{i}
  -  \frac{1}{\sum_{i=1}^{N}\frac{\hlcw(W_{i})}{\hat{p}_{0}(W_{i})}X_{i0}}
  \sum_{i=1}^{N}\frac{\hlcw(W_{i})}{\hat{p}_{0}(W_{i})}X_{i0}Y_{i}.
\end{equation}
When the treatment is binary and $\hat{p}_{k}$ is obtained via a linear
regression, this weighted regression estimator coincides with the usual
(unweighted) regression estimator that regresses $Y_{i}$ onto $D_{i}$ and
$W_{i}$.\footnote{To see this, note that in this case
  $\hat{\lambda}(W_{i})=\hat{p}_{1}(W_{i})\hat{p}_{0}(W_{i})$, so that
  $\hat{\beta}_{\hlcw, 1}=
  \frac{\sum_{i=1}^{N}(1-\hat{p}_{1}(W_{i}))D_{i}Y_{i}}{
    \sum_{i=1}^{N}(1-\hat{p}_{1}(W_{i}))D_{i}} -
  \frac{\sum_{i=1}^{N}\hat{p}_{1}(W_{i})(1-D_{i})Y_{i}}{\sum_{i=1}^{N}\hat{p}_{1}(W_{i})
    (1-D_{i})} = \frac{\sum_{i=1}^{N}(D_{i}-\hat{p}_{1}(W_{i}))Y_{i}}{
    \sum_{i=1}^{N}(D_{{i}}-\hat{p}_{1}(W_{i}))^{2}} $, where the second equality
  uses the least-squares normal equations
  $\sum_{i=1}^{N}X_{i1}=\sum_{i=1}^{N}\hat{p}_1(W_i)$ and
  $\sum_{i}X_{i1}\hat{p}_1(W_i)=\sum_{i=1}^{N}\hat{p}_1(W_i)^2$.}
\Cref{theorem:efficient_estimation} in \Cref{apdx:proofs} shows that the
estimator $\hat{\beta}_{\hlcw}$ is efficient in the sense that it achieves the
semiparametric efficiency bound for estimating $\beta_{\lcw}=\sum_i \lcw(W_i)\tau(W_i)/\sum_i \lcw(W_i)$.
\begin{remark}\label{remark:doubly_robust_cw}
  The estimator $\hat{\beta}_{\hlcw}$ is justified by a parametric model for the
  propensity score. In order to
  guard against misspecification of the propensity score, mirroring the discussion in \Cref{sec:estim-aver-treatm}, it may be attractive
  to instead use a doubly robust version of this estimator that combines
  propensity score weighting with a regression adjustment using an estimate of
  $\mu_{k}$. Another approach is a weighted version of the approach of \textcite{ReZu20}, in which the observations are weighted by $\hlcw$ multiplied by balancing weights (instead of
  the inverse estimated propensity score).\footnote{Under propensity score
    misspecification, $\hlcw$ would generally converge to a probability limit
    $\tilde{\lambda}^{\textnormal{CW}}$ that may be different from $\lcw$. Both
    of these alternative approaches would estimate a weighted average of
    \acp{ATE} weighted by $\tilde{\lambda}^{\textnormal{CW}}$ in this case.} We leave detailed study of
  these approaches to future research.
\end{remark}
\begin{remark}\label{remark:ate_vs_wate}
  Under homoskedasticity, the second and third solutions yield estimates with
  smaller asymptotic variance than the estimator of the unweighted \ac{ATE}.
  These gains in precision are achieved by changing the estimand to a different
  convex average of conditional treatment effects. In particular, covariate
  values $w$ where the propensity score $p_{k}(w)$ is close to zero for some $k$
  will be effectively discarded. In practice, explicitly plotting the treatment
  weights $\lcw$ and $\lambda^{k}$ may help to identify the types of individuals
  who are downweighted by these solutions, and to assess the variation in these
  weights. Plotting them against treatment effect estimates $\hat{\tau}_{k}$ can
  help visually assess the extent to which differences in weighting schemes
  drive differences in between estimates. In particular, the difference between
  the ATE and any weighted \ac{ATE} estimand of the effect of treatment $k$ with
  weights ${\lambda}(W_i)$, normalized such that $E[\lambda(W_i)]=1$ is given by
  $E[\lambda(W_i)\tau_k(W_i)]-E[\tau_k(W_i)] =
  E[\lambda(W_i)\tau_k(W_i)]-E[\lambda(W_i)]E[\tau_k(W_i)]=\cov(\lambda(W_i),
  \tau_{k}(W_i))$. Thus, if the \emph{own} treatment weights $\lambda$ display
  only a weak covariance with \emph{own} treatment effect, the weighting will
  have little effect on the estimand. This is analogous to the observation in
  \Cref{remark:bias_magnitude} that contamination bias reflects the covariance
  between the contamination weights and treatment effects of the \emph{other}
  treatments.
\end{remark}

\section{Applications}\label{sec:applications}

\subsection{Project STAR Application}\label{sec:star}

We first illustrate our framework for analyzing and addressing contamination bias with data from Project STAR, as studied in \textcite{krueger1999experimental}.
The Project STAR \ac{RCT} randomized 11,600 students in 79 public Tennessee elementary schools to one of three types of classes: regular-sized (20--25 students), small (target size 13--17 students), or regular-sized with a teaching aide. The proportion of students randomized to the small class size and teaching aide treatment varied over schools, due to school size and other constraints on classroom organization. Students entering kindergarten in the 1985--1986 school year participated in the experiment through the third grade. Other students entering a participating school in grades 1--3 during these years were similarly randomized between the three class types. We focus on kindergarten effects, where differential attrition and other complications with the experimental analysis are minimal.\footnote{Students in regular-sized classes were randomly reassigned between classrooms with and without a teaching aide after kindergarten, complicating the interpretation of the aide effect in later grades. The randomization of students entering the sample after kindergarten was also complicated by the uneven availability of slots in small and regular-sized classes \parencite{krueger1999experimental}.}

\begin{table}
\begin{threeparttable}
\caption{Project STAR contamination bias and treatment effect estimates}\label{tab:contamination}
\begin{tabular}{@{}l SSSSS@{}}
\toprule
  &\multicolumn{5}{@{}l@{}}{A\@. Treatment effect estimates}\\
  \cmidrule{2-6}
  & {$\hat{\beta}$} & {Own} & {\ac{ATE}} & {\ac{EW}} & {\ac{CW}}\\
  & {(1)} & {(2)} & {(3)} & {(4)} & {(5)}\\
  \cmidrule(lr){2-6}
  \input{./StarA.tex}\\
  \midrule
  &\multicolumn{3}{@{}l}{B\@. Contamination bias estimates}\\
  \cmidrule{2-4}
  & &\multicolumn{2}{c}{Worst-Case Bias}\\
  \cmidrule(lr){3-4}
  & \multicolumn{1}{c}{Bias} & \multicolumn{1}{c}{Negative} & \multicolumn{1}{c}{Positive}\\
  & \multicolumn{1}{c}{(1)} & \multicolumn{1}{c}{(2)} & \multicolumn{1}{c}{(3)}\\
  \midrule
  \input{./StarB_text.tex}\\
  \bottomrule
\end{tabular}
  \begin{tablenotes}
  \item\footnotesize \emph{Notes:} Panel A gives estimates of small class and
    teaching aide treatment effects for the Project STAR kindergarten analysis.
    Col.~1 reports estimates from a partially linear model in
    \cref{eq:uninteracted_linear}, col.~2 reports the own-treatment component of
    the decomposition in \cref{eq:beta_hat}, col.~3 reports the interacted
    regression estimates based on~\cref{eq:ate_linear}, col.~4 reports estimates
    based on the \ac{EW} scheme using one-treatment-at-a-time regressions in
    \cref{eq:1-at-a-time}, and col~5 uses the \ac{CW} scheme based on
    \cref{eq:common_weight_estimate}. Panel B gives the contamination bias
    component of the decomposition in \cref{eq:beta_hat} in col.~1, while
    cols.~2 and 3 reports the smallest (largest) possible contamination bias
    from reordering the conditional \acp{ATE} to be as negatively (positively)
    correlated with the cross-treatment weights as possible. Robust standard
    errors are reported in parentheses. Robust standard errors that assume the
    propensity scores are known are reported in square brackets.
  \end{tablenotes}
\end{threeparttable}
\end{table}

Column 1 of Panel A in \Cref{tab:contamination} reports estimates of kindergarten treatment effects in a sample of 5,868 students initially randomized to the small class size and teaching aide treatments. Specifically, we estimate the partially linear regression (\cref{eq:uninteracted_linear}) where $Y_i$ is student $i$'s test score achievement at the end of kindergarten, $X_{i}=(X_{i1}, X_{i2})$ are indicators for the initial experimental assignment to a small kindergarten class and a regular-sized class with a teaching aide, respectively, and $W_i$ is a vector of school fixed effects. We follow \textcite{krueger1999experimental} in computing $Y_i$ as the average percentile of student $i$'s math, reading, and word recognition score on the Stanford Achievement Test in the experimental sample. As in the original analysis \parencite[column 6 of Table V, panel A]{krueger1999experimental}, we obtain a small class size effect of 5.36 with a heteroskedasticity-robust standard error of 0.78 and a teaching aide effect of 0.18 (standard error: 0.72).\footnote{Our sample and estimates are very similar to---but not exactly the same as---those in \textcite{krueger1999experimental}. We use heteroskedasticity-robust (non-clustered) standard errors throughout this analysis, since the randomization of students to classrooms is at the individual level.}

As discussed in \Cref{sec:example}, treatment assignment probabilities vary
across the schools indicated by the fixed effects in $W_i$. If treatment effects
also vary across schools in a way that covaries with the contamination weights
$\lambda_{k\ell}(W_i)$, we expect the estimated effect of small class sizes to
be partly contaminated by the effect of teaching aides (and vice versa). Panel B
reports the contamination bias part of the decomposition in \cref{eq:beta_hat},
which appears minimal for both treatment arms.

It is useful to decompose the contamination bias further into the standard
deviation of the school-specific treatment effect $\tau_{\ell}(W_{i})$, standard
deviation of the contamination weights, and their correlation, as discussed in
\Cref{remark:bias_magnitude}. \Cref{fig:contamination_weights} in
\Cref{apx:additional_fig} does this graphically, plotting estimates of the
school-specific treatment effects $\tau_\ell(W_i)$ against the contamination
weights $\lambda_{k\ell}(W_i)$ for $\ell\neq k$. As can be seen from
\Cref{fig:contamination_weights}, the variability of school-specific treatment effects is substantial: Adjusting for estimation error, we estimate the standard
deviation of $\tau_k(W_i)$ to be
11.0 for the small class treatment and of 9.1 for the aide
treatment.\footnote{We adjust for estimation error by subtracting the average
  squared standard error from the empirical variance of the treatment effect
  estimates and taking the square root.} Both standard deviations are an order
of magnitude larger than the standard errors in \Cref{tab:contamination}. On the
other hand, the standard deviations for the contamination weights for the small
class and aide treatment are only moderate: $0.14$ and $0.11$, respectively.
Moreover, the correlation between the conditional treatment effects and the
contamination weights is weak:  $0.10$ for the small class effect
estimate and $-0.13$ for the aide effect estimate. The moderate variation in
the contamination weights coupled with weak correlation between the weights and
the treatment effects explains why the contamination bias is small, even though
the treatment effects vary substantially across schools.

Had the experimental design been such that the contamination weights strongly
correlate with the treatment effects, sizable contamination bias could have resulted.
To illustrate this, we compute worst-case (positive and negative) weighted
averages of the estimated $\tau_\ell(W_i)$ by re-ordering them across the computed
cross-treatment weights $\lambda_{k\ell}(W_i)$. This exercise highlights
potential scenarios in which the randomization strata happened to have been
highly correlated with the effect heterogeneity. Columns 2 and 3
in panel B of \Cref{tab:contamination} show that both bounds on possible
contamination bias are an order of magnitude larger than the actual
contamination bias: $[-1.65,1.67]$ for the small class size treatment and
$[-1.53,1.53]$ for the teaching aide treatment.\footnote{The point estimates and
  standard errors in columns 4 and 5 in \Cref{tab:contamination} do not account
  for the fact that the re-ordering is based on estimates of $\tau_{k}(W_i)$
  rather than the true treatment effects. This biases the reported estimates
  away from zero, so that they give an upper bound for the worst-case contamination
  bias.}
Overall, for both treatments, the underlying heterogeneity in this setting makes
substantial contamination bias possible even though actual contamination bias
turns out to be relatively small.

Columns 2--5 of panel A report four treatment effect estimates that are free of
contamination bias. Column 2 gives the own-treatment effect component of the
decomposition in \cref{eq:beta_hat}, netting out the contamination bias estimate
from column 1. This doubles the teaching aide effect estimate, from 0.18 to
0.36, but the estimate remains statistically insignificant with standard errors
of around 0.71; the small classroom estimate moves very little. The remaining
columns report the three solutions to contamination bias discussed in
\Cref{sec:solutions}. Column 3 estimates the unweighted \acp{ATE} of the small
class size and teaching aide treatment, by estimating the interacted regression
specification in \cref{eq:ate_linear}. Column 4 estimates the
one-treatment-at-a-time regressions in \cref{eq:1-at-a-time} for $k=1,2$.
Finally, column 5 runs a weighted regression of $Y_i$ onto $X_i$ using the
\ac{CW} scheme in \cref{eq:common_weight_estimate}.

There turns out to be little difference between these alternative estimates. The small
class size effect varies between 5.2 and 5.6, which is close to the original
estimate. The teaching aide effect varies between 0.01 and 0.26. To understand
this lack of variation, recall from \Cref{remark:ate_vs_wate} that the difference between the unweighted \ac{ATE}
and an estimand that uses weights $\lambda(W_{i})$ is given by the covariance between $\lambda(W_{i})$ and the
conditional \acp{ATE} $\tau_k(W_{i})$. Given the sizable variability in the treatment effect estimates, the covariance will be small only if the correlation
between the weights and the treatment effects is small and if the weights
display limited variability. This turns out to be the case here, as depicted
graphically in \Cref{fig:own_weights} in \Cref{apx:additional_fig}. The figure
shows that the correlations fall below $0.25$ in absolute value for all
weighting schemes, and that the weights only vary between 0.7 and 1.2.

As a consequence of strong overlap, the standard errors
are similar across the columns. Indeed, the efficiency gain of the \ac{EW} scheme relative to the \ac{ATE} based on an efficiency bound comparison using \cref{eq:seb} with $\lambda=\lambda^k$ vs $\lambda=1$ is less than $1.6\%$ for both treatments under homoskedasticity; the gain is even smaller under the \ac{CW} scheme. The reported standard errors, which allow for heteroskedasticity and don't assume known propensity scores, align with this prediction.\footnote{The standard errors reported in
  parentheses in Panel B are valid for the population analogs $\beta_k$ and
  $\beta_{\lcw}$, i.e. $E[\lambda^k(W_i)\tau_k(W_i)]/E[\lambda^k(W_i)]$ and
  $E[\lcw(W_{i})\tau_{k}(W_{i})]/E[\lcw(W_{i})]$. Since these standard errors
  are potentially conservative when viewed as standard errors for $\beta_k$ and
  $\beta_{\lcw}$, the standard error comparison gives an upper bound on the cost
  to estimating the weights.} As discussed in \Cref{remark:prop_score_cost} in
\Cref{sec:effic-cw-estim}, these standard errors are affected by the assumption of known propensity scores, used to derive the weighting schemes underlying the estimates in columns 2 and 3. To gauge the impact of this assumption, we also
report a version of the standard errors computed under the assumption that the
sample treatment probabilities in each school match the true propensity
scores. This changes the standard errors little, showing that there is
minimal cost to estimating the weights.

\subsection{Further Applications}\label{sec:further_application}
We next study the broader relevance of contamination bias using data from eight additional studies with multiple-treatment regressions. These studies were identified by a systematic search of papers in the AEA Data and Code Repository from 2013--2022 (see \Cref{sec:article_search_procedure} for details). Five studies are experiments like Project STAR\@; the remaining three use observational regressions to estimate racial disparities across multiple race groups (which we interpret as descriptive, following \Cref{remark:descriptive_bias}).\footnote{\label{fn:why_race}We focus on observational studies of racial disparities as they often include regressions on multiple minority race ``treatments,'' use publicly available data, and are easily identifiable by a keyword search.} We replicate a single representative specification for each paper, corresponding to the first relevant regression discussed in the paper's introduction.\footnote{``Relevant'' here means a multiple-treatment regression specification with controls, where at least one treatment coefficient was statistically significant. The introduction in  \textcite{cole2013barriers} did not discuss any relevant specifications; we instead pick the first specification with variation in treatment probabilities across strata where our results would be most relevant.} \Cref{tab:paper_list} lists the papers and specifications.

\begin{table}
\begin{threeparttable}
  \caption{Further Applications}\label{tab:paper_list}
    \begin{tabular}{l@{}llcrrl@{}}
      \toprule
      &&&&\multicolumn{2}{c}{Sample size}\\
      \cmidrule(rl){5-6}
       & Journal & Type &  Spec. & Original & Overlap & $\operatorname{sd}(\hat{p}(W))$ \\
      Paper
      & \multicolumn{1}{c}{(1)}
        & \multicolumn{1}{c}{(2)} & \multicolumn{1}{c}{(3)}
                      & \multicolumn{1}{c}{(4)} & \multicolumn{1}{c}{(5)}& \multicolumn{1}{c}{(6)} \\
      \midrule
      \textcite{benhassine2015}\hspace{1ex} & AEJ:AE & Exp. & 5(1) & 11,074 & 6,996 & $0.14^{\dagger}$ \\
      \textcite{cole2013barriers} & AEJ:AE & Exp. & 7(6) & 132 & 73 & $0.10^\dagger$ \\
      \textcite{demel2013} & AEJ:AE & Exp. &2(2) & 520 & 520 & 0.02\\
      \textcite{drexler2014} & AEJ:AE & Exp. & 2(2) & 796 & 796 & 0.05 \\
      \textcite{ddk15} & AER & Exp. & 2A(1) & 9,116 & 8,664 & 0.11 \\
      \textcite{fryerlevitt2013} &AER & Obs. & 3(4) & 8,806 & 6,623 & $0.31^\dagger$ \\
      \textcite{rim2020} & AER:P\&P & Obs. & 2(3) & 4,037 & 620 & $0.24^{\dagger}$ \\
      \textcite{weisburst2019} & AER:P\&P & Obs. & 2A & 7,488 & 7,488 & $0.31^{\dagger}$\\
      \bottomrule
  \end{tabular}
  \begin{tablenotes}
  \item\footnotesize \emph{Notes:} This table summarizes the five experimental
    studies and three observational studies of racial disparities collected from
    a search of the AEA Data and Code Repository from 2013--2022 (See
    \Cref{sec:article_search_procedure} for details of this search). Column 3
    reports the table and panel of the replicated specification with the column
    or row of the specification in parentheses. Column 6 gives the standard
    deviation of the estimated propensity score $\hat{p}_{k}(W_{i})$ for the
    treatment arm $k$ displaying the greatest propensity score variation;
    estimates are computed using a multinomial logit model. The symbol
    $^{\dagger}$ indicates that a corresponding hypothesis test for non-zero
    variation was significant. See \Cref{apx:overlap} for details on the overlap
    sample and tests for propensity score variation.
     \end{tablenotes}
   \end{threeparttable}
\end{table}

We conduct two preliminary analyses of each study before assessing contamination bias and comparing alternative estimators. First, we ensure that the estimation sample satisfies overlap, since otherwise the decomposition in \Cref{theorem:main-result} is typically not identified. If strong overlap fails, we identify a large subset of each analysis sample where it is satisfied.  Columns 4 and 5 of \Cref{tab:paper_list} list the number of observations in the full and overlap samples (the sample sizes are equal if the original estimation sample satisfies overlap). Second, we check for propensity score variation in each of the studies.
In principle, protocol descriptions can reveal whether some regression controls are necessary (and hence generate propensity score variation) or whether the controls are just added to improve precision. In practice, however, this is not always clear from paper descriptions.\footnote{Moreover, some regression specifications are run on a non-random subsample of the full experimental population (due to, e.g., attrition, or in a susample analysis). This could generate propensity score variation even in simple experimental protocols.} Column 6 of \Cref{tab:paper_list} gives a quantitative sense of the variability in the propensity scores by reporting the standard deviation of the estimated propensity score, showing that its variability in the observational studies is substantially higher; the dagger symbol indicates that a hypothesis test for non-zero variation in the population propensity scores was statistically significant.
\Cref{apx:overlap} details the overlap sample construction and these tests. We replicate the analyses from \Cref{tab:contamination} for each of the eight papers in \Cref{apdx_fullresults}; we summarize the takeaways here.

\begin{figure}
  \centering
  \small
  \resizebox{0.9\linewidth}{!}{\input{./bias_t.tex}}
    \vspace{-4ex}\caption{Contamination bias across all applications}\label{fig:contamination_replication}
  \floatfoot{\emph{Notes:} This figure summarizes the analysis of contamination bias in the STAR application and the additional applications in \Cref{tab:paper_list}. The six experimental studies are shown in blue; the three observational studies of racial disparities are shown in orange. Column A shows the absolute value of contamination bias \emph{t}-statistics for each regression coefficient, given by \cref{eq:beta_hat}. Column B shows a normalized version of this decomposition that divides each term by the standard error of the regression coefficient. The darker bar shows the own-treatment effect component, while the lighter bar shows the contamination bias component.}
\end{figure}

\Cref{fig:contamination_replication} summarizes the statistical and practical significance of contamination bias in the estimated effect of each treatment for each specification (as estimated in the overlap sample). Column A shows the absolute value of the contamination bias $t$-statistics for each regression coefficient, obtained from the decomposition in \cref{eq:beta_hat}. In both columns, we sort treatments within papers by this absolute $t$-statistic and sort papers by the maximum absolute $t$-statistic across treatments. Column B shows a normalized version of the decomposition that divides each term by the standard error of the regression coefficient. The darker bar shows the own-treatment effect component of the decomposition, while the lighter bar denotes the contamination bias component (which can be of the same or opposite sign).

The figure shows economically and statistically meaningful contamination bias in two of the three observational studies while showing no evidence for bias in any of the experimental studies. This aligns with the intuition that the large propensity score variability in observational studies generates much larger variability in the contamination weights. Specifications from both the \textcite{demel2013} and \textcite{drexler2014} experiments have some of the smallest contamination bias and also smallest propensity score variation, consistent with the theoretical results that contamination bias requires variation in the contamination weights which in turn requires variation in the propensity scores. On the other hand, the two studies with statistically significant contamination bias (\textcite{fryerlevitt2013} and \textcite{weisburst2019}) also display the greatest variation in propensity scores. Broadly, these results highlight the importance of testing for contamination bias---especially in observational settings where the included covariates are likely to drive sizable variation in propensity scores and hence contamination weights.

\begin{figure}
  \centering
  \small
   \resizebox{0.9\linewidth}{!}{\input{./estimates.tex}}
  \vspace{-4ex}\caption{Treatment effect estimates with using different estimators}\label{fig:estimators_replication}
  \floatfoot{\emph{Notes:} This figure plots estimates of treatment
    effects for each estimator from of \Cref{tab:contamination}, applied to the STAR application and additional
    applications in \Cref{tab:paper_list}.  We normalize each estimate by dividing by the standard error of the regression coefficient. The six experimental studies are
    shown in blue; the three observational studies of racial disparities are
    shown in orange. Each specification includes a line connecting the estimate
    from the regression coefficient and the common-weights (CW) estimator. EW
    stands for the easiest-to-estimate weighting. For the Rim et al.\ application the ATE estimate for the ``Asian'' coefficient equals $-8.4$, and it is not displayed as it falls
    outside the axis limits.}
\end{figure}

\Cref{fig:estimators_replication} plots estimates of the treatment effects for each estimator from \Cref{tab:contamination}, again normalizing by the standard error of the regression coefficient. We include a line between the estimates from OLS regression and from the common-weights (CW) estimator we propose. Among observational  studies, we see substantial variation across the different estimates and a much larger difference between the OLS estimator and the CW estimator. In the experimental papers, the difference is much smaller.\footnote{The same pattern arises when comparing the estimates in the full sample; see \Cref{apdx_fullresults}.} This is consistent with the larger propensity score variability in observational studies magnifying the impact of the choice of weighting scheme.

\section{Conclusion}\label{sec:conclusion}
Regressions with multiple treatments and flexible controls are common across a wide range of empirical settings in economics. We show that such regressions generally fail to estimate a convex weighted average of treatment effects: coefficients on each treatment are generally contaminated by the effects of other treatments. We provide intuition for why the influential result of \textcite{angrist98} fails to generalize to multiple treatments, and show how the contamination bias problem connects to a recent literature studying \ac{DiD} regressions. We then discuss three alternative estimators that are free of this bias.

Our analysis of nine empirical applications finds economically and statistically meaningful contamination bias in observational studies. Contamination bias in experimental studies is more limited, even in papers that display statistically significant variation in the propensity scores.  We also find that the choice among alternative estimators that are free of contamination bias matters more in the observational studies. Overall, our analysis highlights the importance of testing the empirical relevance of theoretical concerns with how regression combines heterogeneous effects---particularly in observational studies.

\setstretch{1.1}
\printbibliography%

\clearpage
\setstretch{1.25}
\appendix
\begin{appendices}
\crefalias{section}{appsec}
\crefalias{subsection}{appsubsec}

\section{Proofs and Additional Results}\label{apdx:proofs}

\subsection{Proof of Proposition~\ref{theorem:main-result}}\label{apdx:general}

We prove a generalization of the \Cref{theorem:main-result} which allows any vector of treatments $X_i$ (which may not be binary or mutually exclusive). We continue to consider the partially linear model in \cref{eq:partially_linear}, and maintain \Cref{ass:model_or_design}, as well as conditional mean-independence of the potential outcomes $E[Y_i(x)\mid X_i, W_i]=E[Y_i(x)\mid W_{i}]$, which extends \Cref{ass:mean_indep}. We also assume that the potential outcomes $Y_i(x)$ are linear in $x$, conditional on $W_i$:
\begin{equation*}
    E[Y_i(x)\mid W_i=w]=E[Y_i(0)\mid W_i=w]+x^\prime\tau(w),
\end{equation*}
for some function $\tau$. This condition holds trivially in the main-text discussion of mutually exclusive binary treatments. More generally, $\tau_k(w)$ corresponds to the conditional average effect of increasing $X_{ik}$ by one unit among observations with $W_i=w$. Although this assumption is not essential, it considerably simplifies the derivations. We continue to define $\tau=E[\tau(W_i)]$ as the average vector of per-unit effects.

We now prove that under these assumptions $\beta_{k}$ is given by the expression in \cref{eq:betak_decomposition}. We further prove that $E[\lambda_{kk}(W_{i})] = 1$ and $E[\lambda_{k\ell}(W_i)]=0$ for $\ell\neq k$ in general, and give a more detailed characterization of the weights in the case of mutually exclusive treatment indicators.

First note that by iterated expectations and conditional mean-independence,
$E[\dbtilde{X}_{ik}Y_{i}]=E[E[\dbtilde{X}_{ik}Y_{i}\mid X_i, W_i]]=E[\dbtilde{X}_{ik}E[Y_{i}(0)\mid
W_{i}]]+E[\dbtilde{X}_{ik}X_{i}'\tau(W_{i})]$. By definition of projection,
$E[\tilde{X}_{i}g(W_{i})]=0$ for all $g\in\mathcal{G}$ \parencite[Theorem
11.1]{vdv98}; thus if \cref{eq:model_based} holds
$E[\dbtilde{X}_{ik}E[Y_{i}(0)\mid W_{i}]]=0$. Similarly, under
\cref{eq:pscore_in_G}, $E[\dbtilde{X}_{ik}\mid W_{i}]=0$, so by iterated
expectations, $E[\dbtilde{X}_{ik}E[Y_{i}(0)\mid W_{i}]]=E[E[\dbtilde{X}_{ik}\mid W_i]E[Y_{i}(0)\mid W_{i}]]=0$. Thus,
\begin{equation*}
  \beta_{k}=\frac{E[\dbtilde{X}_{ik}X_{i}'\tau(W_{i})]}{E[\dbtilde{X}_{ik}^{2}]}
  =\frac{E[\dbtilde{X}_{ik}X_{ik}\tau_{k}(W_{i})]}{E[\dbtilde{X}_{ik}^{2}]}
  +\frac{\sum_{\ell\neq k}E[\dbtilde{X}_{ik}X_{i\ell}\tau_{\ell}(W_{i})]}{E[\dbtilde{X}_{ik}^{2}]}.
\end{equation*}
This proves \cref{eq:betak_decomposition}.

To show that $E[\lambda_{kk}(W_{i})] = 1$ and $E[\lambda_{k\ell}(W_i)]=0$ for $\ell\neq k$ in general, note that
\begin{equation*}
  E[\lambda_{kk}(W_{i})]=\frac{E[\dbtilde{X}_{ik}X_{ik}]}{E[\dbtilde{X}_{ik}^{2}]}
  =1,
\end{equation*}
since $\dbtilde{X}_{i, k}$ is a residual from projecting $X_{ik}$ onto the space
spanned by functions of the form $\tilde{g}(W_{i})+X_{i, -k}'\tilde{\beta}_{-k}$,
so that $E[\dbtilde{X}_{ik}X_{ik}]=E[\dbtilde{X}_{ik}^{2}]$. Furthermore,
$\dbtilde{X}_{i, k}$ must also be orthogonal to $X_{i,-k}$ by definition of
projection, so that
$E[\lambda_{k\ell}(W_{i})]=E[\dbtilde{X}_{ik}X_{i\ell}]/E[\dbtilde{X}_{ik}^{2}]=0$.

Finally, if $X_{i}$ are mutually exclusive treatment indicators, write
$E^{*}[X_{ik}\mid X_{i,-k}, W_{i}]=X_{i,-k}'\tilde{\delta}_{k}+\tilde{g}_{k}(W_{i})$.
Since $X_{ik}X_{i,-k}=0$, we may write
\begin{equation*}
 \lambda_{kk}(W_{i})=\frac{p_{k}(W_{i})(1-\tilde{g}_{k}(W_{i}))}{E[\dbtilde{X}_{ik}^{2}]}=
 \frac{p_{k}(W_{i})(1-E^{*}[X_{ik}\mid X_{i,-k}=0,W_{i}])}{E[\dbtilde{X}_{ik}^{2}]},
\end{equation*}
and, by similar arguments, $\lambda_{k\ell}(W_{i})=-p_{\ell}(W_{i})E^{*}[X_{ik}\mid
X_{i\ell}=1,W_{i}]/E[\dbtilde{X}_{ik}^{2}]$, which yields the second expression
for the weights. It remains to show that $\lambda_{kk}(W_{i})\geq 0$ if
\cref{eq:pscore_in_G} holds and $X_{i}$ consists of mutually exclusive
indicators. To that end, observe that $\lambda_{k\ell}(W_{i})$ is given by the
$(k, \ell)$ element of
\begin{equation*}
  \Lambda(W_{i})=  E[\tilde{X}_{i}\tilde{X}_{i}']^{-1}E[\tilde{X}_{i}X_{i}'\mid W_{i}]
\end{equation*}
If \cref{eq:pscore_in_G} holds, then we can write this as $\Lambda(W_{i})=E[v(W_{i})]^{-1}v(W_{i})$ where $v(W_{i})=E[\tilde{X}_{i}\tilde{X}_{i}'\mid W_{i}]$. If $X$ is a vector of
mutually exclusive indicators, then
$v(W_{i})=\diag(p(W_{i}))-p(W_{i})p(W_{i})'$. Let $v_{-k}(W_{i})$ denote the
submatrix with the $k$th row and column removed, and let
$p_{-k}(W_i)$ denote subvector with the $k$th row removed. Then by the block matrix
inverse formula,
\begin{equation*}
  \lambda_{kk}(W_{i})=\frac{p_{k}(W_{i})(1-p_{k}(W_{i}))
    -E[p_{k}(W_{i})p_{-k}(W_{i})']E[v_{-k}(W_{i})]^{-1}
    p_{-k}(W_{i})p_{k}(W_{i})}{E[p_{k}(W_{i})(1-p_{k}(W_{i}))]
    -E[p_{k}(W_{i})p_{-k}(W_{i})']E[v_{-k}(W_{i})]^{-1}E[p_{k}(W_{i})p_{-k}(W_{i})]}
\end{equation*}
Note $p_{0}(W_{i})=1-\sum_{k=1}^{K}p_{k}(W_{i})$ and
$p_{k}(W_{i})p_{-k}(W_{i})=v_{-k}(W_{i})\iota-p_{0}(W_{i})p_{-k}(W_{i})$, where
$\iota$ denotes a $(K-1)$-vector of ones. Thus, the numerator can be written as
\begin{multline*}
  p_{k}(W_{i})(1-p_{k}(W_{i})) -\iota'p_{-k}(W_{i})p_{k}(W_{i})\\
  +E[p_{0}(W_{i})p_{-k}(W_{i})']E[v_{-k}(W_{i})]^{-1}p_{-k}(W_{i})p_{k}(W_{i})\\
  = p_{k}(W_{i})p_{0}(W_{i})
  +E[p_{0}(W_{i})p_{-k}(W_{i})']E[v_{-k}(W_{i})]^{-1}p_{-k}(W_{i})p_{k}(W_{i}).
\end{multline*}
The eigenvalues of $E[v_{-k}(W_i)]$ are positive because it is a
covariance matrix. Furthermore, the off-diagonal elements of $E[v(W_i)]$
are negative, and hence the off-diagonal elements of $E[v_{-k}(W_i)]$ are also negative.
It therefore follows that $E[v_{-k}(W_i)]$ is an $M$-matrix \parencite[property
$D_{16}$, p.~135]{BePl94}. Hence, all elements of $E[v_{-k}(W_i)]^{-1}$ are
positive \parencite[property $N_{38}$, p.~137]{BePl94}. Thus, both summands in
the above expression are positive, so that $\lambda_{kk}(W_i)\geq 0$.

\subsection{Proof of Proposition~\ref{theorem:seb}}\label{sec:proof_p2}
The parameter of interest $\theta_{\lambda, c}$ depends on the realizations of the controls.
We therefore derive the semiparametric efficiency bound conditional
on the controls; i.e.\ we show that \cref{eq:seb} is almost-surely the variance bound
for estimators that are regular conditional on the controls. Relative to the
earlier results in \textcite{hahn98} and \textcite{HiImRi03}, we need to account
for the fact that the data are no longer
i.i.d.\ once we condition on the controls.

To that end, we use the notion of semiparametric efficiency based on the
convolution theorem of \textcite[Theorem 2.1]{vVWe89} \parencite[see
also][Chapter 3.11]{vVWe96}. We first review the result for convenience.
Consider a model $\{P_{n, \theta}\colon \theta\in\Theta\}$ parametrized by (a
possibly infinite-dimensional) parameter $\theta$. Let $\dot{\mathcal{P}}$
denote a tangent space, a linear subspace of some Hilbert space with an inner
product $\ip{\cdot, \cdot}$. Suppose that the model is \ac{LAN} at $\theta$
relative to a tangent space $\dot{\mathcal{P}}$: for each
$g\in\dot{\mathcal{P}}$, there exists a sequence $\theta_{n}(g)$ such that the
likelihood ratios are asymptotically quadratic,
$dP_{n, \theta_{n}(g)}/dP_{n, \theta}=\Delta_{n, g}-\ip{g, g}/2+o_{P_{n, \theta}}(1)$,
where $(\Delta_{n, g})_{g\in\dot{\mathcal{P}}}$ converges under $P_{n, \theta}$ to
a Gaussian process with covariance kernel $\ip{g_{1}, g_{2}}$. Suppose also that the parameter $\beta_{n}(P_{n, \theta})$ is differentiable: for each $g$,
$\sqrt{n}(\beta_{n}(P_{n, \theta_{n}(g)})-\beta_{n}(P_{n, \theta}))\to \ip{\psi,
  g}$ for some $\psi$ that lies in the completion of $\dot{\mathcal{P}}$. Then
the semiparametric efficiency bound is given by $\ip{\psi, \psi}$: the asymptotic distribution of any regular estimator of this parameter, based on a
sample $\mathbf{S}_{n}\sim P_{n, \theta}$, is given by the
convolution of a random variable $Z\sim\mathcal{N}(0, \ip{\psi, \psi})$ and some
other random variable $U$ that is independent of $Z$.

To apply this result in our setting, we proceed in three steps. First, we define
the tangent space and the probability-one set over which we will prove the
efficiency bound. Next, we verify that the model is \ac{LAN}. Finally, we verify
differentiability and derive the efficient influence function $\psi$.

\paragraph{Step 1}
By the conditional independence assumption in \cref{eq:full_ci}, we can write
the density of the vector $(Y_{i}(0), \dotsc, Y_{i}(K), D_{i})$ (with respect to
some $\sigma$-finite measure) conditional on $W_{i}=w$ as
$f(y_{0}, \dotsc, y_{K}\mid w)\cdot\prod_{k=0}^{K}p_{k}(w)^{\1{d=k}}$, where $f$
denotes the conditional density of the potential outcomes, conditional on the
controls. The density of the observed data
$\mathbf{S}_{N}=\{(Y_{i}, D_{i})\}_{i=1}^{N}$ conditional on
$(W_{1}, \dotsc, W_{N})=(w_{1}, \dotsc, w_{N})$ is given by
$\prod_{i=1}^{N}\prod_{k=0}^{K}(f_{k}(y_{i}\mid
w_{i})p_{k}(w_{i}))^{\1{d_{i}=k}}$, where
$f_{k}(y\mid w)=\int f(y_{k}, y_{-k}\mid w)dy_{-k}$.

Since the propensity scores are known, the model is parametrized by $\theta=f$.
Consider one-dimensional submodels of the form
$f_{k}(y\mid w;t)=f_{k}(y\mid w)(1+t\times s_{k}(y\mid w))$, where the function
$s_{k}$ is bounded and satisfies $\int s_{k}(y\mid w)f_{k}(y\mid w)d y=0$ for
all $w\in\mathcal{W}$ with $\mathcal{W}$ denoting the support of $W_{i}$. For
small enough $t$, we have $f_{k}(y\mid w;t)\geq 0$ by boundedness of $s_{k}$; hence
$f_{k}(y\mid w;t)$ is a well-defined density for $t$ small enough. The joint
log-likelihood, conditional on the controls, is given by
\begin{equation*}
  \sum_{i=1}^{N}\sum_{k=0}^{K}\1{D_{i}=k}(\log f_{k}(Y_{i}\mid w_{i};t) + \log p_{k}(w_{i})).
\end{equation*}
The score at $t=0$ is  $\sum_{i=1}^{N}s(Y_{i}, D_{i}\mid w_{i})$, with
$s(Y_{i}, D_{i}\mid w_{i})=\sum_{k=0}^{K}\1{D_{i}=k}s_{k}(Y_{i}\mid w_{i})$.

This result suggests defining the tangent space to consist of functions
$s(y, d\mid w)=\sum_{k=0}^{K}\1{d=k}s_{k}(y\mid W_i=w)$, such that $s_{k}$ is
bounded and satisfies $\int s_{k}(y\mid w)f_{k}(y\mid w)d y=0$ for all
$w\in\mathcal{W}$. Define the inner product on this space by
$\ip{s_{1}, s_{2}}=E[s_{1}(Y_{i}, D_{i}\mid W_{i})s_{2}(Y_{i}, D_{i}\mid W_{i})]$.
Note this is a marginal (rather than a conditional) expectation, over the
unconditional distribution $(Y_{i}, D_{i}, W_{i})$ of the observed data.

We will prove the efficiency bound on the event $\mathcal{E}$ that (i)
$\frac{1}{N}\sum_{i=1}^{N}E[s(Y_{i}, D_{i}\mid W_{i})^{2}\mid W_{i}]\to
E[s(Y, D_{i}\mid W_{i})^{2}]$, (ii)
$\frac{1}{N}\sum_{i=1}^{N}\lambda(W_{i})\to E[\lambda(W_{i})]$, and (iii)
$\frac{1}{N}\sum_{i=1}^{N}\lambda(W_{i})
\sum_{k=0}^{K}c_{k} \cdot E[Y_{i}(k)s_{k}(Y_{i}(k) \mid W_{i})\mid W_{i}]\to \sum_{k=0}^{K}c_{k}E[\lambda(W_{i})
Y_{i}(k)s_{k}(Y_{i}(k)\mid W_{i})]$. By assumptions of the
\namecref{theorem:efficient_estimation}, these are all averages of functions of
$W_{i}$ with finite absolute moments. Hence, by the law of large numbers, $\mathcal{E}$ is a probability one set.

\paragraph{Step 2}
We verify that the conditions (3.7--12) of Theorem 3.1 in \textcite{McNW00} hold
on the set $\mathcal{E}$ conditional on the controls, with
$\theta_{N}(s)=f(\cdot\mid \cdot;1/\sqrt{N})$. Let
$\alpha_{Ni}=\prod_{k=0}^{K}(f_{k}(Y_{i}\mid w_{i};1/\sqrt{N})/f_{k}(Y_{i}\mid
w_{i}))^{\1{D_{i}=k}}=\prod_{k=0}^{K}(1+s_{k}(Y_{i}\mid
w)/\sqrt{N})^{\1{D_{i}=k}}$ denote the likelihood ratio associated with the
$i$th observation. Since this is bounded by the boundedness of $s_{k}$,
condition (3.7) holds. Also since $(1+ts_{k})^{1/2}$ is continuously
differentiable for $t$ small enough, with derivative $s_{k}/2 \sqrt{1+ts_{k}}$,
it follows from Lemma 7.6 in \textcite{vdv98} that
$N^{-1}\sum_{i=1}^{N}E[\sqrt{N}(\alpha_{Ni}^{1/2}-1)-s(Y_{i}, D_{i}\mid
w_{i})/2\mid W_{i}=w_{i}]^{2}\to 0$ such that the quadratic mean differentiability
condition (3.8) holds. Since $s_{k}$ is bounded, the Lindeberg condition (3.9)
also holds. Next,
$\frac{1}{N}\sum_{i=1}^{N}E[s(Y_{i}, D_{i}\mid W_{i})^{2}\mid W_{i}]$ converges
to $E[s(Y, D_{i}\mid W_{i})^{2}]=\ip{s, s}$ on $\mathcal{E}$ by assumption. Hence,
conditions (3.10) and (3.11) also hold. Since the scores
$\Delta_{N, s}=\frac{1}{\sqrt{N}}\sum_{i=1}^{N}s(Y_{i}, D_{i}\mid w_{i})$ are
exactly linear in $s$, condition (3.12) also holds. It follows that the model is
\ac{LAN} on $\mathcal{E}$.

\paragraph{Step 3}
Write the parameter of interest $\theta_{\lambda, c}$ as
$\beta_{N}(f)=\sum_{i=1}^{N}\lambda(w_{i})\int y\sum_{k=0}^{K}c_{k}f_{k}(y\mid
w_{i})dy/\sum_{i=1}^{N}\lambda(w_{i})$. It follows that
\begin{multline*}
    \sqrt{N}(\beta_{N}(f(\cdot\mid \cdot;1/\sqrt{N}))-\beta_{N}(f)) \\
    =\frac{1}{N^{-1}\sum_{i=1}^{N}\lambda(w_{i})}\frac{1}{\sqrt{N}}\sum_{i=1}^{N}\lambda(w_{i})\int
     y\sum_{k=0}^{K}c_{k}(f_{k}(y\mid
    w_{i};1/\sqrt{N})-f_{k}(y\mid w_{i}))dy\\
    =\frac{1}{N^{-1}\sum_{i=1}^{N}\lambda(w_{i})}\frac{1}{N}\sum_{i=1}^{N}\lambda(w_{i})
    \sum_{k=0}^{K}c_{k} \int
     y s_{k}(y\mid w_{i})f_{k}(y\mid w_{i}) dy,
\end{multline*}
which converges to
$ \sum_{k=0}^{K}c_{k} E[\lambda(W_{i})Y_{i}(k) s_{k}(Y_{i}(k)\mid W_{i})]
/E[\lambda(W_{i})]$ on $\mathcal{E}$ by assumption. We can write this
as $\ip{\psi, s}$, where
\begin{equation*}
  \psi(Y_i, D_i, W_i)=\sum_{k=0}^{K}\1{D_i=k}\lambda(W_i)c_{k}\frac{(Y_i-\mu_{k}(W_i)).
  }{p_{k}(W_i)E[\lambda(W_{i})]}.
\end{equation*}
Observe that $\psi$ is in the model tangent space, with the summands playing the
role of $s_{k}(y\mid w)$ (more precisely, since $\psi$ is unbounded, it lies in
the completion of the tangent space). Hence, the semiparametric efficiency bound
is given by $E[\psi^{2}]$.

\subsection{Efficiency of the CW estimator}\label{sec:effic-cw-estim}

The next result shows that the estimator in  \cref{eq:weighted_regression} is
efficient. We defer its proof to \Cref{sec:proof-prop-refth}.

\begin{proposition}\label{theorem:efficient_estimation}
  Suppose \cref{eq:full_ci} holds in an i.i.d.\ sample of size $N$, with known
  non-degenerate propensity scores $p_{k}(W_{i})$. Let
  $\beta^{*}_{\lcw, k} =
  E[\lcw(W_{i})\tau_{k}(W_{i})]/E[\lcw(W_{i})]$, and
  $\alpha^{*}_{k}=\beta^{*}_{\lcw, k}+E[\lcw(W_{i})\mu_{0}(W_{i})]/E[\lcw(W_{i})]$.
  Suppose that the fourth moments of $\lcw(W_{i})$ and $\mu(W_{i})$ are
  bounded, and that $p_{k}\in\mathcal{G}$,
  $(\mu_{k}(W_{i})-\alpha^{*}_{k})
  \frac{\lcw(W_{i})^{2}}{p_{k'}(W_{i})^{2}}\in\mathcal{G}$, and
  $(\mu_{k}(W_{i})-\alpha^{*}_{k})\frac{\lcw(W_{i})}{p_{k}(W_{i})}\in\mathcal{G}$
  for all $k, k'$. Then, provided it is asymptotically linear and regular,
  $\hat{\beta}_{\hlcw}$ achieves the semiparametric
  efficiency bound for estimating ${\beta}_{\lcw}$, with
  diagonal elements of its asymptotic variance of:
  \begin{multline*}
    \frac{1}{E[\lcw(W_{i})]^{2}}E\left[
      \frac{\lcw(W_{i})^{2}\sigma_{0}^{2}(W_{i})}{p_{0}(W_{i})}
      +      \frac{\lcw(W_{i})^{2}\sigma_{k}^{2}(W_{i})}{p_{k}(W_{i})}\right.\\
    \left.
      +{\lcw}(W_{i})^{2}(\tau_{k}(W_{i})-\beta^{*}_{\lcw,
      k})^{2}
      \left(\sum_{k'=0}^{K}\frac{{\lcw}(W_{i})^{2}}{p_{k}(W_{i})^{3}}-1\right)
    \right].
  \end{multline*}
\end{proposition}
\noindent
This efficiency result doesn't rely on homoskedasticity: under
heteroskedasticity, the estimator $\hat{\beta}_{\hlcw}$ is
still efficient for ${\beta}_{\lcw}$ (although the weighting
$\lcw(W_{i})$ need not be optimal under heteroskedasticity). It
is stated under the high-level condition that
$\hat{\beta}_{\hlcw}$ is regular; the proof uses
calculations from \textcite{newey94ecta} to verify the estimator achieves the
efficiency bound. Primitive regularity conditions will depend on the form of
$\mathcal{G}$ and are omitted for brevity.

\begin{remark}\label{remark:prop_score_cost}
  The asymptotic variance of the estimator $\hat{\beta}_{\lcw}$ is
  larger than the asymptotic variance of the infeasible estimator that replaces
  the estimated weights $\hlcw(W_i)/\hat{p}_{D_i}(W_i)$ in
  \cref{eq:weighted_regression} with the infeasible weights
  $\lcw(W_i)/p_{D_i}(W_i)$. The latter achieves the asymptotic
  variance implied by \Cref{theorem:F-weighting},
\begin{equation}\label{eq:infeasible_var}
    \frac{1}{E[\lcw(W_{i})]^{2}}E\left[
      \frac{\lcw(W_{i})^{2}\sigma_{0}^{2}(W_{i})}{p_{0}(W_{i})}
      +      \frac{\lcw(W_{i})^{2}\sigma_{k}^{2}(W_{i})}{p_{k}(W_{i})}
    \right].
 \end{equation}
 The extra term of the asymptotic variance in
 \Cref{theorem:efficient_estimation} relative to \cref{eq:infeasible_var}
 reflects the cost of having to estimate the weights.\footnote{The extra term
   shows this cost is zero if either there is no treatment effect heterogeneity,
   so that $\tau_{k}(W_i)=\beta^{*}_{\lcw, k}$, or if the treatment assignment
   is completely randomized so that $p_{k}(W_{i})=1/(K+1)$. In the latter case
   $\lambda^*(W_{i})=1/(K+1)^{2}$ so
   $\sum_{k=0}^{K}\lcw(W_{i})^{2}/p(W_{i})^{3}=1$. The extra term can be avoided
   altogether if we interpret $\hat{\beta}_{\hlcw}$ as an estimator of
   ${\beta}_{\hlcw}$. This follows from arguments in \textcite[Lemma
   B.6]{crump2006dealing}.} Analogous term is present in the expression for the
 asymptotic variance of the one-treatment-at-a-time estimator implementing the
 weights from \Cref{cor:angrist}.
\end{remark}

\subsection{Proof of Proposition~\ref{theorem:efficient_estimation}}\label{sec:proof-prop-refth}

We first derive the semiparametric efficiency bound for estimating
$\beta_{\lcw}$ when the propensity scores are not known, using the
same steps, notation, and setup as in the proof of \Cref{theorem:main-result}.
We then verify that the estimator $\hat{\beta}_{\hlcw}$
achieves this bound.

\paragraph{Step 1}

Since the propensity scores are not known, the model is now parametrized by
$\theta=(f, p)$. Consider one-dimensional submodels of the form
$f_{k}(y\mid w;t)=f_{k}(y\mid w)(1+t s_{y, k}(y\mid w))$, and
$p_{k}(w;t)=p_{k}(w)(1+t s_{p, k}(x))$, where the functions $s_{y, k}, s_{p, k}$
are bounded and satisfy $\int s_{y, k}(y\mid w)f_{k}(y\mid w)d y=0$ and
$\sum_{k=0}^{K}p_{k}(w)s_{p, k}(w)=0$ for all $w\in\mathcal{W}$. These
conditions ensure that $f_{k}(y\mid w;t)$ and $p_{k}(w;t)$ are positive for $t$
small enough and that $\sum_{k=0}^{K}p_{k}(w;t)=\sum_{k=0}^{K}p_{k}(w)=1$, so
that the submodel is well-defined. The joint log-likelihood, conditional on the
controls, is given by
\begin{equation*}
  \sum_{i=1}^{N}\sum_{k=0}^{K}\1{D_{i}=k}(\log f_{k}(Y_{i}\mid w_{i};t) + \log
  p_{k}(w_{i}; t)).
\end{equation*}
The score at $t=0$ is given by $\sum_{i=1}^{N}s(Y_{i}, D_{i}\mid w_{i})$, with
$s(Y_{i}, D_{i}\mid w_{i})=\sum_{k=0}^{K}\1{D_{i}=k}(s_{y, k}(Y_{i}\mid
w_{i})+s_{p, k}(w_{i}))$.

In line with this result, we define the tangent space to consist of all
functions $s(y, d\mid w)=\sum_{k=0}^{K}\1{d=k} (s_{y, k}(y\mid w)+s_{p, k}(w))$
such that $s_{y, k}$ and $s_{p, k}$ satisfy the above restrictions. Define the
inner product on this space by the marginal expectation
$\ip{s_{1}, s_{2}}=E[s_{1}(Y_{i}, D_{i}\mid W_{i})s_{2}(Y_{i}, D_{i}\mid W_{i})]$.
We will prove the efficiency bound on the event $\mathcal{E}$ that (i)
$\frac{1}{N}\sum_{i=1}^{N}E[s(Y_{i}, D_{i}\mid W_{i})^{2}\mid W_{i}]\to
E[s(Y, D_{i}\mid W_{i})^{2}]$; (ii)
$N^{-1}\sum_{i}\lcw(W_{i})\to E[\lcw(W_{i})]$;
(iii) $N^{-1}\sum_{i}\lcw(W_{i})
\sum_{k=0}^{K}c_{k}E[Y_{i}(k)\cdot s_{y, k}(Y_{i}\mid W_{i})\mid W_{i}]\to
\sum_{k=0}^{K}c_{k} E[\lcw(W_{i})Y_{i}(k)\cdot \allowbreak s_{y, k} (Y_{i}(k)\mid W_{i})]$;
(iv) $N^{-1}\sum_{i=1}^{N}\lcw(W_{i})^{2}
\sum_{k, k'}c_{k'}\mu_{k'}(W_{i})\frac{s_{p, k}(W_{i})}{p_{k}(W_{i})}\to
E[\lcw(W_{i})^{2} \cdot\allowbreak \sum_{k, k'}c_{k'}\cdot\allowbreak\mu_{k'}(W_{i})\frac{s_{p, k}(W_{i})}{p_{k}(W_{i})} ]$;
(v)
$N^{-1}\sum_{i=1}^{N}\lcw(W_{i})^{2}\sum_{k=0}^{K}\frac{s_{p, k}(W_{i})}{p_{k}(W_{i})}\to
E[\lcw(W_{i})^{2}\sum_{k=0}^{K}\frac{s_{p, k}(W_{i})}{p_{k}(W_{i})}]$; and
(vi)
$\beta_{\lcw} \to \beta^{*}_{\lcw}$. Under the \namecref{theorem:efficient_estimation} assumptions
 and the law of large numbers,
$\mathcal{E}$ is a probability-one set.

\paragraph{Step 2}
We verify that the conditions (3.7--3.12) of Theorem 3.1 in \textcite{McNW00} hold
on the set $\mathcal{E}$ conditional on the controls, with
$\theta_{N}(s)=(f(\cdot\mid \cdot;1/\sqrt{N}), p(\cdot;1/\sqrt{N}))$. Let
$\alpha_{Ni}=\prod_{k=0}^{K}(f_{k}(Y_{i}\mid
w_{i};1/\sqrt{N})p_{k}(w_{i};1/\sqrt{N})/f_{k}(Y_{i}\mid
w_{i})p_{k}(w_{i}))^{\1{D_{i}=k}}= \prod_{k=0}^{K}((1+N^{-1/2}s_{y, k}(Y_{i}\mid
W_{i};N^{-1/2})) (1+N^{-1/2}s_{p, k}(w_{i};1/\sqrt{N})))^{\1{D_{i}=k}}$ denote
the likelihood ratio associated with the $i$th observation. Since this is
bounded by the boundedness of $s_{y, k}, s_{p, k}$, condition (3.7) holds. Also,
since $(1+ts_{p, k})^{1/2}$ and $(1+ts_{y, k})^{1/2}$ are continuously
differentiable for $t$ small enough, it follows from Lemma 7.6 in
\textcite{vdv98} that the quadratic mean differentiability condition (3.8)
holds. Since $s_{k}$ is bounded, the Lindeberg condition (3.9) also holds. Next,
$\frac{1}{N}\sum_{i=1}^{N}E[s(Y_{i}, D_{i}\mid W_{i})^{2}\mid W_{i}]$ converges
to $E[s(Y, D_{i}\mid W_{i})^{2}]=\ip{s, s}$ on $\mathcal{E}$ by assumption.
Hence, conditions (3.10) and (3.11) also hold. Since the scores
$\Delta_{N, s}=\frac{1}{\sqrt{N}}\sum_{i=1}^{N}s(Y_{i}, D_{i}\mid w_{i})$ are
exactly linear in $s$, condition (3.12) also holds. It follows that the model is
\ac{LAN} on $\mathcal{E}$.

\paragraph{Step 3}
Write the parameter of interest, $\beta_{\lcw}$, as
$\beta_{N}(\theta)=\sum_{i=1}^{N}\lcw(w_{i})\int
y\sum_{k=0}^{K}c_{k}f_{k}(y\mid
w_{i})dy/\sum_{i=1}^{N}\lcw(w_{i})$, where
$\lcw(w_{i})=1/\sum_{k=0}^{K}p_{k}(w_{i})^{-1}$. Letting
$\dot{\beta}_{N}(\theta)$ denote the derivative of
$\beta_{N}(\theta(\cdot\mid \cdot;t))$ at $t=0$, we have
\begin{equation*}
  \sqrt{N}(\beta_{N}(\theta(\cdot\mid \cdot;1/\sqrt{N}))-\beta_{N}(\theta))
  =\dot{\beta}_{N}(\theta)+o(1).
\end{equation*}
Let
$h(w)=\lcw(w) \sum_{k=0}^{K}c_{k}\int ys_{y, k}(y\mid
w){f}_{k}(y\mid w)dy$, and $\tilde{h}(W_{i})=\sum_{k'=0}^{K}c_{k'}\mu_{k'}(W_{i})
        -\beta^{*}_{\lcw}$. The derivative may then be written as
\begin{equation*}
  \begin{split}
    \dot{\beta}_{N}(\theta)& =\frac{1}{\sum_{i=1}^{N}\lcw(w_{i})}
    \sum_{i=1}^{N}\left(h(w_{i}) + \lcw(w_{i})^{2}
      \sum_{k=0}^{K}\frac{s_{p, k}(w_{i})}{p_{k}(w_{i})} \left(
        \sum_{k'=0}^{K}c_{k'}\mu_{k'}(w_{i}) -\beta_{N}(\theta)\right)\right)\\
    &\to \frac{1}{E[\lcw_{i}]} E\left[h(W_{i}) +
      (\lcw_{i})^{2}
      \sum_{k=0}^{K}\frac{s_{p, k}(W_{i})}{p_{k}(W_{i})} \left(
        \sum_{k'=0}^{K}c_{k'}\mu_{k'}(W_{i}) -\beta^{*}_{\lcw}\right)\right]\\
    &=\frac{1}{E[\lcw_{i}]} E\left[\lcw_{i}\sum_{k=0}^{K}X_{ki}\left(
      c_{k}\frac{Y_{i}-\mu_{k}(W_{i})}{
        p_{k}(W_{i})} +
      \frac{\lcw_{i}\tilde{h}(W_{i})}{p_{k}(W_{i})^{2}} \right)s(Y_{i}, D_{i}\mid W_{i})\right],
  \end{split}
\end{equation*}
where $\lcw_{i}=\lcw(W_{i})$, the limit on the second line holds on the event $\mathcal{E}$, and the
third line uses
$E[X_{ki}(Y_{i}-\mu_{k}(W_{i}))s(Y_{i}, D_{i}\mid W_{i})\mid W_{i}]=
p_{k}(W_{i})E[Y_{i}(k)s_{y, k}(Y_{i}(k)\mid W_{i})\mid W_{i}]$ and
$E[X_{ki}s(Y_{i}, D_{i}\mid W_{i})\mid W_{i}]=p_{k}(W_{i})s_{p, k}(W_{i})$. Since
for any function $a(W_{i})$, $E[a(W_{i})s(Y_{i}, D_{i}\mid W_{i})]=0$,
subtracting
$\frac{1}{E[\lcw_{i}]}
\sum_{k=0}^{K}E[(\lcw_{i})^{2}
  \frac{\tilde{h}(W_{i})}{p_{k}(W_{i})} s(Y_{i}, D_{i}\mid W_{i})]=0$ from
the preceding display implies
$\sqrt{N}(\beta_{N}(\theta(\cdot\mid
\cdot;1/\sqrt{N}))-\beta_{N}(\theta))=E[\psi(Y_{i}, D_{i}, W_{i})s(Y_{i}, D_{i}\mid
W_{i})]+o(1)$, where
\begin{equation*}
  \psi(Y_{i}, D_{i}, W_{i})=
    \sum_{k=0}^{K}X_{ki}\cdot \left(
    \frac{\lcw_{i}}{E[\lcw_{i}]}c_{k}
    \frac{Y_{i}-\mu_{k}(W_{i})}{p_{k}(W_{i})}
    +   \frac{\lcw_{i}}{E[\lcw_{i}]}\tilde{h}(W_{i})
    \left(\frac{\lcw_{i}}{p_{k}^{2}}-1\right)
  \right).
\end{equation*}
Observe that $\psi$ lies in the completion of the tangent space, with the
expression in parentheses playing the role of
$s_{y, k}(Y_{i}\mid W_{i})+s_{p, k}(W_{i})$. Hence, the semiparametric efficiency
bound is given by $E[\psi^{2}]$, which yields the expression in the statement of
the \nameCref{theorem:efficient_estimation}.

\paragraph{Attainment of the bound}

We derive the result in two steps. First, we show that
\begin{equation}\label{eq:if_star}
  \sqrt{N}(\beta_{\lcw}-\beta^{*}_{\lcw})=\frac{1}{\sqrt{N}}\sum_{i=1}^{N}
  \psi^{*}(W_{i})+o_{p}(1)\, \text{where}\, \psi^{*}(W_{i})=
  \frac{\lcw_{i}}{E[\lcw_{i}]}(\tau(W_{i})-\beta^{*}_{\lcw}).
\end{equation}
Second, we show that
\begin{equation}\label{eq:if_pop}
  \sqrt{N}(\hat{\beta}_{\hlcw}-\beta^{*}_{\lcw})=\frac{1}{\sqrt{N}}\sum_{i=1}^{N}
  \psi(Y_{i}, D_{i}, W_{i})+o_{p}(1),
\end{equation}
where, letting $\epsilon_{ki}=Y_{i}-\mu_{k}(W_{i})$,
\begin{equation*}
  \psi_{k}(Y_{i}, D_{i}, W_{i})=
  \frac{\lcw_{i}}{E[\lcw_{i}]} \left(\frac{
      X_{ki}\epsilon_{ki}}{p_{k}(W_{i})} -\frac{
      X_{0i}\epsilon_{0i}}{p_{k}(W_{i})}+
    (\tau_{k}(W_{i})-\beta^{*}_{\lcw, k}) \lcw_{i}\sum_{k'}
    \frac{X_{k'i}}{p_{k'}(W_{i})^{2}}
  \right).
\end{equation*}
Together, these results imply that the asymptotic variance of
$\hat{\beta}_{\hlcw}$ as an estimator of
${\beta}_{\lcw}$ is given by $\var(\psi-\psi^{*})$, which
coincides with the semiparametric efficiency bound.

\Cref{eq:if_star} follows directly under the assumptions of the
\namecref{theorem:efficient_estimation} by the law of large numbers and the fact
that the variance of
$\lcw_{i}(\tau(W_{i})-\beta^{*}_{\lcw})$ is
bounded. To show \cref{eq:if_pop}, write
$\hat{\beta}_{\hlcw, k}=\hat{\alpha}_{k}-\hat{\alpha}_{0}$,
where $\hat{\alpha}$ is a two-step method of moments estimator based on the
$(K+1)$ dimensional moment condition $E[m(Y_{i}, D_{i}, W_{i}, \alpha^{*}, p)]=0$
with elements
$m_{k}(Y_{i}, D_{i}, W_{i}, \alpha^{*},
p)=\lcw_{i}\frac{X_{ki}}{p_{k}(W_{i})}(Y_{i}-\alpha^{*}_{k})$,
and $\alpha^{*}$ is a $(K+1)$ dimensional vector with elements
$\alpha_{k}^{*}=E[\lcw_{i}\mu_{k}(W_{i})]/E[\lcw_{i}]$.

Consider a one-dimensional path $F_{t}$ such that the distribution of the data
is given by $F_{0}$. Let $p_{k, t}(W_{i})=E_{F_{t}}[X_{ki}\mid W_{i}]$ denote the
propensity score along this path. The derivative of
$E[m_{k}(Y_{i}, D_{i}, W_{i}, \alpha^{*}, p_{t})]$ with respect to $t$ evaluated at
$t=0$ is
\begin{equation*}
  E\left[\frac{\lcw_{i}X_{ki}}{p_{k}(W_{i})}(Y_{i}-\alpha^{*}_{k})\left(
      \lcw_{i}\sum_{k'=0}^{K}\frac{\dot{p}_{k'}(W_{i})}{p_{k'}(W_{i})^{2}}
      -  \frac{\dot{p}_{k}(W_{i})}{p_{k}(W_{i})}\right)\right]\\
  =\sum_{k'=0}^{K}E[\delta_{kk'}(W_{i})'\dot{p}_{k'}(W_{i})]
  ,
\end{equation*}
where $\dot{p}_{k}$ denotes the derivative of $p_{k, t}$ at $t=0$, and
\begin{equation*}
  \delta_{k, k'}(W_{i})=\lcw_{i}(\mu_{k}(W_{i})-\alpha^{*}_{k})\left(
      \frac{\lcw_{i}}{p_{k'}(W_{i})^{2}} -
      \frac{\1{k=k'}}{p_{k}(W_{i})}\right).
\end{equation*}
Under the assumptions of the \namecref{theorem:efficient_estimation},
$\delta_{k, k'}\in\mathcal{G}$. It therefore follows by Proposition 4 in
\textcite{newey94ecta} that the influence function for $\hat{\alpha}_{k}$ is
given by
\begin{multline*}
  \frac{1}{E[\lcw_{i}]} \left(\frac{\lcw_{i}
      X_{ki}}{p_{k}(W_{i})}(Y_{i}-\alpha^{*}_{k})+
    \sum_{k'}\delta_{kk'}(W_{i})(X_{k'i}-p_{k'}(W_{i}))\right)\\
  = \frac{\lcw_{i}}{E[\lcw_{i}]} \left(\frac{
      X_{ki}\epsilon_{ki}}{p_{k}(W_{i})} +
    (\mu_{k}(W_{i})-\alpha^{*}_{k}) \lcw_{i}\sum_{k'} \frac{X_{k'i}}{p_{k'}(W_{i})^{2}}
  \right),
\end{multline*}
which yields \cref{eq:if_pop}.

\section{Connections to the DiD Literature}\label{apdx:did}

In this \namecref{apdx:did} we elaborate on the connections between \Cref{theorem:main-result} and the recent literature studying potential biases from heterogeneous treatment effects in \ac{DiD}  regressions and related specifications \parencite[e.g.][]{goodman2021difference, sun2021estimating,hull2018movers, de2020aer,de2020two,callaway2021difference, borusyak2021revisiting,wooldridge2021mundlak}. We first show how our framework fits a \ac{TWFE} regression with a general treatment specification. We then show how \Cref{theorem:main-result} applies to three particular specifications: a static binary treatment, a dynamic ``event study'' treatment, and a static multivalued treatment (or ``movers regression''). In each case we discuss whether there is a potential for bias---either contamination bias or own-treatment negative weighting---and give a numerical illustration.

Consider a panel of units indexed by $j = 1, \dotsc, n$ which are observed over time periods $t = 1, \dotsc, T$. For simplicity, we assume the panel is balanced such that the sample size is $N = nT$. For an observation $i = (j, t)$, let $J_i=j$ and $T_i=t$ denote the corresponding unit and time period, respectively. In a \ac{TWFE} specification, the controls only comprise these two variables, $W_{i} = (J_i, T_{i})$, and they enter the control function as dummies, $g(W_i)=\alpha+(\1{J_i=2}, \dotsc, \1{J_i=n}, \1{T_i=2}, \dotsc, \1{T_i=T})^\prime\gamma$, with the indicators $\1{J_i=1}$ and $\1{T_i=1}$ omitted to avoid perfect collinearity.

To study these specifications, we follow \textcite{de2020aer, borusyak2021revisiting} in considering the $n$ observed units as fixed, and we condition on their treatment status (results when the units are sampled from a large population are analogous). For each unit $j$, we observe a random $T$-vector of outcomes $Y_j=(Y_{j1}, \dotsc, Y_{jT})$ and a fixed $T$-vector of $(\kk+1)$-valued treatments $\dd_j=(\dd_{j1}, \dotsc, \dd_{jT})$. These treatments are used to construct a vector of $(K+1)$-valued ``treatments states'' $D_j=(D_{j1}, \dotsc, D_{jT})$, with $D_{jt}\in\{0, \dotsc, K\}$. Setting $D_{j}=\dd_{j}$ covers scenarios with static treatments; as we show below, other choices of $D_j$ allows us to cover scenarios with dynamic treatment effects. As in the main text, $X_{jt}$ denotes a $K$-vector of treatment status indicators derived from $D_{jt}$.

We make two assumptions. First, we assume that potential outcomes $Y_{jt}(d_t)$ depend on the $T$-vector of treatments only through the current value $d_t$ of the treatment state, such that $Y_{jt}=Y_{jt}(D_{jt})$.\footnote{This assumption rules out misspecification of the treatment states, such as when there are dynamic effects but $D_{jt}=\dd_{jt}$ only indexes contemporaneous treatment status, as noted in \cref{fn_dd}.} Second, we make a parallel trends assumption by writing the untreated potential outcomes as
\begin{equation*}
    Y_{jt}(0)=\alpha_{j}+\lambda_{t}+\eta_{jt},
\end{equation*}
for fixed $\alpha_j$ and $\lambda_t$, and assuming \begin{equation}\label{eq:parallel_trends_did}
    E[\eta_{jt}]=0.
\end{equation}
Together these expressions imply $E[Y_{jt}(0)]=\alpha_{j}+\lambda_{t}$, which is how parallel trends is sometimes formalized
(c.f.\ Assumption 1 in \textcite{borusyak2021revisiting}; weaker versions of the parallel trends assumption yield analogous results). We do not restrict the dependence of $\eta_{jt}$ across units or time, nor do we make restrictions on the potentially random treatment effects $\tau_{jt, k}=Y_{jt}(k)-Y_{jt}(0)$. Collecting these effects in a vector $\tau_{jt}$, we have
\begin{equation}\label{eq:outcome-equation-did}
    Y_{jt}=X_{jt}'\tau_{jt} + \alpha_{j}+\lambda_{t}+\eta_{jt}.
\end{equation}
This outcome model reduces to a conventional \ac{TWFE} model under the assumption of constant treatment effects: $\tau_{jt}=\beta$ for all $(j, t)$.

Since the only source of randomness are the shocks $\eta_{jt}$ and the treatment effects $\tau_{jt}$, this setup fits into the framework of \Cref{sec:problem} if we interpret the expectation in \cref{eq:partially_linear} as averaging over the joint distribution of $\{\tau_{jt}, \eta_{jt}\}_{j=1,t=1}^{n, T}$. Specifically, $(\beta, g)$ are the minimizers of $N^{-1}\sum_{j=1}^{n}\sum_{t=1}^{T} E_{\tau, \eta}[(Y_{jt}-{X}_{jt}^\prime \tilde{\beta}-\tilde{g}(W_{jt}))^{2}]$, where the subscript on the expectation makes explicit that we only integrate over the joint distribution of $\{\tau_{jt}, \eta_{jt}\}_{j=1,t=1}^{n, T}$. The parallel trends assumption implies $\mu_0(W_i)=\alpha_{J_i}+\lambda_{T_i}$, so that \cref{eq:model_based} in \Cref{ass:model_or_design} holds. In other words, the parallel trend assumption implies that our controls $g(W_{i})$ correctly specify the untreated potential outcome mean. Additionally, \Cref{ass:mean_indep} holds trivially because the treatment vector is non-random.

To make the link to \Cref{theorem:main-result}, note that $\tilde{X}_{jt}=X_{jt}-\bar{X}_{j}-\bar{X}_{t}+\bar{X}$ coincides with the sample residual from regressing $X_{i}$ onto unit and time effects. Here $\bar{X}_{j}=\frac{1}{T}\sum_{t=1}^{T}X_{jt}$, $\bar{X}_{t}=\frac{1}{n}\sum_{j=1}^{n}X_{jt}$, and $\bar{X}=\frac{1}{n}\sum_{j=1}^{n}\bar{X}_{j}$.
We may then write \cref{eq:beta_general} as
\begin{equation}\label{eq:beta_vector}
    \begin{split}
    \beta&=
    \left(\sum_{j=1}^{n}\sum_{t=1}^{T} E_{\tau, \eta}[\tilde{{X}}_{jt}\tilde{{X}}_{jt}']\right)^{-1}
     \sum_{j=1}^{n}\sum_{t=1}^{T}
     E_{\tau, \eta}[\tilde{X}_{jt} Y_{jt}]\\
     &=\left(\sum_{j=1}^{n}\sum_{t=1}^{T} \tilde{X}_{jt}\tilde{X}_{jt}'\right)^{-1}
    \sum_{j=1}^{n}\sum_{t=1}^{T}\tilde{X}_{jt}X_{jt}'E[\tau_{jt}],
    \end{split}
\end{equation}
where the second equality uses \cref{eq:parallel_trends_did,eq:outcome-equation-did}, and the fact that only $\eta_{jt}$ and $\tau_{jt}$ are stochastic. \Cref{theorem:main-result} implies that the coefficient on the $k$th
element on $X_{jt}$ is given by
\begin{equation}\label{eq:betak_decomposition_did}
  \beta_{k}=\sum_{j, t}\lambda_{kk}(j, t)E[\tau_{jt, k}]+
  \sum_{\ell\neq k}\sum_{j, t}\lambda_{k\ell}(j, t)E[\tau_{jt, \ell}]
\end{equation}
where
\begin{equation*}
  \lambda_{kk}(j, t)=\frac{\dbtilde{X}_{jt, k}X_{jt, k}}{
    \sum_{j, t}\dbtilde{X}_{jt, k}^{2}}, \qquad\text{and}\qquad
  \lambda_{k\ell}(j, t)=  \frac{\dbtilde{X}_{jt, k}X_{jt, \ell}}{
    \sum_{j, t}\dbtilde{X}_{jt, k}^{2}},
\end{equation*}
and $\dbtilde{X}_{jt, k}$ is the sample residual from regressing
$\tilde{X}_{jt, k}$ onto the remaining elements of $\tilde{X}_{jt}$. Recall that since we do not assume that \cref{eq:pscore_in_G} holds, it is not guaranteed that $\lambda_{kk}(j, t) \geq 0$.

To unpack this result, we now consider four special cases from the literature.

\paragraph{Static binary treatment}

Consider a \ac{DiD} setting where units adopt (and potentially drop) a binary
treatment at different time periods---as studied by \textcite{de2020aer} and
\textcite{goodman2021difference}. For example, different states
$j$ may choose to roll out a policy in different years and a researcher
wishes to estimate the average effect of this policy using this staggered
adoption. We assume that the treatment is static, setting $D_{jt}=\dd_{jt}$,
with $K=\kk=1$. Since the treatment is binary, $X_{jt}=D_{jt}$ is a scalar with
$\dbtilde{X}_{jt,1}=\tilde{X}_{jt}$, and the second term in
\cref{eq:betak_decomposition_did} drops; the weights on the first term
simplify to
\begin{equation*}
  \lambda_{11}(j, t)=\frac{\tilde{X}_{jt}X_{jt}}{\sum_{j', t'}\tilde{X}_{j't'}^{2}}=
  \frac{(1-\overbar{X}_{j}-\overbar{X}_{t}+\overbar{X})X_{jt}}{\sum_{j', t'}\tilde{X}_{j't'}^{2}},
\end{equation*}
which coincides with the expression in Theorem 1 of \textcite{de2020aer}. These
treatment weights are not guaranteed to be convex since \cref{eq:pscore_in_G}
does not hold.\footnote{Since $E[X_{jt}\mid W_{jt}]=X_{jt}\in\{0,1\}$, if
  \cref{eq:pscore_in_G} held, then the residual $\tilde{X}_{jt}$ must be zero
  (this is true if, e.g., all units have the same treatment adoption date). But
  that would generate a multicollinearity issue, precluding the researcher from
  including unit and time effects in the regression.} In contrast,
\textcite{athey2022design} consider staggered \ac{DiD} regressions where
\cref{eq:pscore_in_G} holds because intervention timing is assumed to be random
(in place of the parallel trends assumption). Under this design-based
assumption, \Cref{theorem:main-result} shows the treatment weights
(corresponding to those in Theorem 1(iv) of \textcite{athey2022design}) are
convex.

The above expression for $\lambda_{11}$ yields a simple necessary and sufficient
condition for convex weights, which is that for units $j$ that are treated in
period $t$, $1-\overbar{X}_{j}-\overbar{X}_{t}+\overbar{X}\geq 0$. In staggered
adoption designs, $\overbar{X}_{t}$ is increasing with $t$. Thus, in staggered
adoption designs, it suffices to check this condition for $t=T$, and for unit
$j$ that adopts the treatment first---that is, to check whether
\begin{equation}\label{eq:convex_staggered}
  1-\max_{j}\overbar{X}_{j}-\overbar{X}_{T}+\overbar{X}\geq 0.
\end{equation}

Condition~\eqref{eq:convex_staggered} holds in the canonical \ac{DiD} case with
a single intervention date, where the first $n_1<n$ units treated in the last
$T_{1}<T$ periods and untreated in the earlier periods $1, \dotsc, T-T_{1}$. The
remaining units are never treated, so that
$D_{jt}=\dd_{jt}=\1{j\leq n_1,t\geq T-T_1}$. This nests the simplest \ac{DiD}
specification where $T=2$ and $T_1=1$. %
In this case, when units in the treatment group are treated,
$1-\overbar{X}_{j}-\overbar{X}_{t}+\overbar{X}=(1-n_{1}/n)(1-T_{1}/T)$ so that
the weights $\lambda_{11}(j, t)$ are non-negative, and
\cref{eq:betak_decomposition_did} simplifies to:
\begin{equation*}
  \beta_{1}=\sum_{j, t}\lambda_{11}(j, t)E[\tau_{jt,1}], \qquad
  \lambda_{11}(j, t)=\frac{(1-\frac{n_{1}}{n})(1-\frac{T_1}{T})X_{jt}}{
    (1-\frac{n_{1}}{n})(1-\frac{T_1}{T})\frac{n_{1}T_1}{nT}}
  =\frac{X_{jt}}{
    n_{1}T_1/N},
\end{equation*}
which is simply the average treatment effect for the $n_{1}T_1$ treated
observations.

However, in presence of multiple treatment adoption dates,
\cref{eq:convex_staggered} may fail. To illustrate, consider a case with three
time periods ($T=3$) and three groups of units: $\mathcal{E}$, $\mathcal{L}$,
and $\mathcal{N}$, with respective sizes $n_{E}$, $n_{L}$, and $n_{N}$. Units
$j\in\mathcal{E}$ are ``early adopters'', and are treated beginning in period
$2$. Units $j\in \mathcal{L}$ are ``late adopters'', and are treated only in
period $3$. Units in the last group are never treated.\footnote{This example is
  a special case of the example discussed in Figure 2 of
  \textcite{goodman2021difference}.} In this case, \cref{eq:convex_staggered}
simplifies to $1-2/3-(n_{E}+n_{L})/n+(2/3n_{E}+1/3n_{L})/n=(n_{N}-n_{L})/3n$,
which is negative if there are more late adopters than never adopters; otherwise,
if $n_{L}<n_{N}$, all weights are positive. Indeed, some algebra shows
\begin{align*}
\lambda_{11}(j,3) &= \frac{n_{E}+2n_{N}}{\kappa}
\qquad j \in \mathcal{L},\\
\lambda_{11}(j,2) &= \frac{n_{N}+2n_{L}}{\kappa}
\qquad j \in \mathcal{E},\\
\lambda_{11}(j,3) &= \frac{n_{N}-n_{L}}{\kappa}
\qquad j \in \mathcal{E},
\end{align*}
where $\kappa=2(n_{E}n_{L}+n_{E}n_{N}+n_{N}n_{L})$
and $\lambda_{11}(j, t)=0$ otherwise.

Condition~\eqref{eq:convex_staggered} is generally quite restrictive. Consider, for
instance, a setting in which no units are treated in the first period
and a fraction $1/T$ of observations adopts the treatment in period
$t=2,\dotsc, T$. Then for the group adopting treatment in period $2$,
\cref{eq:convex_staggered} becomes $(3-T)/2T$, which is negative if $T\geq 4$.
Similarly, condition~\eqref{eq:convex_staggered} fails if there exists an
always-treated group, or if everyone is treated in the last period.

\paragraph{Dynamic binary treatment with staggered adoption} Next, consider an
``event study'' setting in which each unit $j$ starts being treated in period
$A(j)\in\{1,2,\dotsc, T\}\hspace{0.05cm}\cup\hspace{0.05cm}\infty$ and remains
treated thereafter, with $A(j)=\infty$ denoting a unit that is never treated. Thus, $\dd_{jt}=\1{t>A(j)}$, with $\kk=1$. Unlike in previous cases, we allow for dynamic effects by letting $D_{jt}=t-A(j)$ index the number of periods
since the treatment adoption date (breaking with our usual indexing convention
of $D_{jt}\ge 0$), assuming no anticipation effect one period before adoption,
and correspondingly normalizing $D_{jt}=-1$ for the never-treated group. $X_{jt}$ then consists of indicators for all leads and lags relative to the
adoption date:
$X_{jt}=(\1{D_{jt}=-(T-1)}, \dotsc, \1{D_{jt}=-2}, \1{D_{jt}=0},
\dotsc, \1{D_{jt}=T-1})^\prime$, with the indicator for the period just prior to
adoption ($D_{jt}=-1$) excluded. This specification avoids perfect collinearity
when all treatment adoption dates are represented in the data (including the
never-treated group). %
\textcite{sun2021estimating,borusyak2021revisiting} study such ``fully-dynamic''
event study specifications.

Since $X_{jt}$ is now a vector with $K=2(T-1)$, the second contamination bias term in \cref{eq:betak_decomposition_did} will generally be present. As such, \textcite{sun2021estimating} and \textcite{borusyak2021revisiting} study the potential for contamination across estimates of post- and pre-treatment effects (with the latter used in conventional pre-trend specification tests). Furthermore, like in the previous case with static treatment, the own-treatment weights in the first term are potentially negative. While random treatment timing assumptions may solve the issue of negative own treatment weights, contamination bias remains a concern even under such assumptions.

To illustrate the potential for contamination bias, consider again the example
with early, late, and never adopters and $T=3$, except we now allow the
treatment effect to be dynamic. Let $\tau_{jts}=Y_{jt}(s)-Y_{jt}(-\infty)$,
$s\in \{-2,1,0,1\}$ denote the effect on unit $j$ in time period $t$ of adopting
the treatment $s$ periods ago. If $s$ is negative, we interpret this as the
anticipation effect of adopting the treatment $-s$ periods from now. Under our
assumptions $\tau_{jt,-1}=0$, such that there is no anticipation effect
immediately before treatment adoption. To test whether the two-period-ahead
anticipation effect is zero, and whether the effect of the treatment fades out
over time, we let
$X_{jt}=(\1{D_{jt}=-2}, \1{D_{jt}=0}, \1{D_{jt}=T-1}) ^\prime$. Thus, for
instance, $X_{j1}=(1,0,0)^\prime$ for late adopters while
$X_{j2}=(0,1,0)^\prime$ for early adopters. Let
$\tau_{E, ts}=n_{E}^{-1}\sum_{j\in \mathcal{E}}E[\tau_{jts}]$ denote the average
effect among early adopters, and define $\tau_{L, ts}$ similarly. Then some
rather tedious algebra shows that
\begin{equation*}
  \beta=
  \begin{pmatrix}
    \tau_{L, 1, -2}\\
    0\\
    \tau_{E,3, 1}\\
  \end{pmatrix}+\lambda_{E,0}\tau_{E,2,0}+\lambda_{L,0}\tau_{L,3,0},
\end{equation*}
where
\begin{equation*}
  \lambda_{E,0}=
\frac{1}{\zeta
    }  \begin{pmatrix}
    3n_{L}n_{E}+n_{N}n_{E}
    \\
    3n_{L}n_{E}+2n_{N}n_{E}
    \\
    -n_{L}n_{N}
  \end{pmatrix},\qquad
  \lambda_{L, 0}=
  \frac{1}{\zeta}  \begin{pmatrix}
    -3n_{L}n_{E} -n_{N}n_{E}
    \\
    3n_{E}n_{L} + 2n_{N}n_{L}
    \\
    n_{N}n_{L}
  \end{pmatrix},
\end{equation*}
and $\zeta=2(3n_{L}n_{E} + n_{E}n_{N}+n_{L}n_{N})$. In other words, the estimand for the two-period-ahead anticipation effect $\beta_{1}$ equals the anticipation effect for late adopters in period 1 (this is the only group we ever observe two periods before treatment) plus a contamination bias term coming from the effect of the treatment on impact.  Similarly, the estimand for the effect of the treatment one period since adoption, $\beta_{3}$, equals the effect for early adopters in period 3 (this is the only group we ever observe one period after treatment) plus a contamination bias term coming from the effect of the treatment on impact. The estimand for the effect of the treatment upon adoption, $\beta_{0}$, has no contamination bias, and equals a weighted average of the effect for early and late adopters. In this example, the own treatment weights are always positive, but the contamination weights can be large.  For instance, with equal-sized groups, $\lambda_{E,0}=(2/5,1/2,-1/10)'$ and $\lambda_{L,0}=(-2/5,1/2,1/10)'$, so the contamination weights in the estimand $\beta_{1}$ are almost as large as the own treatment weights for $\beta_{2}$.

It is worth noting that if all treated units share a single adoption date then
contamination bias disappears and a \ac{TWFE} regression recovers a vector of average dynamic treatment effects for the treated, in analogy to
the static case discussed above. To show this result, let us set
$A(j)=T_{1}$ for the first $n_{1}$ units, with $A(j)=\infty$ for the remaining
$n_{0}=n-n_{1}$ units. Excluding the indicator just prior to the adoption date,
as well as leads and lags that are always zero for all units, the treatment
vector has length $T-1$:
$X_{jt}=(\1{D_{jt}=-(T_{1}-1)},\dotsc,\1{D_{jt}=-2}, \1{D_{jt}=0}, \dotsc, \1{D_{jt}=T-T_{1}})$.
For the control units, this vector is always zero. For the adopters,
$X_{jt}=e_{t}$ (the $t$th unit vector) if $t\leq T_{1}-2$, $X_{j,T-1}$ is zero, and  $X_{jt}=e_{t-1}$ for $t\geq T_{1}$. We may write this
compactly as $X_{jt}=e_{t}\1{t< T_{1}-1}+e_{t-1}\1{t\geq T_{1}}$ for
$j\leq n_{1}$. Partialling out the unit and time effects therefore yields
\begin{equation*}
  \tilde{X}_{jt}=(\1{j\leq n_{1}}-n_{1}/n)(e_{t}\1{t< T_{1}-1}+e_{t-1}\1{t\geq T_{1}}-\iota_{T-1}/T),
\end{equation*}
where $\iota_{T-1}$ is a $T-1$ vector of ones. Hence,
$\sum_{j=1}^{n}\sum_{t=1}^{n}\tilde{X}_{jt}\tilde{X}_{jt}'=\frac{n_{1}n_{0}}{n}\left(I_{T-1}-\iota_{T-1}\iota_{T-1}'/T\right)$.
By the Woodbury identity, we therefore obtain
\begin{multline*}
  \Lambda(j, t)=\Big(\sum_{j=1}^{n}\sum_{t=1}^{n}\tilde{X}_{jt}\tilde{X}_{jt}'\Big)^{-1}\tilde{X}_{jt}X_{jt}'
  =\frac{n}{n_{1}n_{0}}(I_{T-1}+\iota_{T-1}\iota_{T-1}')\tilde{X}_{jt}X_{jt}'\\
  =\frac{1}{n_{1}}(I_{T-1}+\iota_{T-1}\iota_{T-1}')(X_{jt}-\iota_{T-1}/T)X_{jt}'
  =\frac{1}{n_{1}}X_{jt}X_{jt}'.
\end{multline*}
Hence, by \cref{eq:beta_vector}, \ac{TWFE} regression identifies the
average treatment for the treated,
$\beta=\frac{1}{n_{1}}\sum_{j=1}^{n_{1}}(\tau_{j1,-(T-1)}, \dotsc, \tau_{j, T_{1}-2,-2}, \tau_{jT_{1},1}, \dotsc, \tau_{jT, T-T_{1}})$.
Intuitively, since the contamination weights sum to zero and there is only one
group of adopters, the contamination weights must be identically zero.

\paragraph{Mover regressions: multiple treatments with multiple transitions.}

Finally, consider a ``mover regression'' in a setting with a static multivalued treatment $\dd_{jt}\in\{0, \dotsc, K\}$ with multiple transitions of units between treatment states, leading to multiple treatment paths. We focus on the static treatment case, setting $D_{jt}=\dd_{jt}$. This setting has been studied by \textcite{hull2018movers, de2020two}. Our \Cref{theorem:main-result} shows that such specifications can suffer from two distinct sources of bias: own-treatment negative weighting from multiple transitions and contamination bias from the multiple treatments. As before the former bias disappears under random treatment timing (as in \textcite{athey2022design}), or other assumptions which make \cref{eq:pscore_in_G} hold.

To illustrate this case, consider a setting with $T=3$ periods, $K=3$ treatments, and three groups of units, $\mathcal{E}$, $\mathcal{L}$, and $\mathcal{N}$. Units in the first group start out untreated, move to treatment $2$ in period $1$, and move to treatment $3$ in period $3$. Units in the second group start in treatment $1$, move to being untreated in period $2$, and move to treatment $2$ in period $3$. Units in group $\mathcal{N}$ are never treated. This example is isomorphic to the previous event study example, in that it leads to the same regression specification and the same  \cref{eq:betak_decomposition_did} characterization of regression coefficients. Thus, there are no negative own-treatment weights in this example, but there are potentially large contamination weights depending on the relative group sizes.

\section{Details on the Further Applications}\label{apx:empirical_apps}

This appendix details our procedure for selecting the additional empirical
examples in \Cref{sec:further_application}. We also discuss the implementation
details and provide the full set of results.

\subsection{Article Search Protocol}\label{sec:article_search_procedure}

We scraped the American Economic Association (AEA) website for a list of all published articles across all AEA journals over 2013--2022. This search included all articles from the following journals: \emph{American Economic Review},  \emph{American Economic Review: Insights}, \emph{American Economic Journal: Applied Economics},  \emph{American Economic Journal: Economic Policy},  \emph{American Economic Journal: Macroeconomics},  \emph{American Economic Journal: Microeconomics},  \emph{Journal of Economic Literature} (excluding articles with ``review'' in the title and articles labeled as Front Matter, Doctoral Dissertations, and Annotated Listings),  \emph{Journal of Economic Perspectives}, and \emph{AER/AEA Papers and Proceedings} (excluding articles with ``report'' or ``minutes'' in the title). We limited this search to articles with online replication packages which include at least one data file.\footnote{Here ``data files''  refers to those with any of the following extensions: Stata (\texttt{`dta'}), Excel (\texttt{`xls'} or \texttt{`xlsx'}), Matlab (\texttt{`mat'}), R (\texttt{`rdata'}, \texttt{`rda'}, \texttt{`rds'}), HDFS (\texttt{`h5`}, \texttt{`hdf5'}), Apache (\texttt{`parquet'}, \texttt{`arrow'}), SAS (\texttt{`sas7bdat'}), and delimited files (\texttt{`csv'}, \texttt{`tsv'}).}

We next filtered articles by two keyword searches of titles, abstracts, and main texts:
\begin{itemize}
  \item Experiments (keywords: stratified, random, RCT, experiment).
  \item Racial disparities (keywords: racial/ethnic differences, discrimination, disparities, gaps).
\end{itemize}
We focused on racial disparities as a set of possible examples because these papers typically have three or more categories, and they were easily identifiable based on keywords, giving us a systematic way to identify them.
These searches yielded a total of 1,848 experiments and 67 observational studies on race. To further narrow down experiments, we restricted attention to papers where one of the keywords appears in the paper's title, abstract, or associated tweet.

For each search, we then manually reviewed papers in reverse citation order (as measured by Google Scholar) keeping those which include in the main text a linear regression of some outcome on multiple treatments or race indicators and controls. We ignored specifications where a single treatment or race indicator is interacted with some set of fixed effects or controls, such as event study specifications. We stopped the review when five papers were identified with such a specification, or when we exhausted all papers in the search.

\subsection{Overlap Sample and Propensity Score Variation}\label{apx:overlap}
For each main specification, we identify a subset of the analysis sample with full treatment overlap using the following procedure. First, we define a primary strata variable (when not otherwise obvious from the paper) as the discrete variable with the greatest number of unique levels. In the experimental applications this is always the randomization strata; in the observational applications this is the ``finest'' fixed effect. We then drop observations for the levels of this variable which do not exhibit all levels of the treatment. Finally, in the remaining sample, we drop any additional controls which have no within-treatment variation.

We check for meaningful propensity score variation in each specification with two tests, summarized in \Cref{tab:stat_tests}. Specifically, we compute the Wald and LM tests of the null hypothesis that, in a multinomial logit regression of the treatment on the controls, all coefficients on the controls equal zero. The table gives evidence for statistically significant propensity score variation (at 10\% level) in the Project STAR application, two of the additional experimental applications (Cole et al.\ and Benhassine et al.), and all three observational studies.

\begin{table}[tp]
\begin{threeparttable}
\caption{Tests of Propensity Score Variation}\label{tab:stat_tests}
\begin{tabular}{@{}lrrr rrrr@{}}
  \toprule
  & \multicolumn{3}{c}{Wald}  & \multicolumn{3}{c}{LM}    \\
  \cmidrule(lr){2-4}  \cmidrule(lr){5-7}
  & Statistic & (d.f.) & p-value & Statistic & (d.f.) & p-value \\
  \midrule
  \input{./pscore.tex}\\
  \bottomrule
\end{tabular}
\begin{tablenotes}
\item\footnotesize \emph{Notes:} This table summarizes Wald and Lagrange multiplier tests of the null hypothesis that the coefficients on the controls in a multinomial logit regression of the treatment on the controls all equal zero. The tests allow for clustering in Benhassine et al., Duflo et al., Rim et al., and Weisburst, and for heteroskedasticity in the remaining applications.
  \end{tablenotes}
\end{threeparttable}
\end{table}

\subsection{Full Results}\label{apdx_fullresults}
In \Cref{tab:benhassine_multe}-\ref{tab:weisburst_multe}, we report the
estimated effects for each additional application. Panel A of each table first
reports the $\hat{\beta}$ estimates from the multiple-treatment regression as
reported in the original paper and corresponding standard errors. We also report
the own-treatment effect component from the decomposition in \cref{eq:beta_hat}
along with three alternative estimators: the ATE estimator, the
easiest-to-estimate weighted ATE estimator (EW) and the common-weight (CW) estimator. Panel B reports the difference between $\hat{\beta}$ and these 4 alternative estimators. The $\hat{\beta}$, EW and CW estimators are consistent even without overlap. However, if overlap fails in the full sample, the own-treatment effect component from the decomposition in \cref{eq:beta_hat} may not be identified for all treatments, and the ATE is not identified. If identification of the decomposition fails for the full treatment vector, we subset to the overlap sample, as described in \Cref{apx:overlap} above, and report the full set of estimates from the different estimators.

\clearpage

\begin{landscape}
\begin{table}
\begin{threeparttable}
\caption{Full results: \textcite{benhassine2015}}\label{tab:benhassine_multe}
\small
\begin{tabular}{@{}l SSSSS SSSSS @{}}
  \toprule
  & \multicolumn{5}{c}{Full sample}  & \multicolumn{5}{c}{Overlap}\\
  \cmidrule(lr){2-6}  \cmidrule(lr){7-11}
  A\@. Estimates & {$\hat{\beta}$} & {Own} & {\ac{ATE}} & {\ac{EW}} & {\ac{CW}}&
{$\hat{\beta}$} & {Own} & {\ac{ATE}} & {\ac{EW}} & {\ac{CW}}\\
  \cmidrule(lr){2-6}  \cmidrule(lr){7-11}
  \input{./benhassineA.tex}\\
  \midrule
  \multicolumn{2}{@{}l}{B\@. Bias}\\
  \input{./benhassineB.tex}\\
  \bottomrule
\end{tabular}
\begin{tablenotes}
\item\footnotesize\emph{Notes:} This table reports estimates from the Benhassine
  et al.\ application, as described in \Cref{apdx_fullresults}. The regression
  specification comes from column 1 of Table 5 in \textcite{benhassine2015}.
  Standard errors clustered by school sector are reported in parentheses.
  Standard errors assuming known propensity scores are reported in square
  brackets.
  \end{tablenotes}
\end{threeparttable}
\end{table}
\end{landscape}

\begin{landscape}
\begin{table}
\begin{threeparttable}
\caption{Full results: \textcite{cole2013barriers}}\label{tab:cole_multe}
\scriptsize
\begin{tabular}{@{}l SSSSS SSSSS @{}}
  \toprule
  & \multicolumn{5}{c}{Full sample}  & \multicolumn{5}{c}{Overlap}\\
  \cmidrule(lr){2-6}  \cmidrule(lr){7-11}
  A\@. Estimates & {$\hat{\beta}$} & {Own} & {\ac{ATE}} & {\ac{EW}} & {\ac{CW}}&
{$\hat{\beta}$} & {Own} & {\ac{ATE}} & {\ac{EW}} & {\ac{CW}}\\
  \cmidrule(lr){2-6}  \cmidrule(lr){7-11}
  \input{./coleA.tex}\\
  \midrule
  \multicolumn{2}{@{}l}{B\@. Bias}\\
  \input{./coleB.tex}\\
  \bottomrule
\end{tabular}
\begin{tablenotes}\item\footnotesize
  \emph{Notes:} This table reports estimates from the Cole et a.\ application,
  as described in \Cref{apdx_fullresults}. The regression specification comes
  from column 6 of Table 7 in \textcite{cole2013barriers}. Robust standard
  errors are reported in parentheses.
  Standard errors assuming known propensity scores are reported in square
  brackets.
\end{tablenotes}
\end{threeparttable}
\end{table}
\end{landscape}

\begin{table}
\begin{threeparttable}
\caption{Full results: \textcite{demel2013}}\label{tab:demel_multe}
\begin{tabular}{@{}l SSSSS@{}}
\toprule
  & \multicolumn{5}{c}{Full sample}\\
  \cmidrule(lr){2-6}
  A\@. Estimates & {$\hat{\beta}$} & {Own} & {\ac{ATE}} & {\ac{EW}} & {\ac{CW}}\\
  \cmidrule(lr){2-6}
  \input{./demelA.tex}\\
  \midrule
  \multicolumn{2}{@{}l}{B\@. Bias}\\
  \input{./demelB.tex}\\
  \bottomrule
\end{tabular}
\begin{tablenotes}\item\footnotesize
  \emph{Notes:} This table reports all results from the \textcite{demel2013}
  application, as described in \Cref{apdx_fullresults}. The regression
  specification comes from column 2 of Table 2 in \textcite{demel2013}. Robust
  standard errors are reported in parentheses. Standard errors assuming known
  propensity scores are reported in square brackets.
\end{tablenotes}
\end{threeparttable}
\end{table}

\begin{table}
\begin{threeparttable}
\caption{Full results: \textcite{drexler2014}}\label{tab:drexler_multe}
\begin{tabular}{@{}l SSSSS@{}}
\toprule
  & \multicolumn{5}{c}{Full sample}\\
  \cmidrule(lr){2-6}
  A\@. Estimates & {$\hat{\beta}$} & {Own} & {\ac{ATE}} & {\ac{EW}} & {\ac{CW}}\\
  \cmidrule(lr){2-6}
  \input{./drexlerA.tex}\\[1em]
  \midrule
  \multicolumn{2}{@{}l}{B\@. Bias}\\
  \input{./drexlerB.tex}\\
  \bottomrule
\end{tabular}
\begin{tablenotes}\item\footnotesize
  \emph{Notes:} This table reports estimates from the \textcite{drexler2014}
  application, as described in \Cref{apdx_fullresults}. The regression
  specification comes from row 2 of Table 2 in \textcite{drexler2014}. Robust
  standard errors are reported in parentheses.
  Standard errors assuming known propensity scores are reported in square
  brackets.
   \end{tablenotes}
 \end{threeparttable}
\end{table}

\begin{landscape}
\begin{table}
\begin{threeparttable}
\caption{Full results: \textcite{ddk15}}\label{tab:duflo_multe}
\begin{tabular}{@{}l SSS@{}SS SSSSS @{}}
  \toprule
  & \multicolumn{5}{c}{Full sample}  & \multicolumn{5}{c}{Overlap}\\
  \cmidrule(lr){2-6}  \cmidrule(lr){7-11}
  A\@. Estimates & {$\hat{\beta}$} & {Own} & {\ac{ATE}} & {\ac{EW}} & {\ac{CW}}&
{$\hat{\beta}$} & {Own} & {\ac{ATE}} & {\ac{EW}} & {\ac{CW}}\\
  \cmidrule(lr){2-6}  \cmidrule(lr){7-11}
  \input{./dufloA.tex}\\
  \midrule
  \multicolumn{2}{@{}l}{B\@. Bias}\\
  \input{./dufloB.tex}\\
  \bottomrule
\end{tabular}
\begin{tablenotes}\item\footnotesize\emph{Notes:} This table reports estimates
  from the \textcite{ddk15} application, as described in \Cref{apdx_fullresults}.
  The regression specification comes from column 1 of Table 2, panel A in
  \textcite{ddk15}. Standard errors clustered by school reported in parentheses.
  Standard errors assuming known propensity scores are reported in square
  brackets.
  \end{tablenotes}
\end{threeparttable}
\end{table}
\end{landscape}

\begin{landscape}
\begin{table}
\begin{threeparttable}
\caption{Full results: \textcite{fryerlevitt2013}}\label{tab:fryerlevitt_multe}
\scriptsize
\begin{tabular}{@{}l SSSSS SSSSS @{}}
  \toprule
  & \multicolumn{5}{c}{Full sample}  & \multicolumn{5}{c}{Overlap}\\
  \cmidrule(lr){2-6}  \cmidrule(lr){7-11}
  A\@. Estimates & {$\hat{\beta}$} & {Own} & {\ac{ATE}} & {\ac{EW}} & {\ac{CW}}&
{$\hat{\beta}$} & {Own} & {\ac{ATE}} & {\ac{EW}} & {\ac{CW}}\\
  \cmidrule(lr){2-6}  \cmidrule(lr){7-11}
  \input{./fryer_levittA.tex}\\
  \midrule
  \multicolumn{2}{@{}l}{B\@. Bias}\\
  \input{./fryer_levittB.tex}\\
  \bottomrule
\end{tabular}
\begin{tablenotes}\item\footnotesize\emph{Notes:}
  This table reports estimates from the \textcite{fryerlevitt2013} application, as
  described in \Cref{apdx_fullresults}. The regression specification comes from
  column 4 of Table 3 in \textcite{fryerlevitt2013}. Robust standard errors are
  reported in parentheses. Standard errors assuming known propensity scores are
  reported in square brackets.
\end{tablenotes}
\end{threeparttable}
\end{table}
\end{landscape}

\begin{landscape}
\begin{table}
\begin{threeparttable}
\caption{Full results: \textcite{rim2020}}\label{tab:rim_multe}
\begin{tabular}{@{}l @{}SSS@{}SS SSSSS @{}}
  \toprule
  & \multicolumn{5}{c}{Full sample}  & \multicolumn{5}{c}{Overlap}\\
  \cmidrule(lr){2-6}  \cmidrule(lr){7-11}
  A\@. Estimates & {$\hat{\beta}$} & {Own} & {\ac{ATE}} & {\ac{EW}} & {\ac{CW}}&
{$\hat{\beta}$} & {Own} & {\ac{ATE}} & {\ac{EW}} & {\ac{CW}}\\
  \cmidrule(lr){2-6}  \cmidrule(lr){7-11}
  \input{./rimA.tex}\\
  \midrule
  \multicolumn{2}{@{}l}{B\@. Bias}\\
  \input{./rimB.tex}\\
  \bottomrule
\end{tabular}
\begin{tablenotes}\item\footnotesize\emph{Notes:} This table reports estimates
  from the \textcite{rim2020} application, as described in
  \Cref{apdx_fullresults}. The regression specification comes from column 3 of
  Table 2 in \textcite{rim2020}. Standard errors clustered by cohort are
  reported in parentheses. Standard errors assuming known propensity scores are
  reported in square brackets.
   \end{tablenotes}
 \end{threeparttable}
\end{table}
\end{landscape}

\begin{table}
\begin{threeparttable}
\caption{Full results: \textcite{weisburst2019}}\label{tab:weisburst_multe}
\begin{tabular}{@{}l SSSSS@{}}
\toprule
  & \multicolumn{5}{c}{Full sample}\\
  \cmidrule(lr){2-6}
  A\@. Estimates & {$\hat{\beta}$} & {Own} & {\ac{ATE}} & {\ac{EW}} & {\ac{CW}}\\
  \cmidrule(lr){2-6}
  \input{./weisburstA.tex}\\
  \midrule
  \multicolumn{2}{@{}l}{B\@. Bias}\\
  \input{./weisburstB.tex}\\
  \bottomrule
\end{tabular}
\begin{tablenotes}\item\footnotesize
  \emph{Notes:} This table reports all results from the \textcite{weisburst2019}
  application, as described in \Cref{apdx_fullresults}. The regression
  specification comes from Table 2, panel A in \textcite{weisburst2019}.
  Standard errors clustered by police beat are reported in parentheses. Standard
  errors that assume the propensity scores are known are reported in square
  brackets.
   \end{tablenotes}
 \end{threeparttable}
\end{table}

\clearpage

\section{Additional Figures}\label{apx:additional_fig}

\begin{figure}[H]
  \centering
  \includegraphics[width=0.5\linewidth]{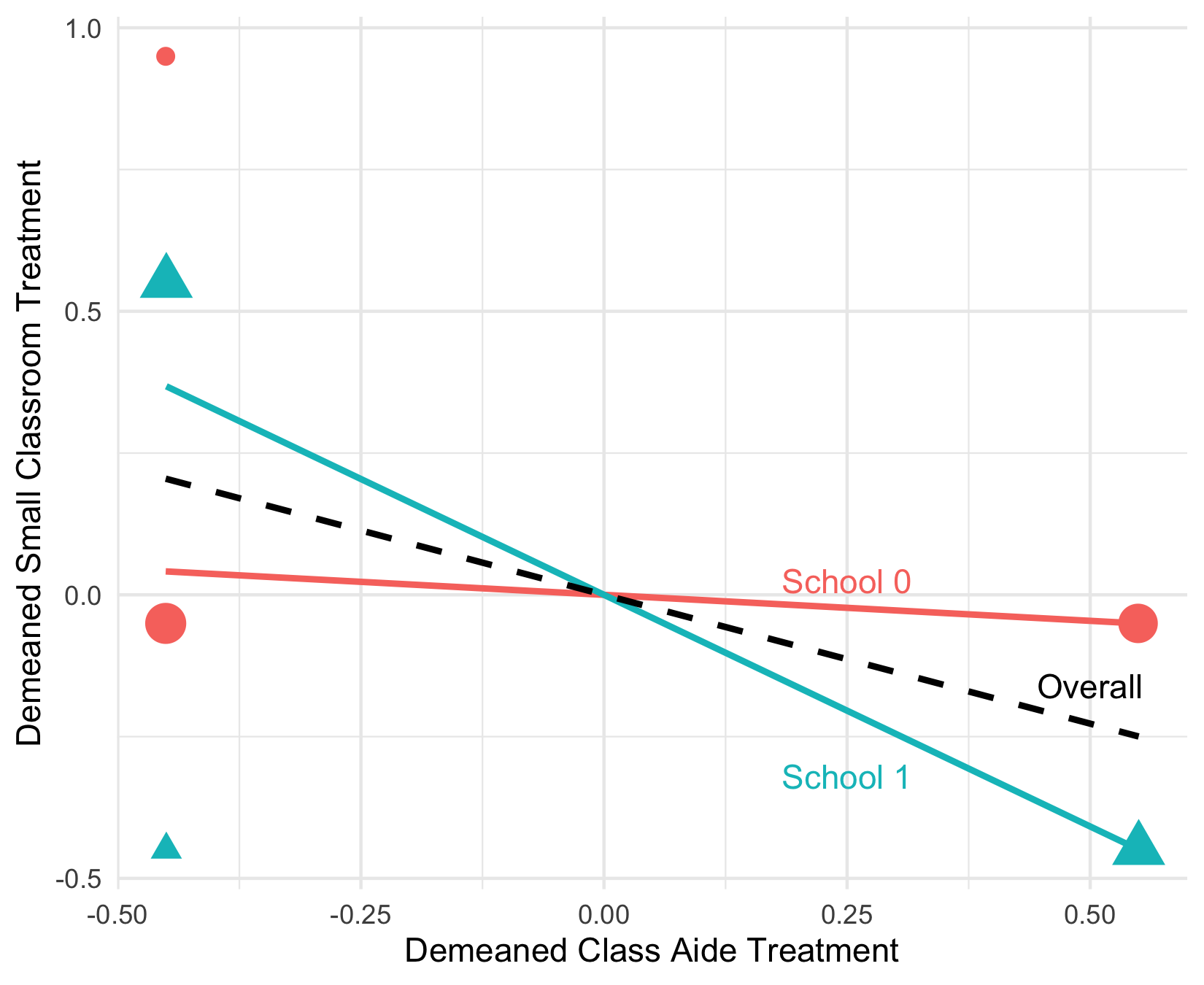}
  \caption{Regression of Small Classroom Treatment on Class Aide Treatment}\label{fig:example_misspecification}
  \floatfoot{\emph{Note:} This figure plots values of the demeaned class aide
    treatment ($\tilde{X}_{2i}$, the $x$-axis) against values of the demeaned
    small classroom treatment ($\tilde{X}_{1i}$, the $y$-axis) in our numerical
    example from \Cref{sec:illustr-intu}. The size of the points corresponds to
    the density of observations. The solid red and blue lines mark the
    within-school regression of the two residualized treatments, while the
    dashed black line is the overall regression line. The residuals from this
    line give $\dbtilde{X}_{i1}$.}
\end{figure}

\begin{figure}
  \centering
  \small
  \resizebox{0.9\linewidth}{!}{\input{./cont_weight.tex}}
    \caption{Project STAR contamination weights.}\label{fig:contamination_weights}
    \floatfoot{\emph{Notes:} This figure shows correlations between estimated
      school-specific treatment effects and contamination weights. Panel A
      depicts the correlation between the estimated teaching aide treatment
      effects by school against the estimated contamination weight for the small
      class estimate. Panel B gives the correlation between the estimated small
      class treatment effects by school against the estimated contamination
      weight for the teaching aide estimate. Correlations are reported on each
      panel. The size of the points is proportional to the number of students
      enrolled in each school.}
\end{figure}

\begin{figure}
  \centering \small
  \resizebox{0.9\linewidth}{!}{\input{./own_weight.tex}}
  \vspace{-1ex}\caption{Project STAR treatment weights}\label{fig:own_weights}
  \floatfoot{\emph{Notes:} This figure shows correlations between estimated
    school-specific treatment effects and the weights used by different
    estimators. Panel A gives the correlations for the small class treatment,
    and Panel B gives them for the teaching aide treatment. The first row plots
    the own treatment weights from the contamination bias decomposition in
    \cref{eq:beta_hat}. The second row gives plots the \ac{EW} scheme from
    \Cref{cor:angrist}, and the third row gives the \ac{CW} scheme from
    \Cref{theorem:F-weighting}. Correlations are reported on each panel. The
    size of the points is proportional to the number of students enrolled in
    each school.}
\end{figure}

\end{appendices}
\end{document}

%% file: starA.tex
Small & 5.357 & 5.202 & 5.561 & 5.295 & 5.577\\
 & (0.778) & (0.778) & (0.763) & (0.775) & (0.764)\\
 &  &  & [0.744] & [0.743] & [0.742]\\
Aide & 0.177 & 0.360 & 0.070 & 0.263 & 0.011\\
 & (0.720) & (0.714) & (0.708) & (0.715) & (0.712)\\
 &  &  & [0.694] & [0.691] & [0.695]\\
Number of controls & \multicolumn{1}{r}{77} &  &  &  & \\
Sample size & \multicolumn{1}{r}{5,868} &  &  &  & 

%% file: starB_text.tex
Small class size & 0.155 & -1.654 & 1.670\\
 & (0.160) & (0.185) & (0.187)\\
Teaching aide & -0.183 & -1.529 & 1.530\\
 & (0.149) & (0.176) & (0.177)

%% file: bias_t.tex
\begin{tikzpicture}[x=1pt,y=1pt]
\definecolor{fillColor}{RGB}{255,255,255}
\path[use as bounding box,fill=fillColor,fill opacity=0.00] (0,0) rectangle (448.07,549.25);
\begin{scope}
\path[clip] (  0.00,  0.00) rectangle (448.07,549.25);
\definecolor{drawColor}{RGB}{255,255,255}
\definecolor{fillColor}{RGB}{255,255,255}

\path[draw=drawColor,line width= 0.6pt,line join=round,line cap=round,fill=fillColor] ( -0.00,  0.00) rectangle (448.07,549.25);
\end{scope}
\begin{scope}
\path[clip] (156.13,466.00) rectangle (297.10,525.97);
\definecolor{fillColor}{RGB}{255,255,255}

\path[fill=fillColor] (156.13,466.00) rectangle (297.10,525.97);
\definecolor{drawColor}{RGB}{238,228,218}

\path[draw=drawColor,line width= 0.6pt,dash pattern=on 4pt off 4pt ,line join=round] (156.13,474.57) --
	(297.10,474.57);

\path[draw=drawColor,line width= 0.6pt,dash pattern=on 4pt off 4pt ,line join=round] (156.13,488.85) --
	(297.10,488.85);

\path[draw=drawColor,line width= 0.6pt,dash pattern=on 4pt off 4pt ,line join=round] (156.13,503.13) --
	(297.10,503.13);

\path[draw=drawColor,line width= 0.6pt,dash pattern=on 4pt off 4pt ,line join=round] (156.13,517.40) --
	(297.10,517.40);

\path[draw=drawColor,line width= 0.6pt,dash pattern=on 4pt off 4pt ,line join=round] (162.54,466.00) --
	(162.54,525.97);

\path[draw=drawColor,line width= 0.6pt,dash pattern=on 4pt off 4pt ,line join=round] (188.25,466.00) --
	(188.25,525.97);

\path[draw=drawColor,line width= 0.6pt,dash pattern=on 4pt off 4pt ,line join=round] (213.97,466.00) --
	(213.97,525.97);

\path[draw=drawColor,line width= 0.6pt,dash pattern=on 4pt off 4pt ,line join=round] (239.69,466.00) --
	(239.69,525.97);

\path[draw=drawColor,line width= 0.6pt,dash pattern=on 4pt off 4pt ,line join=round] (265.41,466.00) --
	(265.41,525.97);

\path[draw=drawColor,line width= 0.6pt,dash pattern=on 4pt off 4pt ,line join=round] (291.12,466.00) --
	(291.12,525.97);
\definecolor{fillColor}{RGB}{230,159,0}

\path[fill=fillColor] (162.54,468.14) rectangle (166.80,480.99);

\path[fill=fillColor] (162.54,482.42) rectangle (175.65,495.27);

\path[fill=fillColor] (162.54,496.70) rectangle (241.49,509.55);

\path[fill=fillColor] (162.54,510.98) rectangle (290.69,523.83);
\end{scope}
\begin{scope}
\path[clip] (156.13,415.31) rectangle (297.10,461.00);
\definecolor{fillColor}{RGB}{255,255,255}

\path[fill=fillColor] (156.13,415.31) rectangle (297.10,461.00);
\definecolor{drawColor}{RGB}{238,228,218}

\path[draw=drawColor,line width= 0.6pt,dash pattern=on 4pt off 4pt ,line join=round] (156.13,423.88) --
	(297.10,423.88);

\path[draw=drawColor,line width= 0.6pt,dash pattern=on 4pt off 4pt ,line join=round] (156.13,438.16) --
	(297.10,438.16);

\path[draw=drawColor,line width= 0.6pt,dash pattern=on 4pt off 4pt ,line join=round] (156.13,452.44) --
	(297.10,452.44);

\path[draw=drawColor,line width= 0.6pt,dash pattern=on 4pt off 4pt ,line join=round] (162.54,415.31) --
	(162.54,461.00);

\path[draw=drawColor,line width= 0.6pt,dash pattern=on 4pt off 4pt ,line join=round] (188.25,415.31) --
	(188.25,461.00);

\path[draw=drawColor,line width= 0.6pt,dash pattern=on 4pt off 4pt ,line join=round] (213.97,415.31) --
	(213.97,461.00);

\path[draw=drawColor,line width= 0.6pt,dash pattern=on 4pt off 4pt ,line join=round] (239.69,415.31) --
	(239.69,461.00);

\path[draw=drawColor,line width= 0.6pt,dash pattern=on 4pt off 4pt ,line join=round] (265.41,415.31) --
	(265.41,461.00);

\path[draw=drawColor,line width= 0.6pt,dash pattern=on 4pt off 4pt ,line join=round] (291.12,415.31) --
	(291.12,461.00);
\definecolor{fillColor}{RGB}{230,159,0}

\path[fill=fillColor] (162.54,417.45) rectangle (162.77,430.30);

\path[fill=fillColor] (162.54,431.73) rectangle (211.78,444.58);

\path[fill=fillColor] (162.54,446.01) rectangle (277.68,458.86);
\end{scope}
\begin{scope}
\path[clip] (156.13,336.07) rectangle (297.10,410.31);
\definecolor{fillColor}{RGB}{255,255,255}

\path[fill=fillColor] (156.13,336.07) rectangle (297.10,410.31);
\definecolor{drawColor}{RGB}{238,228,218}

\path[draw=drawColor,line width= 0.6pt,dash pattern=on 4pt off 4pt ,line join=round] (156.13,344.63) --
	(297.10,344.63);

\path[draw=drawColor,line width= 0.6pt,dash pattern=on 4pt off 4pt ,line join=round] (156.13,358.91) --
	(297.10,358.91);

\path[draw=drawColor,line width= 0.6pt,dash pattern=on 4pt off 4pt ,line join=round] (156.13,373.19) --
	(297.10,373.19);

\path[draw=drawColor,line width= 0.6pt,dash pattern=on 4pt off 4pt ,line join=round] (156.13,387.47) --
	(297.10,387.47);

\path[draw=drawColor,line width= 0.6pt,dash pattern=on 4pt off 4pt ,line join=round] (156.13,401.75) --
	(297.10,401.75);

\path[draw=drawColor,line width= 0.6pt,dash pattern=on 4pt off 4pt ,line join=round] (162.54,336.07) --
	(162.54,410.31);

\path[draw=drawColor,line width= 0.6pt,dash pattern=on 4pt off 4pt ,line join=round] (188.25,336.07) --
	(188.25,410.31);

\path[draw=drawColor,line width= 0.6pt,dash pattern=on 4pt off 4pt ,line join=round] (213.97,336.07) --
	(213.97,410.31);

\path[draw=drawColor,line width= 0.6pt,dash pattern=on 4pt off 4pt ,line join=round] (239.69,336.07) --
	(239.69,410.31);

\path[draw=drawColor,line width= 0.6pt,dash pattern=on 4pt off 4pt ,line join=round] (265.41,336.07) --
	(265.41,410.31);

\path[draw=drawColor,line width= 0.6pt,dash pattern=on 4pt off 4pt ,line join=round] (291.12,336.07) --
	(291.12,410.31);
\definecolor{fillColor}{RGB}{86,180,233}

\path[fill=fillColor] (162.54,338.21) rectangle (169.55,351.06);

\path[fill=fillColor] (162.54,352.49) rectangle (190.41,365.34);

\path[fill=fillColor] (162.54,366.76) rectangle (191.10,379.61);

\path[fill=fillColor] (162.54,381.04) rectangle (198.67,393.89);

\path[fill=fillColor] (162.54,395.32) rectangle (242.86,408.17);
\end{scope}
\begin{scope}
\path[clip] (156.13,299.65) rectangle (297.10,331.07);
\definecolor{fillColor}{RGB}{255,255,255}

\path[fill=fillColor] (156.13,299.65) rectangle (297.10,331.07);
\definecolor{drawColor}{RGB}{238,228,218}

\path[draw=drawColor,line width= 0.6pt,dash pattern=on 4pt off 4pt ,line join=round] (156.13,308.22) --
	(297.10,308.22);

\path[draw=drawColor,line width= 0.6pt,dash pattern=on 4pt off 4pt ,line join=round] (156.13,322.50) --
	(297.10,322.50);

\path[draw=drawColor,line width= 0.6pt,dash pattern=on 4pt off 4pt ,line join=round] (162.54,299.65) --
	(162.54,331.07);

\path[draw=drawColor,line width= 0.6pt,dash pattern=on 4pt off 4pt ,line join=round] (188.25,299.65) --
	(188.25,331.07);

\path[draw=drawColor,line width= 0.6pt,dash pattern=on 4pt off 4pt ,line join=round] (213.97,299.65) --
	(213.97,331.07);

\path[draw=drawColor,line width= 0.6pt,dash pattern=on 4pt off 4pt ,line join=round] (239.69,299.65) --
	(239.69,331.07);

\path[draw=drawColor,line width= 0.6pt,dash pattern=on 4pt off 4pt ,line join=round] (265.41,299.65) --
	(265.41,331.07);

\path[draw=drawColor,line width= 0.6pt,dash pattern=on 4pt off 4pt ,line join=round] (291.12,299.65) --
	(291.12,331.07);
\definecolor{fillColor}{RGB}{86,180,233}

\path[fill=fillColor] (162.54,301.80) rectangle (192.10,314.65);

\path[fill=fillColor] (162.54,316.07) rectangle (229.42,328.92);
\end{scope}
\begin{scope}
\path[clip] (156.13,263.24) rectangle (297.10,294.65);
\definecolor{fillColor}{RGB}{255,255,255}

\path[fill=fillColor] (156.13,263.24) rectangle (297.10,294.65);
\definecolor{drawColor}{RGB}{238,228,218}

\path[draw=drawColor,line width= 0.6pt,dash pattern=on 4pt off 4pt ,line join=round] (156.13,271.81) --
	(297.10,271.81);

\path[draw=drawColor,line width= 0.6pt,dash pattern=on 4pt off 4pt ,line join=round] (156.13,286.09) --
	(297.10,286.09);

\path[draw=drawColor,line width= 0.6pt,dash pattern=on 4pt off 4pt ,line join=round] (162.54,263.24) --
	(162.54,294.65);

\path[draw=drawColor,line width= 0.6pt,dash pattern=on 4pt off 4pt ,line join=round] (188.25,263.24) --
	(188.25,294.65);

\path[draw=drawColor,line width= 0.6pt,dash pattern=on 4pt off 4pt ,line join=round] (213.97,263.24) --
	(213.97,294.65);

\path[draw=drawColor,line width= 0.6pt,dash pattern=on 4pt off 4pt ,line join=round] (239.69,263.24) --
	(239.69,294.65);

\path[draw=drawColor,line width= 0.6pt,dash pattern=on 4pt off 4pt ,line join=round] (265.41,263.24) --
	(265.41,294.65);

\path[draw=drawColor,line width= 0.6pt,dash pattern=on 4pt off 4pt ,line join=round] (291.12,263.24) --
	(291.12,294.65);
\definecolor{fillColor}{RGB}{86,180,233}

\path[fill=fillColor] (162.54,265.38) rectangle (212.29,278.23);

\path[fill=fillColor] (162.54,279.66) rectangle (225.43,292.51);
\end{scope}
\begin{scope}
\path[clip] (156.13,198.27) rectangle (297.10,258.24);
\definecolor{fillColor}{RGB}{255,255,255}

\path[fill=fillColor] (156.13,198.27) rectangle (297.10,258.24);
\definecolor{drawColor}{RGB}{238,228,218}

\path[draw=drawColor,line width= 0.6pt,dash pattern=on 4pt off 4pt ,line join=round] (156.13,206.84) --
	(297.10,206.84);

\path[draw=drawColor,line width= 0.6pt,dash pattern=on 4pt off 4pt ,line join=round] (156.13,221.12) --
	(297.10,221.12);

\path[draw=drawColor,line width= 0.6pt,dash pattern=on 4pt off 4pt ,line join=round] (156.13,235.40) --
	(297.10,235.40);

\path[draw=drawColor,line width= 0.6pt,dash pattern=on 4pt off 4pt ,line join=round] (156.13,249.68) --
	(297.10,249.68);

\path[draw=drawColor,line width= 0.6pt,dash pattern=on 4pt off 4pt ,line join=round] (162.54,198.27) --
	(162.54,258.24);

\path[draw=drawColor,line width= 0.6pt,dash pattern=on 4pt off 4pt ,line join=round] (188.25,198.27) --
	(188.25,258.24);

\path[draw=drawColor,line width= 0.6pt,dash pattern=on 4pt off 4pt ,line join=round] (213.97,198.27) --
	(213.97,258.24);

\path[draw=drawColor,line width= 0.6pt,dash pattern=on 4pt off 4pt ,line join=round] (239.69,198.27) --
	(239.69,258.24);

\path[draw=drawColor,line width= 0.6pt,dash pattern=on 4pt off 4pt ,line join=round] (265.41,198.27) --
	(265.41,258.24);

\path[draw=drawColor,line width= 0.6pt,dash pattern=on 4pt off 4pt ,line join=round] (291.12,198.27) --
	(291.12,258.24);
\definecolor{fillColor}{RGB}{86,180,233}

\path[fill=fillColor] (162.54,200.42) rectangle (187.94,213.27);

\path[fill=fillColor] (162.54,214.69) rectangle (197.47,227.54);

\path[fill=fillColor] (162.54,228.97) rectangle (211.46,241.82);

\path[fill=fillColor] (162.54,243.25) rectangle (222.84,256.10);
\end{scope}
\begin{scope}
\path[clip] (156.13,147.58) rectangle (297.10,193.27);
\definecolor{fillColor}{RGB}{255,255,255}

\path[fill=fillColor] (156.13,147.58) rectangle (297.10,193.27);
\definecolor{drawColor}{RGB}{238,228,218}

\path[draw=drawColor,line width= 0.6pt,dash pattern=on 4pt off 4pt ,line join=round] (156.13,156.15) --
	(297.10,156.15);

\path[draw=drawColor,line width= 0.6pt,dash pattern=on 4pt off 4pt ,line join=round] (156.13,170.43) --
	(297.10,170.43);

\path[draw=drawColor,line width= 0.6pt,dash pattern=on 4pt off 4pt ,line join=round] (156.13,184.71) --
	(297.10,184.71);

\path[draw=drawColor,line width= 0.6pt,dash pattern=on 4pt off 4pt ,line join=round] (162.54,147.58) --
	(162.54,193.27);

\path[draw=drawColor,line width= 0.6pt,dash pattern=on 4pt off 4pt ,line join=round] (188.25,147.58) --
	(188.25,193.27);

\path[draw=drawColor,line width= 0.6pt,dash pattern=on 4pt off 4pt ,line join=round] (213.97,147.58) --
	(213.97,193.27);

\path[draw=drawColor,line width= 0.6pt,dash pattern=on 4pt off 4pt ,line join=round] (239.69,147.58) --
	(239.69,193.27);

\path[draw=drawColor,line width= 0.6pt,dash pattern=on 4pt off 4pt ,line join=round] (265.41,147.58) --
	(265.41,193.27);

\path[draw=drawColor,line width= 0.6pt,dash pattern=on 4pt off 4pt ,line join=round] (291.12,147.58) --
	(291.12,193.27);
\definecolor{fillColor}{RGB}{86,180,233}

\path[fill=fillColor] (162.54,149.73) rectangle (173.92,162.58);

\path[fill=fillColor] (162.54,164.00) rectangle (196.98,176.85);

\path[fill=fillColor] (162.54,178.28) rectangle (198.67,191.13);
\end{scope}
\begin{scope}
\path[clip] (156.13, 96.89) rectangle (297.10,142.58);
\definecolor{fillColor}{RGB}{255,255,255}

\path[fill=fillColor] (156.13, 96.89) rectangle (297.10,142.58);
\definecolor{drawColor}{RGB}{238,228,218}

\path[draw=drawColor,line width= 0.6pt,dash pattern=on 4pt off 4pt ,line join=round] (156.13,105.46) --
	(297.10,105.46);

\path[draw=drawColor,line width= 0.6pt,dash pattern=on 4pt off 4pt ,line join=round] (156.13,119.74) --
	(297.10,119.74);

\path[draw=drawColor,line width= 0.6pt,dash pattern=on 4pt off 4pt ,line join=round] (156.13,134.02) --
	(297.10,134.02);

\path[draw=drawColor,line width= 0.6pt,dash pattern=on 4pt off 4pt ,line join=round] (162.54, 96.89) --
	(162.54,142.58);

\path[draw=drawColor,line width= 0.6pt,dash pattern=on 4pt off 4pt ,line join=round] (188.25, 96.89) --
	(188.25,142.58);

\path[draw=drawColor,line width= 0.6pt,dash pattern=on 4pt off 4pt ,line join=round] (213.97, 96.89) --
	(213.97,142.58);

\path[draw=drawColor,line width= 0.6pt,dash pattern=on 4pt off 4pt ,line join=round] (239.69, 96.89) --
	(239.69,142.58);

\path[draw=drawColor,line width= 0.6pt,dash pattern=on 4pt off 4pt ,line join=round] (265.41, 96.89) --
	(265.41,142.58);

\path[draw=drawColor,line width= 0.6pt,dash pattern=on 4pt off 4pt ,line join=round] (291.12, 96.89) --
	(291.12,142.58);
\definecolor{fillColor}{RGB}{230,159,0}

\path[fill=fillColor] (162.54, 99.04) rectangle (182.23,111.89);

\path[fill=fillColor] (162.54,113.31) rectangle (184.27,126.16);

\path[fill=fillColor] (162.54,127.59) rectangle (191.38,140.44);
\end{scope}
\begin{scope}
\path[clip] (156.13, 31.93) rectangle (297.10, 91.89);
\definecolor{fillColor}{RGB}{255,255,255}

\path[fill=fillColor] (156.13, 31.93) rectangle (297.10, 91.89);
\definecolor{drawColor}{RGB}{238,228,218}

\path[draw=drawColor,line width= 0.6pt,dash pattern=on 4pt off 4pt ,line join=round] (156.13, 40.49) --
	(297.10, 40.49);

\path[draw=drawColor,line width= 0.6pt,dash pattern=on 4pt off 4pt ,line join=round] (156.13, 54.77) --
	(297.10, 54.77);

\path[draw=drawColor,line width= 0.6pt,dash pattern=on 4pt off 4pt ,line join=round] (156.13, 69.05) --
	(297.10, 69.05);

\path[draw=drawColor,line width= 0.6pt,dash pattern=on 4pt off 4pt ,line join=round] (156.13, 83.33) --
	(297.10, 83.33);

\path[draw=drawColor,line width= 0.6pt,dash pattern=on 4pt off 4pt ,line join=round] (162.54, 31.93) --
	(162.54, 91.89);

\path[draw=drawColor,line width= 0.6pt,dash pattern=on 4pt off 4pt ,line join=round] (188.25, 31.93) --
	(188.25, 91.89);

\path[draw=drawColor,line width= 0.6pt,dash pattern=on 4pt off 4pt ,line join=round] (213.97, 31.93) --
	(213.97, 91.89);

\path[draw=drawColor,line width= 0.6pt,dash pattern=on 4pt off 4pt ,line join=round] (239.69, 31.93) --
	(239.69, 91.89);

\path[draw=drawColor,line width= 0.6pt,dash pattern=on 4pt off 4pt ,line join=round] (265.41, 31.93) --
	(265.41, 91.89);

\path[draw=drawColor,line width= 0.6pt,dash pattern=on 4pt off 4pt ,line join=round] (291.12, 31.93) --
	(291.12, 91.89);
\definecolor{fillColor}{RGB}{86,180,233}

\path[fill=fillColor] (162.54, 34.07) rectangle (162.54, 46.92);

\path[fill=fillColor] (162.54, 48.35) rectangle (163.22, 61.20);

\path[fill=fillColor] (162.54, 62.62) rectangle (163.77, 75.47);

\path[fill=fillColor] (162.54, 76.90) rectangle (163.84, 89.75);
\end{scope}
\begin{scope}
\path[clip] (302.10,466.00) rectangle (443.07,525.97);
\definecolor{fillColor}{RGB}{255,255,255}

\path[fill=fillColor] (302.10,466.00) rectangle (443.07,525.97);
\definecolor{drawColor}{RGB}{238,228,218}

\path[draw=drawColor,line width= 0.6pt,dash pattern=on 4pt off 4pt ,line join=round] (302.10,474.57) --
	(443.07,474.57);

\path[draw=drawColor,line width= 0.6pt,dash pattern=on 4pt off 4pt ,line join=round] (302.10,488.85) --
	(443.07,488.85);

\path[draw=drawColor,line width= 0.6pt,dash pattern=on 4pt off 4pt ,line join=round] (302.10,503.13) --
	(443.07,503.13);

\path[draw=drawColor,line width= 0.6pt,dash pattern=on 4pt off 4pt ,line join=round] (302.10,517.40) --
	(443.07,517.40);

\path[draw=drawColor,line width= 0.6pt,dash pattern=on 4pt off 4pt ,line join=round] (335.93,466.00) --
	(335.93,525.97);

\path[draw=drawColor,line width= 0.6pt,dash pattern=on 4pt off 4pt ,line join=round] (371.75,466.00) --
	(371.75,525.97);

\path[draw=drawColor,line width= 0.6pt,dash pattern=on 4pt off 4pt ,line join=round] (407.57,466.00) --
	(407.57,525.97);
\definecolor{fillColor}{RGB}{230,159,0}

\path[fill=fillColor] (335.93,468.14) rectangle (436.67,480.99);

\path[fill=fillColor] (335.93,482.42) rectangle (431.03,495.27);

\path[fill=fillColor] (335.93,496.70) rectangle (370.21,509.55);

\path[fill=fillColor] (335.93,510.98) rectangle (394.64,523.83);
\definecolor{fillColor}{RGB}{230,159,0}

\path[fill=fillColor,fill opacity=0.50] (335.38,468.14) rectangle (335.93,480.99);

\path[fill=fillColor,fill opacity=0.50] (334.49,482.42) rectangle (335.93,495.27);

\path[fill=fillColor,fill opacity=0.50] (335.93,496.70) rectangle (377.92,509.55);

\path[fill=fillColor,fill opacity=0.50] (335.93,510.98) rectangle (411.01,523.83);
\end{scope}
\begin{scope}
\path[clip] (302.10,415.31) rectangle (443.07,461.00);
\definecolor{fillColor}{RGB}{255,255,255}

\path[fill=fillColor] (302.10,415.31) rectangle (443.07,461.00);
\definecolor{drawColor}{RGB}{238,228,218}

\path[draw=drawColor,line width= 0.6pt,dash pattern=on 4pt off 4pt ,line join=round] (302.10,423.88) --
	(443.07,423.88);

\path[draw=drawColor,line width= 0.6pt,dash pattern=on 4pt off 4pt ,line join=round] (302.10,438.16) --
	(443.07,438.16);

\path[draw=drawColor,line width= 0.6pt,dash pattern=on 4pt off 4pt ,line join=round] (302.10,452.44) --
	(443.07,452.44);

\path[draw=drawColor,line width= 0.6pt,dash pattern=on 4pt off 4pt ,line join=round] (335.93,415.31) --
	(335.93,461.00);

\path[draw=drawColor,line width= 0.6pt,dash pattern=on 4pt off 4pt ,line join=round] (371.75,415.31) --
	(371.75,461.00);

\path[draw=drawColor,line width= 0.6pt,dash pattern=on 4pt off 4pt ,line join=round] (407.57,415.31) --
	(407.57,461.00);
\definecolor{fillColor}{RGB}{230,159,0}

\path[fill=fillColor] (335.93,417.45) rectangle (360.77,430.30);

\path[fill=fillColor] (334.01,431.73) rectangle (335.93,444.58);

\path[fill=fillColor] (308.51,446.01) rectangle (335.93,458.86);
\definecolor{fillColor}{RGB}{230,159,0}

\path[fill=fillColor,fill opacity=0.50] (335.93,417.45) rectangle (360.78,430.30);

\path[fill=fillColor,fill opacity=0.50] (335.93,431.73) rectangle (346.83,444.58);

\path[fill=fillColor,fill opacity=0.50] (335.93,446.01) rectangle (364.90,458.86);
\end{scope}
\begin{scope}
\path[clip] (302.10,336.07) rectangle (443.07,410.31);
\definecolor{fillColor}{RGB}{255,255,255}

\path[fill=fillColor] (302.10,336.07) rectangle (443.07,410.31);
\definecolor{drawColor}{RGB}{238,228,218}

\path[draw=drawColor,line width= 0.6pt,dash pattern=on 4pt off 4pt ,line join=round] (302.10,344.63) --
	(443.07,344.63);

\path[draw=drawColor,line width= 0.6pt,dash pattern=on 4pt off 4pt ,line join=round] (302.10,358.91) --
	(443.07,358.91);

\path[draw=drawColor,line width= 0.6pt,dash pattern=on 4pt off 4pt ,line join=round] (302.10,373.19) --
	(443.07,373.19);

\path[draw=drawColor,line width= 0.6pt,dash pattern=on 4pt off 4pt ,line join=round] (302.10,387.47) --
	(443.07,387.47);

\path[draw=drawColor,line width= 0.6pt,dash pattern=on 4pt off 4pt ,line join=round] (302.10,401.75) --
	(443.07,401.75);

\path[draw=drawColor,line width= 0.6pt,dash pattern=on 4pt off 4pt ,line join=round] (335.93,336.07) --
	(335.93,410.31);

\path[draw=drawColor,line width= 0.6pt,dash pattern=on 4pt off 4pt ,line join=round] (371.75,336.07) --
	(371.75,410.31);

\path[draw=drawColor,line width= 0.6pt,dash pattern=on 4pt off 4pt ,line join=round] (407.57,336.07) --
	(407.57,410.31);
\definecolor{fillColor}{RGB}{86,180,233}

\path[fill=fillColor] (335.93,338.21) rectangle (336.00,351.06);

\path[fill=fillColor] (335.93,352.49) rectangle (353.29,365.34);

\path[fill=fillColor] (335.93,366.76) rectangle (340.83,379.61);

\path[fill=fillColor] (335.93,381.04) rectangle (349.04,393.89);

\path[fill=fillColor] (335.93,395.32) rectangle (345.16,408.17);
\definecolor{fillColor}{RGB}{86,180,233}

\path[fill=fillColor,fill opacity=0.50] (335.93,338.21) rectangle (336.59,351.06);

\path[fill=fillColor,fill opacity=0.50] (331.87,352.49) rectangle (335.93,365.34);

\path[fill=fillColor,fill opacity=0.50] (331.20,366.76) rectangle (335.93,379.61);

\path[fill=fillColor,fill opacity=0.50] (335.93,381.04) rectangle (354.86,393.89);

\path[fill=fillColor,fill opacity=0.50] (327.39,395.32) rectangle (335.93,408.17);
\end{scope}
\begin{scope}
\path[clip] (302.10,299.65) rectangle (443.07,331.07);
\definecolor{fillColor}{RGB}{255,255,255}

\path[fill=fillColor] (302.10,299.65) rectangle (443.07,331.07);
\definecolor{drawColor}{RGB}{238,228,218}

\path[draw=drawColor,line width= 0.6pt,dash pattern=on 4pt off 4pt ,line join=round] (302.10,308.22) --
	(443.07,308.22);

\path[draw=drawColor,line width= 0.6pt,dash pattern=on 4pt off 4pt ,line join=round] (302.10,322.50) --
	(443.07,322.50);

\path[draw=drawColor,line width= 0.6pt,dash pattern=on 4pt off 4pt ,line join=round] (335.93,299.65) --
	(335.93,331.07);

\path[draw=drawColor,line width= 0.6pt,dash pattern=on 4pt off 4pt ,line join=round] (371.75,299.65) --
	(371.75,331.07);

\path[draw=drawColor,line width= 0.6pt,dash pattern=on 4pt off 4pt ,line join=round] (407.57,299.65) --
	(407.57,331.07);
\definecolor{fillColor}{RGB}{86,180,233}

\path[fill=fillColor] (335.93,301.80) rectangle (349.51,314.65);

\path[fill=fillColor] (335.93,316.07) rectangle (377.62,328.92);
\definecolor{fillColor}{RGB}{86,180,233}

\path[fill=fillColor,fill opacity=0.50] (335.11,301.80) rectangle (335.93,314.65);

\path[fill=fillColor,fill opacity=0.50] (334.13,316.07) rectangle (335.93,328.92);
\end{scope}
\begin{scope}
\path[clip] (302.10,263.24) rectangle (443.07,294.65);
\definecolor{fillColor}{RGB}{255,255,255}

\path[fill=fillColor] (302.10,263.24) rectangle (443.07,294.65);
\definecolor{drawColor}{RGB}{238,228,218}

\path[draw=drawColor,line width= 0.6pt,dash pattern=on 4pt off 4pt ,line join=round] (302.10,271.81) --
	(443.07,271.81);

\path[draw=drawColor,line width= 0.6pt,dash pattern=on 4pt off 4pt ,line join=round] (302.10,286.09) --
	(443.07,286.09);

\path[draw=drawColor,line width= 0.6pt,dash pattern=on 4pt off 4pt ,line join=round] (335.93,263.24) --
	(335.93,294.65);

\path[draw=drawColor,line width= 0.6pt,dash pattern=on 4pt off 4pt ,line join=round] (371.75,263.24) --
	(371.75,294.65);

\path[draw=drawColor,line width= 0.6pt,dash pattern=on 4pt off 4pt ,line join=round] (407.57,263.24) --
	(407.57,294.65);
\definecolor{fillColor}{RGB}{86,180,233}

\path[fill=fillColor] (335.93,265.38) rectangle (431.70,278.23);

\path[fill=fillColor] (335.93,279.66) rectangle (343.08,292.51);
\definecolor{fillColor}{RGB}{86,180,233}

\path[fill=fillColor,fill opacity=0.50] (335.93,265.38) rectangle (434.55,278.23);

\path[fill=fillColor,fill opacity=0.50] (332.29,279.66) rectangle (335.93,292.51);
\end{scope}
\begin{scope}
\path[clip] (302.10,198.27) rectangle (443.07,258.24);
\definecolor{fillColor}{RGB}{255,255,255}

\path[fill=fillColor] (302.10,198.27) rectangle (443.07,258.24);
\definecolor{drawColor}{RGB}{238,228,218}

\path[draw=drawColor,line width= 0.6pt,dash pattern=on 4pt off 4pt ,line join=round] (302.10,206.84) --
	(443.07,206.84);

\path[draw=drawColor,line width= 0.6pt,dash pattern=on 4pt off 4pt ,line join=round] (302.10,221.12) --
	(443.07,221.12);

\path[draw=drawColor,line width= 0.6pt,dash pattern=on 4pt off 4pt ,line join=round] (302.10,235.40) --
	(443.07,235.40);

\path[draw=drawColor,line width= 0.6pt,dash pattern=on 4pt off 4pt ,line join=round] (302.10,249.68) --
	(443.07,249.68);

\path[draw=drawColor,line width= 0.6pt,dash pattern=on 4pt off 4pt ,line join=round] (335.93,198.27) --
	(335.93,258.24);

\path[draw=drawColor,line width= 0.6pt,dash pattern=on 4pt off 4pt ,line join=round] (371.75,198.27) --
	(371.75,258.24);

\path[draw=drawColor,line width= 0.6pt,dash pattern=on 4pt off 4pt ,line join=round] (407.57,198.27) --
	(407.57,258.24);
\definecolor{fillColor}{RGB}{86,180,233}

\path[fill=fillColor] (335.93,200.42) rectangle (400.02,213.27);

\path[fill=fillColor] (335.93,214.69) rectangle (364.50,227.54);

\path[fill=fillColor] (335.93,228.97) rectangle (368.85,241.82);

\path[fill=fillColor] (335.93,243.25) rectangle (397.94,256.10);
\definecolor{fillColor}{RGB}{86,180,233}

\path[fill=fillColor,fill opacity=0.50] (335.93,200.42) rectangle (405.63,213.27);

\path[fill=fillColor,fill opacity=0.50] (335.93,214.69) rectangle (369.62,227.54);

\path[fill=fillColor,fill opacity=0.50] (335.93,228.97) rectangle (376.55,241.82);

\path[fill=fillColor,fill opacity=0.50] (322.94,243.25) rectangle (335.93,256.10);
\end{scope}
\begin{scope}
\path[clip] (302.10,147.58) rectangle (443.07,193.27);
\definecolor{fillColor}{RGB}{255,255,255}

\path[fill=fillColor] (302.10,147.58) rectangle (443.07,193.27);
\definecolor{drawColor}{RGB}{238,228,218}

\path[draw=drawColor,line width= 0.6pt,dash pattern=on 4pt off 4pt ,line join=round] (302.10,156.15) --
	(443.07,156.15);

\path[draw=drawColor,line width= 0.6pt,dash pattern=on 4pt off 4pt ,line join=round] (302.10,170.43) --
	(443.07,170.43);

\path[draw=drawColor,line width= 0.6pt,dash pattern=on 4pt off 4pt ,line join=round] (302.10,184.71) --
	(443.07,184.71);

\path[draw=drawColor,line width= 0.6pt,dash pattern=on 4pt off 4pt ,line join=round] (335.93,147.58) --
	(335.93,193.27);

\path[draw=drawColor,line width= 0.6pt,dash pattern=on 4pt off 4pt ,line join=round] (371.75,147.58) --
	(371.75,193.27);

\path[draw=drawColor,line width= 0.6pt,dash pattern=on 4pt off 4pt ,line join=round] (407.57,147.58) --
	(407.57,193.27);
\definecolor{fillColor}{RGB}{86,180,233}

\path[fill=fillColor] (335.93,149.73) rectangle (348.36,162.58);

\path[fill=fillColor] (329.07,164.00) rectangle (335.93,176.85);

\path[fill=fillColor] (335.93,178.28) rectangle (369.26,191.13);
\definecolor{fillColor}{RGB}{86,180,233}

\path[fill=fillColor,fill opacity=0.50] (335.93,149.73) rectangle (350.57,162.58);

\path[fill=fillColor,fill opacity=0.50] (335.93,164.00) rectangle (343.02,176.85);

\path[fill=fillColor,fill opacity=0.50] (329.86,178.28) rectangle (335.93,191.13);
\end{scope}
\begin{scope}
\path[clip] (302.10, 96.89) rectangle (443.07,142.58);
\definecolor{fillColor}{RGB}{255,255,255}

\path[fill=fillColor] (302.10, 96.89) rectangle (443.07,142.58);
\definecolor{drawColor}{RGB}{238,228,218}

\path[draw=drawColor,line width= 0.6pt,dash pattern=on 4pt off 4pt ,line join=round] (302.10,105.46) --
	(443.07,105.46);

\path[draw=drawColor,line width= 0.6pt,dash pattern=on 4pt off 4pt ,line join=round] (302.10,119.74) --
	(443.07,119.74);

\path[draw=drawColor,line width= 0.6pt,dash pattern=on 4pt off 4pt ,line join=round] (302.10,134.02) --
	(443.07,134.02);

\path[draw=drawColor,line width= 0.6pt,dash pattern=on 4pt off 4pt ,line join=round] (335.93, 96.89) --
	(335.93,142.58);

\path[draw=drawColor,line width= 0.6pt,dash pattern=on 4pt off 4pt ,line join=round] (371.75, 96.89) --
	(371.75,142.58);

\path[draw=drawColor,line width= 0.6pt,dash pattern=on 4pt off 4pt ,line join=round] (407.57, 96.89) --
	(407.57,142.58);
\definecolor{fillColor}{RGB}{230,159,0}

\path[fill=fillColor] (335.93, 99.04) rectangle (366.87,111.89);

\path[fill=fillColor] (335.93,113.31) rectangle (385.42,126.16);

\path[fill=fillColor] (335.93,127.59) rectangle (344.04,140.44);
\definecolor{fillColor}{RGB}{230,159,0}

\path[fill=fillColor,fill opacity=0.50] (335.93, 99.04) rectangle (369.59,111.89);

\path[fill=fillColor,fill opacity=0.50] (335.93,113.31) rectangle (391.30,126.16);

\path[fill=fillColor,fill opacity=0.50] (333.69,127.59) rectangle (335.93,140.44);
\end{scope}
\begin{scope}
\path[clip] (302.10, 31.93) rectangle (443.07, 91.89);
\definecolor{fillColor}{RGB}{255,255,255}

\path[fill=fillColor] (302.10, 31.93) rectangle (443.07, 91.89);
\definecolor{drawColor}{RGB}{238,228,218}

\path[draw=drawColor,line width= 0.6pt,dash pattern=on 4pt off 4pt ,line join=round] (302.10, 40.49) --
	(443.07, 40.49);

\path[draw=drawColor,line width= 0.6pt,dash pattern=on 4pt off 4pt ,line join=round] (302.10, 54.77) --
	(443.07, 54.77);

\path[draw=drawColor,line width= 0.6pt,dash pattern=on 4pt off 4pt ,line join=round] (302.10, 69.05) --
	(443.07, 69.05);

\path[draw=drawColor,line width= 0.6pt,dash pattern=on 4pt off 4pt ,line join=round] (302.10, 83.33) --
	(443.07, 83.33);

\path[draw=drawColor,line width= 0.6pt,dash pattern=on 4pt off 4pt ,line join=round] (335.93, 31.93) --
	(335.93, 91.89);

\path[draw=drawColor,line width= 0.6pt,dash pattern=on 4pt off 4pt ,line join=round] (371.75, 31.93) --
	(371.75, 91.89);

\path[draw=drawColor,line width= 0.6pt,dash pattern=on 4pt off 4pt ,line join=round] (407.57, 31.93) --
	(407.57, 91.89);
\definecolor{fillColor}{RGB}{86,180,233}

\path[fill=fillColor] (335.93, 34.07) rectangle (379.25, 46.92);

\path[fill=fillColor] (335.93, 48.35) rectangle (392.48, 61.20);

\path[fill=fillColor] (335.93, 62.62) rectangle (341.97, 75.47);

\path[fill=fillColor] (335.93, 76.90) rectangle (432.61, 89.75);
\definecolor{fillColor}{RGB}{86,180,233}

\path[fill=fillColor,fill opacity=0.50] (335.93, 34.07) rectangle (335.93, 46.92);

\path[fill=fillColor,fill opacity=0.50] (335.82, 48.35) rectangle (335.93, 61.20);

\path[fill=fillColor,fill opacity=0.50] (335.93, 62.62) rectangle (342.29, 75.47);

\path[fill=fillColor,fill opacity=0.50] (335.76, 76.90) rectangle (335.93, 89.75);
\end{scope}
\begin{scope}
\path[clip] (156.13,525.97) rectangle (297.10,544.25);
\definecolor{fillColor}{RGB}{255,255,255}

\path[fill=fillColor] (156.13,525.97) rectangle (297.10,544.25);
\definecolor{drawColor}{RGB}{85,85,85}

\node[text=drawColor,anchor=base,inner sep=0pt, outer sep=0pt, scale=  0.80] at (226.62,532.36) {A: Cont.~bias $t$-statistic, $\frac{|\hat{\beta}_{cb}|}{se(\hat{\beta}_{cb})}$};
\end{scope}
\begin{scope}
\path[clip] (302.10,525.97) rectangle (443.07,544.25);
\definecolor{fillColor}{RGB}{255,255,255}

\path[fill=fillColor] (302.10,525.97) rectangle (443.07,544.25);
\definecolor{drawColor}{RGB}{85,85,85}

\node[text=drawColor,anchor=base,inner sep=0pt, outer sep=0pt, scale=  0.80] at (372.59,532.36) {B: Decomposition: $\frac{\hat{\beta}}{se(\hat{\beta})}=\frac{\hat{\beta}_{own}}{se(\hat{\beta})}+\frac{\hat{\beta}_{cb}}{se(\hat{\beta})}$};
\end{scope}
\begin{scope}
\path[clip] ( 19.15,466.00) rectangle ( 87.84,525.97);
\definecolor{fillColor}{RGB}{255,255,255}

\path[fill=fillColor] ( 19.15,466.00) rectangle ( 87.84,525.97);
\definecolor{drawColor}{RGB}{230,159,0}

\node[text=drawColor,anchor=base west,inner sep=0pt, outer sep=0pt, scale=  0.80] at ( 29.15,520.46) {Fryer \& Levitt};
\end{scope}
\begin{scope}
\path[clip] ( 19.15,415.31) rectangle ( 87.84,461.00);
\definecolor{fillColor}{RGB}{255,255,255}

\path[fill=fillColor] ( 19.15,415.31) rectangle ( 87.84,461.00);
\definecolor{drawColor}{RGB}{230,159,0}

\node[text=drawColor,anchor=base west,inner sep=0pt, outer sep=0pt, scale=  0.80] at ( 29.15,455.49) {Weisburst};
\end{scope}
\begin{scope}
\path[clip] ( 19.15,336.07) rectangle ( 87.84,410.31);
\definecolor{fillColor}{RGB}{255,255,255}

\path[fill=fillColor] ( 19.15,336.07) rectangle ( 87.84,410.31);
\definecolor{drawColor}{RGB}{86,180,233}

\node[text=drawColor,anchor=base west,inner sep=0pt, outer sep=0pt, scale=  0.80] at ( 29.15,404.80) {Cole et al};
\end{scope}
\begin{scope}
\path[clip] ( 19.15,299.65) rectangle ( 87.84,331.07);
\definecolor{fillColor}{RGB}{255,255,255}

\path[fill=fillColor] ( 19.15,299.65) rectangle ( 87.84,331.07);
\definecolor{drawColor}{RGB}{86,180,233}

\node[text=drawColor,anchor=base west,inner sep=0pt, outer sep=0pt, scale=  0.80] at ( 29.15,325.56) {Drexler et al};
\end{scope}
\begin{scope}
\path[clip] ( 19.15,263.24) rectangle ( 87.84,294.65);
\definecolor{fillColor}{RGB}{255,255,255}

\path[fill=fillColor] ( 19.15,263.24) rectangle ( 87.84,294.65);
\definecolor{drawColor}{RGB}{86,180,233}

\node[text=drawColor,anchor=base west,inner sep=0pt, outer sep=0pt, scale=  0.80] at ( 29.15,289.14) {STAR};
\end{scope}
\begin{scope}
\path[clip] ( 19.15,198.27) rectangle ( 87.84,258.24);
\definecolor{fillColor}{RGB}{255,255,255}

\path[fill=fillColor] ( 19.15,198.27) rectangle ( 87.84,258.24);
\definecolor{drawColor}{RGB}{86,180,233}

\node[text=drawColor,anchor=base west,inner sep=0pt, outer sep=0pt, scale=  0.80] at ( 29.15,252.73) {Benhassine et al};
\end{scope}
\begin{scope}
\path[clip] ( 19.15,147.58) rectangle ( 87.84,193.27);
\definecolor{fillColor}{RGB}{255,255,255}

\path[fill=fillColor] ( 19.15,147.58) rectangle ( 87.84,193.27);
\definecolor{drawColor}{RGB}{86,180,233}

\node[text=drawColor,anchor=base west,inner sep=0pt, outer sep=0pt, scale=  0.80] at ( 29.15,187.76) {Duflo et al};
\end{scope}
\begin{scope}
\path[clip] ( 19.15, 96.89) rectangle ( 87.84,142.58);
\definecolor{fillColor}{RGB}{255,255,255}

\path[fill=fillColor] ( 19.15, 96.89) rectangle ( 87.84,142.58);
\definecolor{drawColor}{RGB}{230,159,0}

\node[text=drawColor,anchor=base west,inner sep=0pt, outer sep=0pt, scale=  0.80] at ( 29.15,137.07) {Rim et al};
\end{scope}
\begin{scope}
\path[clip] ( 19.15, 31.93) rectangle ( 87.84, 91.89);
\definecolor{fillColor}{RGB}{255,255,255}

\path[fill=fillColor] ( 19.15, 31.93) rectangle ( 87.84, 91.89);
\definecolor{drawColor}{RGB}{86,180,233}

\node[text=drawColor,anchor=base west,inner sep=0pt, outer sep=0pt, scale=  0.80] at ( 29.15, 86.38) {de Mel et al};
\end{scope}
\begin{scope}
\path[clip] (  0.00,  0.00) rectangle (448.07,549.25);
\definecolor{drawColor}{RGB}{130,106,80}

\path[draw=drawColor,line width= 0.6pt,line join=round] (156.13, 31.93) --
	(297.10, 31.93);
\end{scope}
\begin{scope}
\path[clip] (  0.00,  0.00) rectangle (448.07,549.25);
\definecolor{drawColor}{RGB}{130,106,80}

\path[draw=drawColor,line width= 0.6pt,line join=round] (162.54, 29.43) --
	(162.54, 31.93);

\path[draw=drawColor,line width= 0.6pt,line join=round] (188.25, 29.43) --
	(188.25, 31.93);

\path[draw=drawColor,line width= 0.6pt,line join=round] (213.97, 29.43) --
	(213.97, 31.93);

\path[draw=drawColor,line width= 0.6pt,line join=round] (239.69, 29.43) --
	(239.69, 31.93);

\path[draw=drawColor,line width= 0.6pt,line join=round] (265.41, 29.43) --
	(265.41, 31.93);

\path[draw=drawColor,line width= 0.6pt,line join=round] (291.12, 29.43) --
	(291.12, 31.93);
\end{scope}
\begin{scope}
\path[clip] (  0.00,  0.00) rectangle (448.07,549.25);
\definecolor{drawColor}{RGB}{85,85,85}

\node[text=drawColor,anchor=base,inner sep=0pt, outer sep=0pt, scale=  0.80] at (162.54, 21.92) {0.0};

\node[text=drawColor,anchor=base,inner sep=0pt, outer sep=0pt, scale=  0.80] at (188.25, 21.92) {0.5};

\node[text=drawColor,anchor=base,inner sep=0pt, outer sep=0pt, scale=  0.80] at (213.97, 21.92) {1.0};

\node[text=drawColor,anchor=base,inner sep=0pt, outer sep=0pt, scale=  0.80] at (239.69, 21.92) {1.5};

\node[text=drawColor,anchor=base,inner sep=0pt, outer sep=0pt, scale=  0.80] at (265.41, 21.92) {2.0};

\node[text=drawColor,anchor=base,inner sep=0pt, outer sep=0pt, scale=  0.80] at (291.12, 21.92) {2.5};
\end{scope}
\begin{scope}
\path[clip] (  0.00,  0.00) rectangle (448.07,549.25);
\definecolor{drawColor}{RGB}{130,106,80}

\path[draw=drawColor,line width= 0.6pt,line join=round] (302.10, 31.93) --
	(443.07, 31.93);
\end{scope}
\begin{scope}
\path[clip] (  0.00,  0.00) rectangle (448.07,549.25);
\definecolor{drawColor}{RGB}{130,106,80}

\path[draw=drawColor,line width= 0.6pt,line join=round] (335.93, 29.43) --
	(335.93, 31.93);

\path[draw=drawColor,line width= 0.6pt,line join=round] (371.75, 29.43) --
	(371.75, 31.93);

\path[draw=drawColor,line width= 0.6pt,line join=round] (407.57, 29.43) --
	(407.57, 31.93);
\end{scope}
\begin{scope}
\path[clip] (  0.00,  0.00) rectangle (448.07,549.25);
\definecolor{drawColor}{RGB}{85,85,85}

\node[text=drawColor,anchor=base,inner sep=0pt, outer sep=0pt, scale=  0.80] at (335.93, 21.92) {0.0};

\node[text=drawColor,anchor=base,inner sep=0pt, outer sep=0pt, scale=  0.80] at (371.75, 21.92) {2.5};

\node[text=drawColor,anchor=base,inner sep=0pt, outer sep=0pt, scale=  0.80] at (407.57, 21.92) {5.0};
\end{scope}
\begin{scope}
\path[clip] (  0.00,  0.00) rectangle (448.07,549.25);
\definecolor{drawColor}{RGB}{130,106,80}

\path[draw=drawColor,line width= 0.6pt,line join=round] (156.13,466.00) --
	(156.13,525.97);
\end{scope}
\begin{scope}
\path[clip] (  0.00,  0.00) rectangle (448.07,549.25);
\definecolor{drawColor}{RGB}{85,85,85}

\node[text=drawColor,anchor=base east,inner sep=0pt, outer sep=0pt, scale=  0.64] at (151.63,472.37) {Asian};

\node[text=drawColor,anchor=base east,inner sep=0pt, outer sep=0pt, scale=  0.64] at (151.63,486.64) {Hispanic};

\node[text=drawColor,anchor=base east,inner sep=0pt, outer sep=0pt, scale=  0.64] at (151.63,500.92) {Other};

\node[text=drawColor,anchor=base east,inner sep=0pt, outer sep=0pt, scale=  0.64] at (151.63,515.20) {Black};
\end{scope}
\begin{scope}
\path[clip] (  0.00,  0.00) rectangle (448.07,549.25);
\definecolor{drawColor}{RGB}{130,106,80}

\path[draw=drawColor,line width= 0.6pt,line join=round] (153.63,474.57) --
	(156.13,474.57);

\path[draw=drawColor,line width= 0.6pt,line join=round] (153.63,488.85) --
	(156.13,488.85);

\path[draw=drawColor,line width= 0.6pt,line join=round] (153.63,503.13) --
	(156.13,503.13);

\path[draw=drawColor,line width= 0.6pt,line join=round] (153.63,517.40) --
	(156.13,517.40);
\end{scope}
\begin{scope}
\path[clip] (  0.00,  0.00) rectangle (448.07,549.25);
\definecolor{drawColor}{RGB}{130,106,80}

\path[draw=drawColor,line width= 0.6pt,line join=round] (156.13,415.31) --
	(156.13,461.00);
\end{scope}
\begin{scope}
\path[clip] (  0.00,  0.00) rectangle (448.07,549.25);
\definecolor{drawColor}{RGB}{85,85,85}

\node[text=drawColor,anchor=base east,inner sep=0pt, outer sep=0pt, scale=  0.64] at (151.63,421.68) {Other};

\node[text=drawColor,anchor=base east,inner sep=0pt, outer sep=0pt, scale=  0.64] at (151.63,435.95) {Black};

\node[text=drawColor,anchor=base east,inner sep=0pt, outer sep=0pt, scale=  0.64] at (151.63,450.23) {Hispanic};
\end{scope}
\begin{scope}
\path[clip] (  0.00,  0.00) rectangle (448.07,549.25);
\definecolor{drawColor}{RGB}{130,106,80}

\path[draw=drawColor,line width= 0.6pt,line join=round] (153.63,423.88) --
	(156.13,423.88);

\path[draw=drawColor,line width= 0.6pt,line join=round] (153.63,438.16) --
	(156.13,438.16);

\path[draw=drawColor,line width= 0.6pt,line join=round] (153.63,452.44) --
	(156.13,452.44);
\end{scope}
\begin{scope}
\path[clip] (  0.00,  0.00) rectangle (448.07,549.25);
\definecolor{drawColor}{RGB}{130,106,80}

\path[draw=drawColor,line width= 0.6pt,line join=round] (156.13,336.07) --
	(156.13,410.31);
\end{scope}
\begin{scope}
\path[clip] (  0.00,  0.00) rectangle (448.07,549.25);
\definecolor{drawColor}{RGB}{85,85,85}

\node[text=drawColor,anchor=base east,inner sep=0pt, outer sep=0pt, scale=  0.64] at (151.63,342.43) {Hindu \& Group};

\node[text=drawColor,anchor=base east,inner sep=0pt, outer sep=0pt, scale=  0.64] at (151.63,356.71) {Group only};

\node[text=drawColor,anchor=base east,inner sep=0pt, outer sep=0pt, scale=  0.64] at (151.63,370.99) {Muslim only};

\node[text=drawColor,anchor=base east,inner sep=0pt, outer sep=0pt, scale=  0.64] at (151.63,385.26) {Muslim \& Group};

\node[text=drawColor,anchor=base east,inner sep=0pt, outer sep=0pt, scale=  0.64] at (151.63,399.54) {Hindu only};
\end{scope}
\begin{scope}
\path[clip] (  0.00,  0.00) rectangle (448.07,549.25);
\definecolor{drawColor}{RGB}{130,106,80}

\path[draw=drawColor,line width= 0.6pt,line join=round] (153.63,344.63) --
	(156.13,344.63);

\path[draw=drawColor,line width= 0.6pt,line join=round] (153.63,358.91) --
	(156.13,358.91);

\path[draw=drawColor,line width= 0.6pt,line join=round] (153.63,373.19) --
	(156.13,373.19);

\path[draw=drawColor,line width= 0.6pt,line join=round] (153.63,387.47) --
	(156.13,387.47);

\path[draw=drawColor,line width= 0.6pt,line join=round] (153.63,401.75) --
	(156.13,401.75);
\end{scope}
\begin{scope}
\path[clip] (  0.00,  0.00) rectangle (448.07,549.25);
\definecolor{drawColor}{RGB}{130,106,80}

\path[draw=drawColor,line width= 0.6pt,line join=round] (156.13,299.65) --
	(156.13,331.07);
\end{scope}
\begin{scope}
\path[clip] (  0.00,  0.00) rectangle (448.07,549.25);
\definecolor{drawColor}{RGB}{85,85,85}

\node[text=drawColor,anchor=base east,inner sep=0pt, outer sep=0pt, scale=  0.64] at (151.63,306.02) {Standard Accounting};

\node[text=drawColor,anchor=base east,inner sep=0pt, outer sep=0pt, scale=  0.64] at (151.63,320.30) {Rule-of-Thumb};
\end{scope}
\begin{scope}
\path[clip] (  0.00,  0.00) rectangle (448.07,549.25);
\definecolor{drawColor}{RGB}{130,106,80}

\path[draw=drawColor,line width= 0.6pt,line join=round] (153.63,308.22) --
	(156.13,308.22);

\path[draw=drawColor,line width= 0.6pt,line join=round] (153.63,322.50) --
	(156.13,322.50);
\end{scope}
\begin{scope}
\path[clip] (  0.00,  0.00) rectangle (448.07,549.25);
\definecolor{drawColor}{RGB}{130,106,80}

\path[draw=drawColor,line width= 0.6pt,line join=round] (156.13,263.24) --
	(156.13,294.65);
\end{scope}
\begin{scope}
\path[clip] (  0.00,  0.00) rectangle (448.07,549.25);
\definecolor{drawColor}{RGB}{85,85,85}

\node[text=drawColor,anchor=base east,inner sep=0pt, outer sep=0pt, scale=  0.64] at (151.63,269.61) {Small};

\node[text=drawColor,anchor=base east,inner sep=0pt, outer sep=0pt, scale=  0.64] at (151.63,283.88) {Aide};
\end{scope}
\begin{scope}
\path[clip] (  0.00,  0.00) rectangle (448.07,549.25);
\definecolor{drawColor}{RGB}{130,106,80}

\path[draw=drawColor,line width= 0.6pt,line join=round] (153.63,271.81) --
	(156.13,271.81);

\path[draw=drawColor,line width= 0.6pt,line join=round] (153.63,286.09) --
	(156.13,286.09);
\end{scope}
\begin{scope}
\path[clip] (  0.00,  0.00) rectangle (448.07,549.25);
\definecolor{drawColor}{RGB}{130,106,80}

\path[draw=drawColor,line width= 0.6pt,line join=round] (156.13,198.27) --
	(156.13,258.24);
\end{scope}
\begin{scope}
\path[clip] (  0.00,  0.00) rectangle (448.07,549.25);
\definecolor{drawColor}{RGB}{85,85,85}

\node[text=drawColor,anchor=base east,inner sep=0pt, outer sep=0pt, scale=  0.64] at (151.63,204.64) {LCT to mothers};

\node[text=drawColor,anchor=base east,inner sep=0pt, outer sep=0pt, scale=  0.64] at (151.63,218.92) {CCTs to mothers};

\node[text=drawColor,anchor=base east,inner sep=0pt, outer sep=0pt, scale=  0.64] at (151.63,233.19) {CCTs to fathers};

\node[text=drawColor,anchor=base east,inner sep=0pt, outer sep=0pt, scale=  0.64] at (151.63,247.47) {LCT to fathers};
\end{scope}
\begin{scope}
\path[clip] (  0.00,  0.00) rectangle (448.07,549.25);
\definecolor{drawColor}{RGB}{130,106,80}

\path[draw=drawColor,line width= 0.6pt,line join=round] (153.63,206.84) --
	(156.13,206.84);

\path[draw=drawColor,line width= 0.6pt,line join=round] (153.63,221.12) --
	(156.13,221.12);

\path[draw=drawColor,line width= 0.6pt,line join=round] (153.63,235.40) --
	(156.13,235.40);

\path[draw=drawColor,line width= 0.6pt,line join=round] (153.63,249.68) --
	(156.13,249.68);
\end{scope}
\begin{scope}
\path[clip] (  0.00,  0.00) rectangle (448.07,549.25);
\definecolor{drawColor}{RGB}{130,106,80}

\path[draw=drawColor,line width= 0.6pt,line join=round] (156.13,147.58) --
	(156.13,193.27);
\end{scope}
\begin{scope}
\path[clip] (  0.00,  0.00) rectangle (448.07,549.25);
\definecolor{drawColor}{RGB}{85,85,85}

\node[text=drawColor,anchor=base east,inner sep=0pt, outer sep=0pt, scale=  0.64] at (151.63,153.95) {Both};

\node[text=drawColor,anchor=base east,inner sep=0pt, outer sep=0pt, scale=  0.64] at (151.63,168.23) {HIV education};

\node[text=drawColor,anchor=base east,inner sep=0pt, outer sep=0pt, scale=  0.64] at (151.63,182.50) {Educ.~subsity};
\end{scope}
\begin{scope}
\path[clip] (  0.00,  0.00) rectangle (448.07,549.25);
\definecolor{drawColor}{RGB}{130,106,80}

\path[draw=drawColor,line width= 0.6pt,line join=round] (153.63,156.15) --
	(156.13,156.15);

\path[draw=drawColor,line width= 0.6pt,line join=round] (153.63,170.43) --
	(156.13,170.43);

\path[draw=drawColor,line width= 0.6pt,line join=round] (153.63,184.71) --
	(156.13,184.71);
\end{scope}
\begin{scope}
\path[clip] (  0.00,  0.00) rectangle (448.07,549.25);
\definecolor{drawColor}{RGB}{130,106,80}

\path[draw=drawColor,line width= 0.6pt,line join=round] (156.13, 96.89) --
	(156.13,142.58);
\end{scope}
\begin{scope}
\path[clip] (  0.00,  0.00) rectangle (448.07,549.25);
\definecolor{drawColor}{RGB}{85,85,85}

\node[text=drawColor,anchor=base east,inner sep=0pt, outer sep=0pt, scale=  0.64] at (151.63,103.26) {Asian};

\node[text=drawColor,anchor=base east,inner sep=0pt, outer sep=0pt, scale=  0.64] at (151.63,117.54) {Black};

\node[text=drawColor,anchor=base east,inner sep=0pt, outer sep=0pt, scale=  0.64] at (151.63,131.81) {Hispanic};
\end{scope}
\begin{scope}
\path[clip] (  0.00,  0.00) rectangle (448.07,549.25);
\definecolor{drawColor}{RGB}{130,106,80}

\path[draw=drawColor,line width= 0.6pt,line join=round] (153.63,105.46) --
	(156.13,105.46);

\path[draw=drawColor,line width= 0.6pt,line join=round] (153.63,119.74) --
	(156.13,119.74);

\path[draw=drawColor,line width= 0.6pt,line join=round] (153.63,134.02) --
	(156.13,134.02);
\end{scope}
\begin{scope}
\path[clip] (  0.00,  0.00) rectangle (448.07,549.25);
\definecolor{drawColor}{RGB}{130,106,80}

\path[draw=drawColor,line width= 0.6pt,line join=round] (156.13, 31.93) --
	(156.13, 91.89);
\end{scope}
\begin{scope}
\path[clip] (  0.00,  0.00) rectangle (448.07,549.25);
\definecolor{drawColor}{RGB}{85,85,85}

\node[text=drawColor,anchor=base east,inner sep=0pt, outer sep=0pt, scale=  0.64] at (151.63, 38.29) {Rs 20,000};

\node[text=drawColor,anchor=base east,inner sep=0pt, outer sep=0pt, scale=  0.64] at (151.63, 52.57) {Rs 10,000};

\node[text=drawColor,anchor=base east,inner sep=0pt, outer sep=0pt, scale=  0.64] at (151.63, 66.85) {Info and Reimburse};

\node[text=drawColor,anchor=base east,inner sep=0pt, outer sep=0pt, scale=  0.64] at (151.63, 81.12) {Rs 40,000};
\end{scope}
\begin{scope}
\path[clip] (  0.00,  0.00) rectangle (448.07,549.25);
\definecolor{drawColor}{RGB}{130,106,80}

\path[draw=drawColor,line width= 0.6pt,line join=round] (153.63, 40.49) --
	(156.13, 40.49);

\path[draw=drawColor,line width= 0.6pt,line join=round] (153.63, 54.77) --
	(156.13, 54.77);

\path[draw=drawColor,line width= 0.6pt,line join=round] (153.63, 69.05) --
	(156.13, 69.05);

\path[draw=drawColor,line width= 0.6pt,line join=round] (153.63, 83.33) --
	(156.13, 83.33);
\end{scope}
\end{tikzpicture}

%% file: pscore.tex
Project STAR & 308.9 & (154) & 0.000 & 302.9 & (154) & 0.000\\
Benhassine et al. & 207.2 & (159) & 0.006 & 217.2 & (194) & 0.121\\
Cole et al. & 22.7 & (39) & 0.983 & 70.3 & (54) & 0.067\\
de Mel et al. & 0.9 & (392) & 1.000 & 1.1 & (392) & 1.000\\
Drexler et al. & 12.4 & (14) & 0.574 & 12.6 & (14) & 0.555\\
Duflo et al. & 109.6 & (254) & 1.000 & 94.5 & (258) & 1.000\\
Fryer and Levitt & 3947.6 & (630) & 0.000 & 4164.0 & (681) & 0.000\\
Rim et al. & 1403.5 & (88) & 0.000 & 233.0 & (234) & 0.506\\
Weisburst & 2350.0 & (69) & 0.000 & 223.2 & (48) & 0.000

%% file: benhassineA.tex
LCT to fathers & 0.074 &  &  & 0.089 & 0.056 & 0.067 & 0.084 & 0.078 & 0.076 & 0.061\\
 & (0.016) &  &  & (0.017) & (0.018) & (0.019) & (0.024) & (0.015) & (0.020) & (0.020)\\
 &  &  &  & [0.012] & [0.011] &  &  & [0.014] & [0.014] & [0.012]\\
LCT to mothers & 0.078 &  &  & 0.067 & 0.071 & 0.081 & 0.075 & 0.079 & 0.074 & 0.068\\
 & (0.014) &  &  & (0.013) & (0.017) & (0.017) & (0.017) & (0.014) & (0.015) & (0.017)\\
 &  &  &  & [0.009] & [0.011] &  &  & [0.012] & [0.011] & [0.012]\\
CCTs to fathers & 0.055 &  &  & 0.062 & 0.041 & 0.047 & 0.038 & 0.033 & 0.039 & 0.038\\
 & (0.014) &  &  & (0.013) & (0.018) & (0.016) & (0.015) & (0.014) & (0.016) & (0.017)\\
 &  &  &  & [0.009] & [0.012] &  &  & [0.012] & [0.012] & [0.012]\\
CCTs to mothers & 0.053 &  &  & 0.045 & 0.040 & 0.039 & 0.033 & 0.042 & 0.041 & 0.040\\
 & (0.013) &  &  & (0.013) & (0.018) & (0.017) & (0.016) & (0.015) & (0.017) & (0.018)\\
 &  &  &  & [0.011] & [0.013] &  &  & [0.014] & [0.013] & [0.013]\\
Number of controls & \multicolumn{1}{r}{57} &  &  &  &  & \multicolumn{1}{r}{26} &  &  &  & \\
Sample size & \multicolumn{1}{r}{11,074} &  &  &  &  & \multicolumn{1}{r}{ 6,996} &  &  &  & 

%% file: benhassineB.tex
LCT to fathers &  &  &  & -0.016 & 0.018 &  & -0.018 & -0.011 & -0.009 & 0.006\\
 &  &  &  & (0.010) & (0.018) &  & (0.015) & (0.016) & (0.010) & (0.019)\\
LCT to mothers &  &  &  & 0.012 & 0.007 &  & 0.007 & 0.002 & 0.007 & 0.014\\
 &  &  &  & (0.009) & (0.016) &  & (0.013) & (0.011) & (0.010) & (0.015)\\
CCTs to fathers &  &  &  & -0.007 & 0.014 &  & 0.009 & 0.013 & 0.007 & 0.009\\
 &  &  &  & (0.005) & (0.015) &  & (0.009) & (0.010) & (0.006) & (0.015)\\
CCTs to mothers &  &  &  & 0.008 & 0.013 &  & 0.006 & -0.003 & -0.002 & -0.001\\
 &  &  &  & (0.007) & (0.015) &  & (0.009) & (0.009) & (0.006) & (0.015)

%% file: coleA.tex
Muslim only & 0.160 &  &  & 0.095 & 0.033 & 0.001 & 0.038 & -0.012 & -0.036 & 0.010\\
 & (0.086) &  &  & (0.086) & (0.094) & (0.111) & (0.138) & (0.109) & (0.120) & (0.093)\\
 &  &  &  & [0.079] & [0.098] &  &  & [0.109] & [0.121] & [0.104]\\
Hindu only & 0.121 &  &  & 0.058 & 0.062 & 0.006 & 0.075 & 0.080 & 0.060 & 0.076\\
 & (0.089) &  &  & (0.088) & (0.101) & (0.116) & (0.123) & (0.106) & (0.116) & (0.096)\\
 &  &  &  & [0.062] & [0.100] &  &  & [0.097] & [0.080] & [0.092]\\
Group only & 0.239 &  &  & 0.229 & 0.103 & 0.107 & 0.140 & 0.158 & 0.093 & 0.071\\
 & (0.097) &  &  & (0.098) & (0.112) & (0.115) & (0.130) & (0.086) & (0.106) & (0.108)\\
 &  &  &  & [0.076] & [0.097] &  &  & [0.082] & [0.099] & [0.091]\\
Muslim \& Group & 0.169 &  &  & 0.092 & -0.094 & -0.109 & -0.075 & -0.096 & -0.075 & -0.088\\
 & (0.087) &  &  & (0.083) & (0.079) & (0.082) & (0.074) & (0.080) & (0.070) & (0.075)\\
 &  &  &  & [0.038] & [0.076] &  &  & [0.078] & [0.062] & [0.072]\\
Hindu \& Group & 0.018 &  &  & -0.052 & -0.027 & -0.004 & -0.000 & -0.034 & 0.000 & -0.021\\
 & (0.080) &  &  & (0.075) & (0.096) & (0.094) & (0.093) & (0.094) & (0.087) & (0.094)\\
 &  &  &  & [0.056] & [0.089] &  &  & [0.090] & [0.075] & [0.086]\\
Number of controls & \multicolumn{1}{r}{13} &  &  &  &  & \multicolumn{1}{r}{ 3} &  &  &  & \\
Sample size & \multicolumn{1}{r}{132} &  &  &  &  & \multicolumn{1}{r}{ 73} &  &  &  & 

%% file: coleB.tex
Muslim only &  &  &  & 0.065 & 0.127 &  & -0.037 & 0.014 & 0.038 & -0.009\\
 &  &  &  & (0.044) & (0.073) &  & (0.066) & (0.060) & (0.061) & (0.061)\\
Hindu only &  &  &  & 0.063 & 0.059 &  & -0.069 & -0.075 & -0.054 & -0.071\\
 &  &  &  & (0.050) & (0.083) &  & (0.044) & (0.085) & (0.041) & (0.081)\\
Group only &  &  &  & 0.010 & 0.136 &  & -0.033 & -0.050 & 0.014 & 0.036\\
 &  &  &  & (0.060) & (0.103) &  & (0.060) & (0.081) & (0.064) & (0.102)\\
Muslim \& Group &  &  &  & 0.077 & 0.263 &  & -0.033 & -0.013 & -0.033 & -0.021\\
 &  &  &  & (0.056) & (0.091) &  & (0.048) & (0.063) & (0.047) & (0.060)\\
Hindu \& Group &  &  &  & 0.071 & 0.046 &  & -0.004 & 0.030 & -0.004 & 0.016\\
 &  &  &  & (0.048) & (0.080) &  & (0.028) & (0.056) & (0.036) & (0.061)

%% file: demelA.tex
Info and Reimburse & -0.010 & -0.010 & -0.010 & -0.010 & -0.010\\
 & (0.023) & (0.014) & (0.007) & (0.012) & (0.007)\\
 &  &  & [0.000] & [0.000] & [0.000]\\
Rs 10,000 & 0.134 & 0.134 & 0.135 & 0.134 & 0.135\\
 & (0.034) & (0.032) & (0.017) & (0.027) & (0.017)\\
 &  &  & [0.000] & [0.000] & [0.000]\\
Rs 20,000 & 0.105 & 0.105 & 0.104 & 0.105 & 0.104\\
 & (0.035) & (0.030) & (0.017) & (0.026) & (0.017)\\
 &  &  & [0.008] & [0.009] & [0.007]\\
Rs 40,000 & 0.273 & 0.273 & 0.269 & 0.272 & 0.270\\
 & (0.041) & (0.038) & (0.020) & (0.033) & (0.020)\\
 &  &  & [0.000] & [0.000] & [0.000]\\
Number of controls & \multicolumn{1}{r}{98} &  &  &  & \\
Sample size & \multicolumn{1}{r}{520} &  &  &  & 

%% file: demelB.tex
Info and Reimburse &  & -0.001 & -0.001 & -0.001 & -0.000\\
 &  & (0.022) & (0.022) & (0.020) & (0.022)\\
Rs 10,000 &  & -0.000 & -0.001 & 0.000 & -0.001\\
 &  & (0.019) & (0.029) & (0.020) & (0.029)\\
Rs 20,000 &  & -0.000 & 0.000 & -0.000 & 0.000\\
 &  & (0.021) & (0.030) & (0.023) & (0.030)\\
Rs 40,000 &  & -0.000 & 0.004 & 0.001 & 0.003\\
 &  & (0.019) & (0.035) & (0.024) & (0.035)

%% file: drexlerA.tex
Standard Accounting & 0.036 & 0.038 & 0.040 & 0.037 & 0.040\\
 & (0.041) & (0.041) & (0.040) & (0.041) & (0.040)\\
 &  &  & [0.040] & [0.040] & [0.040]\\
Rule-of-Thumb & 0.109 & 0.114 & 0.113 & 0.112 & 0.113\\
 & (0.039) & (0.039) & (0.039) & (0.039) & (0.039)\\
 &  &  & [0.039] & [0.039] & [0.039]\\
Number of controls & \multicolumn{1}{r}{ 7} &  &  &  & \\
Sample size & \multicolumn{1}{r}{796} &  &  &  & 

%% file: drexlerB.tex
Standard Accounting &  & -0.002 & -0.004 & -0.001 & -0.004\\
 &  & (0.004) & (0.005) & (0.003) & (0.005)\\
Rule-of-Thumb &  & -0.005 & -0.004 & -0.004 & -0.004\\
 &  & (0.004) & (0.005) & (0.003) & (0.005)

%% file: dufloA.tex
Educ.~subsity & -0.031 &  &  & -0.036 & -0.029 & -0.024 & -0.029 & -0.025 & -0.032 & -0.027\\
 & (0.012) &  &  & (0.011) & (0.011) & (0.013) & (0.012) & (0.007) & (0.011) & (0.010)\\
 &  &  &  & [0.000] & [0.000] &  &  & [0.001] & [0.001] & [0.001]\\
HIV education & 0.003 &  &  & 0.009 & 0.002 & -0.000 & 0.005 & 0.003 & 0.005 & -0.000\\
 & (0.011) &  &  & (0.009) & (0.012) & (0.011) & (0.010) & (0.007) & (0.010) & (0.011)\\
 &  &  &  & [0.000] & [0.001] &  &  & [0.001] & [0.001] & [0.001]\\
Both & -0.016 &  &  & -0.019 & -0.020 & -0.012 & -0.010 & -0.007 & -0.009 & -0.012\\
 & (0.012) &  &  & (0.010) & (0.011) & (0.012) & (0.010) & (0.007) & (0.010) & (0.010)\\
 &  &  &  & [0.000] & [0.000] &  &  & [0.001] & [0.001] & [0.001]\\
Number of controls & \multicolumn{1}{r}{86} &  &  &  &  & \multicolumn{1}{r}{79} &  &  &  & \\
Sample size & \multicolumn{1}{r}{9,116} &  &  &  &  & \multicolumn{1}{r}{8,664} &  &  &  & 

%% file: dufloB.tex
Educ.~subsity &  &  &  & 0.005 & -0.002 &  & 0.005 & 0.001 & 0.008 & 0.003\\
 &  &  &  & (0.008) & (0.012) &  & (0.008) & (0.011) & (0.007) & (0.011)\\
HIV education &  &  &  & -0.006 & 0.001 &  & -0.005 & -0.003 & -0.006 & 0.000\\
 &  &  &  & (0.007) & (0.011) &  & (0.008) & (0.010) & (0.007) & (0.011)\\
Both &  &  &  & 0.003 & 0.004 &  & -0.002 & -0.005 & -0.003 & -0.000\\
 &  &  &  & (0.008) & (0.013) &  & (0.008) & (0.011) & (0.008) & (0.012)

%% file: fryer_levittA.tex
Black & -0.213 & -0.182 &  & -0.193 & -0.202 & -0.191 & -0.150 & -0.231 & -0.171 & -0.195\\
 & (0.032) & (0.035) &  & (0.034) & (0.065) & (0.037) & (0.041) & (0.038) & (0.040) & (0.059)\\
 &  &  &  & [0.031] & [0.045] &  &  & [0.037] & [0.035] & [0.043]\\
Hispanic & -0.249 &  &  & -0.257 & -0.171 & -0.209 & -0.212 & -0.196 & -0.220 & -0.171\\
 & (0.028) &  &  & (0.030) & (0.046) & (0.032) & (0.035) & (0.033) & (0.034) & (0.045)\\
 &  &  &  & [0.028] & [0.039] &  &  & [0.033] & [0.031] & [0.039]\\
Asian & -0.294 &  &  & -0.324 & -0.330 & -0.275 & -0.276 & -0.150 & -0.283 & -0.317\\
 & (0.035) &  &  & (0.038) & (0.085) & (0.039) & (0.043) & (0.058) & (0.043) & (0.082)\\
 &  &  &  & [0.033] & [0.057] &  &  & [0.056] & [0.036] & [0.055]\\
Other & -0.132 &  &  & -0.116 & -0.127 & -0.127 & -0.104 & -0.084 & -0.105 & -0.105\\
 & (0.038) &  &  & (0.039) & (0.046) & (0.043) & (0.045) & (0.035) & (0.044) & (0.047)\\
 &  &  &  & [0.029] & [0.035] &  &  & [0.034] & [0.031] & [0.035]\\
Number of controls & \multicolumn{1}{r}{176} &  &  &  &  & \multicolumn{1}{r}{127} &  &  &  & \\
Sample size & \multicolumn{1}{r}{8,806} &  &  &  &  & \multicolumn{1}{r}{6,623} &  &  &  & 

%% file: fryer_levittB.tex
Black &  & -0.031 &  & -0.020 & -0.011 &  & -0.042 & 0.040 & -0.020 & 0.004\\
 &  & (0.016) &  & (0.013) & (0.056) &  & (0.017) & (0.028) & (0.014) & (0.048)\\
Hispanic &  &  &  & 0.008 & -0.077 &  & 0.003 & -0.013 & 0.011 & -0.038\\
 &  &  &  & (0.009) & (0.038) &  & (0.013) & (0.021) & (0.011) & (0.035)\\
Asian &  &  &  & 0.030 & 0.036 &  & 0.001 & -0.124 & 0.009 & 0.043\\
 &  &  &  & (0.018) & (0.074) &  & (0.018) & (0.057) & (0.016) & (0.068)\\
Other &  &  &  & -0.015 & -0.005 &  & -0.023 & -0.043 & -0.023 & -0.022\\
 &  &  &  & (0.013) & (0.048) &  & (0.015) & (0.038) & (0.014) & (0.049)

%% file: rimA.tex
Black & -4.059 &  &  & -3.907 & -3.786 & -4.441 & -3.969 & 8.071 & -3.199 & -3.266\\
 & (1.107) &  &  & (1.210) & (1.597) & (1.149) & (1.059) & (11.922) & (1.039) & (1.403)\\
 &  &  &  & [0.393] & [0.747] &  &  & [3.991] & [0.537] & [0.628]\\
Hispanic & -1.119 &  &  & -0.837 & 1.290 & -0.658 & -0.908 & 2.927 & -0.879 & -1.099\\
 & (0.731) &  &  & (0.698) & (3.949) & (1.603) & (1.461) & (3.403) & (1.446) & (2.460)\\
 &  &  &  & [0.142] & [0.637] &  &  & [2.150] & [0.305] & [0.620]\\
Asian & -2.536 &  &  & -2.117 & -4.375 & -3.383 & -3.110 & -8.439 & -3.633 & -3.685\\
 & (0.978) &  &  & (1.206) & (2.896) & (1.440) & (1.114) & (3.606) & (0.930) & (1.824)\\
 &  &  &  & [0.314] & [0.384] &  &  & [1.685] & [0.351] & [0.638]\\
Number of controls & \multicolumn{1}{r}{268} &  &  &  &  & \multicolumn{1}{r}{ 35} &  &  &  & \\
Sample size & \multicolumn{1}{r}{4,037} &  &  &  &  & \multicolumn{1}{r}{  620} &  &  &  & 

%% file: rimB.tex
Black &  &  &  & -0.152 & -0.274 &  & -0.472 & -12.513 & -1.243 & -1.175\\
 &  &  &  & (0.406) & (1.902) &  & (1.117) & (12.089) & (1.277) & (1.210)\\
Hispanic &  &  &  & -0.282 & -2.409 &  & 0.250 & -3.584 & 0.222 & 0.442\\
 &  &  &  & (0.212) & (3.813) &  & (0.446) & (3.269) & (0.344) & (1.154)\\
Asian &  &  &  & -0.418 & 1.839 &  & -0.273 & 5.056 & 0.249 & 0.302\\
 &  &  &  & (0.632) & (2.804) &  & (0.713) & (3.259) & (0.842) & (1.445)

%% file: weisburstA.tex
Black & 0.172 & -0.037 & 0.342 & 0.109 & 0.246\\
 & (0.274) & (0.305) & (0.396) & (0.267) & (0.292)\\
 &  &  & [0.323] & [0.152] & [0.178]\\
Hispanic & 0.043 & -0.754 & -0.330 & -0.496 & -0.466\\
 & (0.394) & (0.404) & (0.395) & (0.341) & (0.289)\\
 &  &  & [0.312] & [0.221] & [0.169]\\
Other & 1.130 & 1.130 & 0.223 & 1.244 & 0.106\\
 & (0.652) & (0.654) & (0.622) & (0.679) & (0.712)\\
 &  &  & [0.394] & [0.347] & [0.566]\\
Number of controls & \multicolumn{1}{r}{256} &  &  &  & \\
Sample size & \multicolumn{1}{r}{7,488} &  &  &  & 

%% file: weisburstB.tex
Black &  & 0.209 & -0.169 & 0.063 & -0.074\\
 &  & (0.218) & (0.337) & (0.190) & (0.264)\\
Hispanic &  & 0.797 & 0.373 & 0.539 & 0.508\\
 &  & (0.356) & (0.390) & (0.310) & (0.330)\\
Other &  & 0.001 & 0.907 & -0.113 & 1.025\\
 &  & (0.125) & (0.340) & (0.120) & (0.578)

%% file: cont_weight.tex
\begin{tikzpicture}[x=1pt,y=1pt]
\definecolor{fillColor}{RGB}{255,255,255}
\path[use as bounding box,fill=fillColor,fill opacity=0.00] (0,0) rectangle (448.07,260.17);
\begin{scope}
\path[clip] (  0.00,  0.00) rectangle (448.07,260.17);
\definecolor{drawColor}{RGB}{255,255,255}
\definecolor{fillColor}{RGB}{255,255,255}

\path[draw=drawColor,line width= 0.6pt,line join=round,line cap=round,fill=fillColor] (  0.00,  0.00) rectangle (448.07,260.17);
\end{scope}
\begin{scope}
\path[clip] ( 36.22, 30.40) rectangle (237.15,238.11);
\definecolor{fillColor}{RGB}{255,255,255}

\path[fill=fillColor] ( 36.22, 30.40) rectangle (237.15,238.11);
\definecolor{drawColor}{RGB}{238,228,218}

\path[draw=drawColor,line width= 0.6pt,dash pattern=on 4pt off 4pt ,line join=round] ( 36.22, 53.80) --
	(237.15, 53.80);

\path[draw=drawColor,line width= 0.6pt,dash pattern=on 4pt off 4pt ,line join=round] ( 36.22, 88.36) --
	(237.15, 88.36);

\path[draw=drawColor,line width= 0.6pt,dash pattern=on 4pt off 4pt ,line join=round] ( 36.22,122.91) --
	(237.15,122.91);

\path[draw=drawColor,line width= 0.6pt,dash pattern=on 4pt off 4pt ,line join=round] ( 36.22,157.46) --
	(237.15,157.46);

\path[draw=drawColor,line width= 0.6pt,dash pattern=on 4pt off 4pt ,line join=round] ( 36.22,192.02) --
	(237.15,192.02);

\path[draw=drawColor,line width= 0.6pt,dash pattern=on 4pt off 4pt ,line join=round] ( 36.22,226.57) --
	(237.15,226.57);

\path[draw=drawColor,line width= 0.6pt,dash pattern=on 4pt off 4pt ,line join=round] ( 37.30, 30.40) --
	( 37.30,238.11);

\path[draw=drawColor,line width= 0.6pt,dash pattern=on 4pt off 4pt ,line join=round] ( 84.22, 30.40) --
	( 84.22,238.11);

\path[draw=drawColor,line width= 0.6pt,dash pattern=on 4pt off 4pt ,line join=round] (131.14, 30.40) --
	(131.14,238.11);

\path[draw=drawColor,line width= 0.6pt,dash pattern=on 4pt off 4pt ,line join=round] (178.05, 30.40) --
	(178.05,238.11);

\path[draw=drawColor,line width= 0.6pt,dash pattern=on 4pt off 4pt ,line join=round] (224.97, 30.40) --
	(224.97,238.11);
\definecolor{drawColor}{RGB}{86,180,233}
\definecolor{fillColor}{RGB}{86,180,233}

\path[draw=drawColor,line width= 0.4pt,line join=round,line cap=round,fill=fillColor] (154.33,135.59) circle (  3.30);

\path[draw=drawColor,line width= 0.4pt,line join=round,line cap=round,fill=fillColor] (146.24,126.21) circle (  3.52);

\path[draw=drawColor,line width= 0.4pt,line join=round,line cap=round,fill=fillColor] (136.23,148.19) circle (  3.38);

\path[draw=drawColor,line width= 0.4pt,line join=round,line cap=round,fill=fillColor] (112.06,173.01) circle (  3.66);

\path[draw=drawColor,line width= 0.4pt,line join=round,line cap=round,fill=fillColor] (175.86,159.09) circle (  3.72);

\path[draw=drawColor,line width= 0.4pt,line join=round,line cap=round,fill=fillColor] (116.55, 60.09) circle (  4.44);

\path[draw=drawColor,line width= 0.4pt,line join=round,line cap=round,fill=fillColor] ( 72.72,132.07) circle (  5.47);

\path[draw=drawColor,line width= 0.4pt,line join=round,line cap=round,fill=fillColor] (135.32, 80.78) circle (  4.72);

\path[draw=drawColor,line width= 0.4pt,line join=round,line cap=round,fill=fillColor] (124.17,176.76) circle (  5.85);

\path[draw=drawColor,line width= 0.4pt,line join=round,line cap=round,fill=fillColor] ( 91.16, 81.59) circle (  3.78);

\path[draw=drawColor,line width= 0.4pt,line join=round,line cap=round,fill=fillColor] (147.75,160.83) circle (  3.38);

\path[draw=drawColor,line width= 0.4pt,line join=round,line cap=round,fill=fillColor] ( 82.28,158.32) circle (  3.14);

\path[draw=drawColor,line width= 0.4pt,line join=round,line cap=round,fill=fillColor] (138.80, 78.66) circle (  4.24);

\path[draw=drawColor,line width= 0.4pt,line join=round,line cap=round,fill=fillColor] (115.46, 71.15) circle (  5.40);

\path[draw=drawColor,line width= 0.4pt,line join=round,line cap=round,fill=fillColor] (120.78,155.11) circle (  3.45);

\path[draw=drawColor,line width= 0.4pt,line join=round,line cap=round,fill=fillColor] (136.49,192.61) circle (  2.96);

\path[draw=drawColor,line width= 0.4pt,line join=round,line cap=round,fill=fillColor] (144.81, 93.16) circle (  6.78);

\path[draw=drawColor,line width= 0.4pt,line join=round,line cap=round,fill=fillColor] (146.60,135.59) circle (  3.30);

\path[draw=drawColor,line width= 0.4pt,line join=round,line cap=round,fill=fillColor] (116.94,145.13) circle (  3.45);

\path[draw=drawColor,line width= 0.4pt,line join=round,line cap=round,fill=fillColor] ( 84.94, 78.05) circle (  4.02);

\path[draw=drawColor,line width= 0.4pt,line join=round,line cap=round,fill=fillColor] (155.50,151.57) circle (  3.52);

\path[draw=drawColor,line width= 0.4pt,line join=round,line cap=round,fill=fillColor] ( 76.65, 83.92) circle (  4.13);

\path[draw=drawColor,line width= 0.4pt,line join=round,line cap=round,fill=fillColor] (136.26, 66.55) circle (  5.40);

\path[draw=drawColor,line width= 0.4pt,line join=round,line cap=round,fill=fillColor] (126.44, 56.40) circle (  4.29);

\path[draw=drawColor,line width= 0.4pt,line join=round,line cap=round,fill=fillColor] (156.14,106.01) circle (  4.02);

\path[draw=drawColor,line width= 0.4pt,line join=round,line cap=round,fill=fillColor] (117.08,182.89) circle (  3.78);

\path[draw=drawColor,line width= 0.4pt,line join=round,line cap=round,fill=fillColor] (150.86,149.86) circle (  2.64);

\path[draw=drawColor,line width= 0.4pt,line join=round,line cap=round,fill=fillColor] (120.92,145.42) circle (  3.66);

\path[draw=drawColor,line width= 0.4pt,line join=round,line cap=round,fill=fillColor] (148.71, 85.04) circle (  4.44);

\path[draw=drawColor,line width= 0.4pt,line join=round,line cap=round,fill=fillColor] (143.45,149.77) circle (  5.98);

\path[draw=drawColor,line width= 0.4pt,line join=round,line cap=round,fill=fillColor] (108.91,132.26) circle (  5.29);

\path[draw=drawColor,line width= 0.4pt,line join=round,line cap=round,fill=fillColor] (121.51,171.81) circle (  1.43);

\path[draw=drawColor,line width= 0.4pt,line join=round,line cap=round,fill=fillColor] ( 96.48,176.21) circle (  3.52);

\path[draw=drawColor,line width= 0.4pt,line join=round,line cap=round,fill=fillColor] (126.19, 59.02) circle (  4.49);

\path[draw=drawColor,line width= 0.4pt,line join=round,line cap=round,fill=fillColor] (138.79,168.77) circle (  3.78);

\path[draw=drawColor,line width= 0.4pt,line join=round,line cap=round,fill=fillColor] (119.84, 84.89) circle (  4.29);

\path[draw=drawColor,line width= 0.4pt,line join=round,line cap=round,fill=fillColor] ( 74.98, 65.39) circle (  4.63);

\path[draw=drawColor,line width= 0.4pt,line join=round,line cap=round,fill=fillColor] ( 94.39, 78.82) circle (  3.96);

\path[draw=drawColor,line width= 0.4pt,line join=round,line cap=round,fill=fillColor] (140.91,168.27) circle (  3.72);

\path[draw=drawColor,line width= 0.4pt,line join=round,line cap=round,fill=fillColor] (160.35, 65.15) circle (  5.21);

\path[draw=drawColor,line width= 0.4pt,line join=round,line cap=round,fill=fillColor] ( 94.05, 80.16) circle (  5.21);

\path[draw=drawColor,line width= 0.4pt,line join=round,line cap=round,fill=fillColor] (122.20,154.11) circle (  6.14);

\path[draw=drawColor,line width= 0.4pt,line join=round,line cap=round,fill=fillColor] (145.17,137.61) circle (  5.58);

\path[draw=drawColor,line width= 0.4pt,line join=round,line cap=round,fill=fillColor] (122.90, 71.20) circle (  3.52);

\path[draw=drawColor,line width= 0.4pt,line join=round,line cap=round,fill=fillColor] (118.55,146.62) circle (  5.44);

\path[draw=drawColor,line width= 0.4pt,line join=round,line cap=round,fill=fillColor] (138.77, 72.97) circle (  3.85);

\path[draw=drawColor,line width= 0.4pt,line join=round,line cap=round,fill=fillColor] (167.38,223.22) circle (  4.63);

\path[draw=drawColor,line width= 0.4pt,line join=round,line cap=round,fill=fillColor] (162.53, 73.59) circle (  4.58);

\path[draw=drawColor,line width= 0.4pt,line join=round,line cap=round,fill=fillColor] (117.64,158.56) circle (  3.66);

\path[draw=drawColor,line width= 0.4pt,line join=round,line cap=round,fill=fillColor] (158.05,151.57) circle (  3.52);

\path[draw=drawColor,line width= 0.4pt,line join=round,line cap=round,fill=fillColor] (135.61, 62.08) circle (  5.44);

\path[draw=drawColor,line width= 0.4pt,line join=round,line cap=round,fill=fillColor] (173.63, 88.55) circle (  5.72);

\path[draw=drawColor,line width= 0.4pt,line join=round,line cap=round,fill=fillColor] (124.13,141.64) circle (  5.85);

\path[draw=drawColor,line width= 0.4pt,line join=round,line cap=round,fill=fillColor] (138.64, 66.56) circle (  5.79);

\path[draw=drawColor,line width= 0.4pt,line join=round,line cap=round,fill=fillColor] (140.93,126.58) circle (  4.72);

\path[draw=drawColor,line width= 0.4pt,line join=round,line cap=round,fill=fillColor] (113.37,121.57) circle (  3.05);

\path[draw=drawColor,line width= 0.4pt,line join=round,line cap=round,fill=fillColor] (144.09,153.41) circle (  3.05);

\path[draw=drawColor,line width= 0.4pt,line join=round,line cap=round,fill=fillColor] (139.33,168.81) circle (  4.58);

\path[draw=drawColor,line width= 0.4pt,line join=round,line cap=round,fill=fillColor] (167.93,127.36) circle (  3.30);

\path[draw=drawColor,line width= 0.4pt,line join=round,line cap=round,fill=fillColor] (107.36,196.60) circle (  6.56);

\path[draw=drawColor,line width= 0.4pt,line join=round,line cap=round,fill=fillColor] (166.87, 56.00) circle (  5.55);

\path[draw=drawColor,line width= 0.4pt,line join=round,line cap=round,fill=fillColor] (103.82, 63.29) circle (  6.70);

\path[draw=drawColor,line width= 0.4pt,line join=round,line cap=round,fill=fillColor] (210.57,146.05) circle (  3.78);

\path[draw=drawColor,line width= 0.4pt,line join=round,line cap=round,fill=fillColor] (139.98, 75.15) circle (  3.45);

\path[draw=drawColor,line width= 0.4pt,line join=round,line cap=round,fill=fillColor] (136.13, 70.26) circle (  3.96);

\path[draw=drawColor,line width= 0.4pt,line join=round,line cap=round,fill=fillColor] (153.64, 67.12) circle (  4.44);

\path[draw=drawColor,line width= 0.4pt,line join=round,line cap=round,fill=fillColor] (145.62,222.40) circle (  6.70);

\path[draw=drawColor,line width= 0.4pt,line join=round,line cap=round,fill=fillColor] (141.72,228.67) circle (  4.76);

\path[draw=drawColor,line width= 0.4pt,line join=round,line cap=round,fill=fillColor] (181.10,126.35) circle (  3.72);

\path[draw=drawColor,line width= 0.4pt,line join=round,line cap=round,fill=fillColor] ( 85.22,144.10) circle (  5.10);

\path[draw=drawColor,line width= 0.4pt,line join=round,line cap=round,fill=fillColor] (119.59,208.81) circle (  4.13);

\path[draw=drawColor,line width= 0.4pt,line join=round,line cap=round,fill=fillColor] (114.92,134.27) circle (  2.36);

\path[draw=drawColor,line width= 0.4pt,line join=round,line cap=round,fill=fillColor] (131.12,145.30) circle (  5.58);

\path[draw=drawColor,line width= 0.4pt,line join=round,line cap=round,fill=fillColor] (133.06,115.50) circle (  6.23);

\path[draw=drawColor,line width= 0.4pt,line join=round,line cap=round,fill=fillColor] (160.64,167.49) circle (  3.59);

\path[draw=drawColor,line width= 0.4pt,line join=round,line cap=round,fill=fillColor] (159.52,208.77) circle (  4.89);

\path[draw=drawColor,line width= 0.4pt,line join=round,line cap=round,fill=fillColor] (127.06, 72.76) circle (  4.13);

\path[draw=drawColor,line width= 0.4pt,line join=round,line cap=round,fill=fillColor] (126.22, 49.03) circle (  5.44);
\definecolor{drawColor}{RGB}{85,85,85}

\node[text=drawColor,anchor=base west,inner sep=0pt, outer sep=0pt, scale=  0.85] at ( 42.58, 36.28) {Correlation: $ 0.10$};
\end{scope}
\begin{scope}
\path[clip] (242.15, 30.40) rectangle (443.07,238.11);
\definecolor{fillColor}{RGB}{255,255,255}

\path[fill=fillColor] (242.15, 30.40) rectangle (443.07,238.11);
\definecolor{drawColor}{RGB}{238,228,218}

\path[draw=drawColor,line width= 0.6pt,dash pattern=on 4pt off 4pt ,line join=round] (242.15, 53.80) --
	(443.07, 53.80);

\path[draw=drawColor,line width= 0.6pt,dash pattern=on 4pt off 4pt ,line join=round] (242.15, 88.36) --
	(443.07, 88.36);

\path[draw=drawColor,line width= 0.6pt,dash pattern=on 4pt off 4pt ,line join=round] (242.15,122.91) --
	(443.07,122.91);

\path[draw=drawColor,line width= 0.6pt,dash pattern=on 4pt off 4pt ,line join=round] (242.15,157.46) --
	(443.07,157.46);

\path[draw=drawColor,line width= 0.6pt,dash pattern=on 4pt off 4pt ,line join=round] (242.15,192.02) --
	(443.07,192.02);

\path[draw=drawColor,line width= 0.6pt,dash pattern=on 4pt off 4pt ,line join=round] (242.15,226.57) --
	(443.07,226.57);

\path[draw=drawColor,line width= 0.6pt,dash pattern=on 4pt off 4pt ,line join=round] (243.23, 30.40) --
	(243.23,238.11);

\path[draw=drawColor,line width= 0.6pt,dash pattern=on 4pt off 4pt ,line join=round] (290.15, 30.40) --
	(290.15,238.11);

\path[draw=drawColor,line width= 0.6pt,dash pattern=on 4pt off 4pt ,line join=round] (337.07, 30.40) --
	(337.07,238.11);

\path[draw=drawColor,line width= 0.6pt,dash pattern=on 4pt off 4pt ,line join=round] (383.98, 30.40) --
	(383.98,238.11);

\path[draw=drawColor,line width= 0.6pt,dash pattern=on 4pt off 4pt ,line join=round] (430.90, 30.40) --
	(430.90,238.11);
\definecolor{drawColor}{RGB}{86,180,233}
\definecolor{fillColor}{RGB}{86,180,233}

\path[draw=drawColor,line width= 0.4pt,line join=round,line cap=round,fill=fillColor] (334.47,125.72) circle (  3.30);

\path[draw=drawColor,line width= 0.4pt,line join=round,line cap=round,fill=fillColor] (346.35,131.35) circle (  3.52);

\path[draw=drawColor,line width= 0.4pt,line join=round,line cap=round,fill=fillColor] (337.14,102.78) circle (  3.38);

\path[draw=drawColor,line width= 0.4pt,line join=round,line cap=round,fill=fillColor] (333.72,121.39) circle (  3.66);

\path[draw=drawColor,line width= 0.4pt,line join=round,line cap=round,fill=fillColor] (433.94,134.30) circle (  3.72);

\path[draw=drawColor,line width= 0.4pt,line join=round,line cap=round,fill=fillColor] (338.67,114.51) circle (  4.44);

\path[draw=drawColor,line width= 0.4pt,line join=round,line cap=round,fill=fillColor] (330.92,177.85) circle (  5.47);

\path[draw=drawColor,line width= 0.4pt,line join=round,line cap=round,fill=fillColor] (357.39,126.19) circle (  4.72);

\path[draw=drawColor,line width= 0.4pt,line join=round,line cap=round,fill=fillColor] (382.86,125.31) circle (  5.85);

\path[draw=drawColor,line width= 0.4pt,line join=round,line cap=round,fill=fillColor] (319.76,139.80) circle (  3.78);

\path[draw=drawColor,line width= 0.4pt,line join=round,line cap=round,fill=fillColor] (405.76,120.91) circle (  3.38);

\path[draw=drawColor,line width= 0.4pt,line join=round,line cap=round,fill=fillColor] (351.65,111.33) circle (  3.14);

\path[draw=drawColor,line width= 0.4pt,line join=round,line cap=round,fill=fillColor] (348.64,132.17) circle (  4.24);

\path[draw=drawColor,line width= 0.4pt,line join=round,line cap=round,fill=fillColor] (306.73, 53.24) circle (  5.40);

\path[draw=drawColor,line width= 0.4pt,line join=round,line cap=round,fill=fillColor] (333.13,107.94) circle (  3.45);

\path[draw=drawColor,line width= 0.4pt,line join=round,line cap=round,fill=fillColor] (344.93,129.40) circle (  2.96);

\path[draw=drawColor,line width= 0.4pt,line join=round,line cap=round,fill=fillColor] (382.74,120.92) circle (  6.78);

\path[draw=drawColor,line width= 0.4pt,line join=round,line cap=round,fill=fillColor] (348.53,125.72) circle (  3.30);

\path[draw=drawColor,line width= 0.4pt,line join=round,line cap=round,fill=fillColor] (307.16,111.43) circle (  3.45);

\path[draw=drawColor,line width= 0.4pt,line join=round,line cap=round,fill=fillColor] (251.28,114.25) circle (  4.02);

\path[draw=drawColor,line width= 0.4pt,line join=round,line cap=round,fill=fillColor] (381.61,116.43) circle (  3.52);

\path[draw=drawColor,line width= 0.4pt,line join=round,line cap=round,fill=fillColor] (307.76,114.92) circle (  4.13);

\path[draw=drawColor,line width= 0.4pt,line join=round,line cap=round,fill=fillColor] (323.23, 53.98) circle (  5.40);

\path[draw=drawColor,line width= 0.4pt,line join=round,line cap=round,fill=fillColor] (341.55,119.91) circle (  4.29);

\path[draw=drawColor,line width= 0.4pt,line join=round,line cap=round,fill=fillColor] (338.49,120.76) circle (  4.02);

\path[draw=drawColor,line width= 0.4pt,line join=round,line cap=round,fill=fillColor] (368.67,128.94) circle (  3.78);

\path[draw=drawColor,line width= 0.4pt,line join=round,line cap=round,fill=fillColor] (388.25,125.58) circle (  2.64);

\path[draw=drawColor,line width= 0.4pt,line join=round,line cap=round,fill=fillColor] (324.47,135.35) circle (  3.66);

\path[draw=drawColor,line width= 0.4pt,line join=round,line cap=round,fill=fillColor] (351.38,137.19) circle (  4.44);

\path[draw=drawColor,line width= 0.4pt,line join=round,line cap=round,fill=fillColor] (386.72,123.22) circle (  5.98);

\path[draw=drawColor,line width= 0.4pt,line join=round,line cap=round,fill=fillColor] (347.70,189.32) circle (  5.29);

\path[draw=drawColor,line width= 0.4pt,line join=round,line cap=round,fill=fillColor] (304.84,114.40) circle (  1.43);

\path[draw=drawColor,line width= 0.4pt,line join=round,line cap=round,fill=fillColor] (338.36,138.03) circle (  3.52);

\path[draw=drawColor,line width= 0.4pt,line join=round,line cap=round,fill=fillColor] (310.57,109.10) circle (  4.49);

\path[draw=drawColor,line width= 0.4pt,line join=round,line cap=round,fill=fillColor] (357.70,125.39) circle (  3.78);

\path[draw=drawColor,line width= 0.4pt,line join=round,line cap=round,fill=fillColor] (354.87,115.44) circle (  4.29);

\path[draw=drawColor,line width= 0.4pt,line join=round,line cap=round,fill=fillColor] (355.54,126.29) circle (  4.63);

\path[draw=drawColor,line width= 0.4pt,line join=round,line cap=round,fill=fillColor] (302.93,120.03) circle (  3.96);

\path[draw=drawColor,line width= 0.4pt,line join=round,line cap=round,fill=fillColor] (375.33,129.51) circle (  3.72);

\path[draw=drawColor,line width= 0.4pt,line join=round,line cap=round,fill=fillColor] (375.31, 55.59) circle (  5.21);

\path[draw=drawColor,line width= 0.4pt,line join=round,line cap=round,fill=fillColor] (331.17, 53.52) circle (  5.21);

\path[draw=drawColor,line width= 0.4pt,line join=round,line cap=round,fill=fillColor] (325.45,127.78) circle (  6.14);

\path[draw=drawColor,line width= 0.4pt,line join=round,line cap=round,fill=fillColor] (331.89,196.44) circle (  5.58);

\path[draw=drawColor,line width= 0.4pt,line join=round,line cap=round,fill=fillColor] (355.69,141.87) circle (  3.52);

\path[draw=drawColor,line width= 0.4pt,line join=round,line cap=round,fill=fillColor] (363.02,198.70) circle (  5.44);

\path[draw=drawColor,line width= 0.4pt,line join=round,line cap=round,fill=fillColor] (334.30,119.66) circle (  3.85);

\path[draw=drawColor,line width= 0.4pt,line join=round,line cap=round,fill=fillColor] (336.85,177.51) circle (  4.63);

\path[draw=drawColor,line width= 0.4pt,line join=round,line cap=round,fill=fillColor] (390.15,115.72) circle (  4.58);

\path[draw=drawColor,line width= 0.4pt,line join=round,line cap=round,fill=fillColor] (341.39,139.17) circle (  3.66);

\path[draw=drawColor,line width= 0.4pt,line join=round,line cap=round,fill=fillColor] (358.25,116.43) circle (  3.52);

\path[draw=drawColor,line width= 0.4pt,line join=round,line cap=round,fill=fillColor] (344.33, 53.08) circle (  5.44);

\path[draw=drawColor,line width= 0.4pt,line join=round,line cap=round,fill=fillColor] (402.51, 57.20) circle (  5.72);

\path[draw=drawColor,line width= 0.4pt,line join=round,line cap=round,fill=fillColor] (362.60,110.93) circle (  5.85);

\path[draw=drawColor,line width= 0.4pt,line join=round,line cap=round,fill=fillColor] (385.55, 57.20) circle (  5.79);

\path[draw=drawColor,line width= 0.4pt,line join=round,line cap=round,fill=fillColor] (382.93, 65.61) circle (  4.72);

\path[draw=drawColor,line width= 0.4pt,line join=round,line cap=round,fill=fillColor] (299.46,119.72) circle (  3.05);

\path[draw=drawColor,line width= 0.4pt,line join=round,line cap=round,fill=fillColor] (381.78,130.87) circle (  3.05);

\path[draw=drawColor,line width= 0.4pt,line join=round,line cap=round,fill=fillColor] (369.49, 77.90) circle (  4.58);

\path[draw=drawColor,line width= 0.4pt,line join=round,line cap=round,fill=fillColor] (422.94,131.62) circle (  3.30);

\path[draw=drawColor,line width= 0.4pt,line join=round,line cap=round,fill=fillColor] (371.55,160.08) circle (  6.56);

\path[draw=drawColor,line width= 0.4pt,line join=round,line cap=round,fill=fillColor] (362.48, 57.52) circle (  5.55);

\path[draw=drawColor,line width= 0.4pt,line join=round,line cap=round,fill=fillColor] (326.98,126.30) circle (  6.70);

\path[draw=drawColor,line width= 0.4pt,line join=round,line cap=round,fill=fillColor] (393.65,125.62) circle (  3.78);

\path[draw=drawColor,line width= 0.4pt,line join=round,line cap=round,fill=fillColor] (334.12,141.75) circle (  3.45);

\path[draw=drawColor,line width= 0.4pt,line join=round,line cap=round,fill=fillColor] (384.16,140.14) circle (  3.96);

\path[draw=drawColor,line width= 0.4pt,line join=round,line cap=round,fill=fillColor] (335.03,126.28) circle (  4.44);

\path[draw=drawColor,line width= 0.4pt,line join=round,line cap=round,fill=fillColor] (313.62,161.87) circle (  6.70);

\path[draw=drawColor,line width= 0.4pt,line join=round,line cap=round,fill=fillColor] (333.36,171.41) circle (  4.76);

\path[draw=drawColor,line width= 0.4pt,line join=round,line cap=round,fill=fillColor] (370.63,115.82) circle (  3.72);

\path[draw=drawColor,line width= 0.4pt,line join=round,line cap=round,fill=fillColor] (320.31,183.91) circle (  5.10);

\path[draw=drawColor,line width= 0.4pt,line join=round,line cap=round,fill=fillColor] (394.94,170.62) circle (  4.13);

\path[draw=drawColor,line width= 0.4pt,line join=round,line cap=round,fill=fillColor] (329.66,113.06) circle (  2.36);

\path[draw=drawColor,line width= 0.4pt,line join=round,line cap=round,fill=fillColor] (330.99,210.52) circle (  5.58);

\path[draw=drawColor,line width= 0.4pt,line join=round,line cap=round,fill=fillColor] (357.20,125.91) circle (  6.23);

\path[draw=drawColor,line width= 0.4pt,line join=round,line cap=round,fill=fillColor] (325.20,117.22) circle (  3.59);

\path[draw=drawColor,line width= 0.4pt,line join=round,line cap=round,fill=fillColor] (342.33,183.15) circle (  4.89);

\path[draw=drawColor,line width= 0.4pt,line join=round,line cap=round,fill=fillColor] (326.92,108.67) circle (  4.13);

\path[draw=drawColor,line width= 0.4pt,line join=round,line cap=round,fill=fillColor] (363.00, 39.84) circle (  5.44);
\definecolor{drawColor}{RGB}{85,85,85}

\node[text=drawColor,anchor=base west,inner sep=0pt, outer sep=0pt, scale=  0.85] at (249.17, 36.28) {Correlation: $-0.13$};
\end{scope}
\begin{scope}
\path[clip] ( 36.22,238.11) rectangle (237.15,255.17);
\definecolor{fillColor}{RGB}{255,255,255}

\path[fill=fillColor] ( 36.22,238.11) rectangle (237.15,255.17);
\definecolor{drawColor}{RGB}{85,85,85}

\node[text=drawColor,anchor=base,inner sep=0pt, outer sep=0pt, scale=  0.80] at (136.68,243.88) {A: Small class treatment};
\end{scope}
\begin{scope}
\path[clip] (242.15,238.11) rectangle (443.07,255.17);
\definecolor{fillColor}{RGB}{255,255,255}

\path[fill=fillColor] (242.15,238.11) rectangle (443.07,255.17);
\definecolor{drawColor}{RGB}{85,85,85}

\node[text=drawColor,anchor=base,inner sep=0pt, outer sep=0pt, scale=  0.80] at (342.61,243.88) {B: Aide treatment};
\end{scope}
\begin{scope}
\path[clip] (  0.00,  0.00) rectangle (448.07,260.17);
\definecolor{drawColor}{RGB}{130,106,80}

\path[draw=drawColor,line width= 0.6pt,line join=round] ( 36.22, 30.40) --
	(237.15, 30.40);
\end{scope}
\begin{scope}
\path[clip] (  0.00,  0.00) rectangle (448.07,260.17);
\definecolor{drawColor}{RGB}{130,106,80}

\path[draw=drawColor,line width= 0.6pt,line join=round] ( 37.30, 27.90) --
	( 37.30, 30.40);

\path[draw=drawColor,line width= 0.6pt,line join=round] ( 84.22, 27.90) --
	( 84.22, 30.40);

\path[draw=drawColor,line width= 0.6pt,line join=round] (131.14, 27.90) --
	(131.14, 30.40);

\path[draw=drawColor,line width= 0.6pt,line join=round] (178.05, 27.90) --
	(178.05, 30.40);

\path[draw=drawColor,line width= 0.6pt,line join=round] (224.97, 27.90) --
	(224.97, 30.40);
\end{scope}
\begin{scope}
\path[clip] (  0.00,  0.00) rectangle (448.07,260.17);
\definecolor{drawColor}{RGB}{85,85,85}

\node[text=drawColor,anchor=base,inner sep=0pt, outer sep=0pt, scale=  0.80] at ( 37.30, 20.39) {-40};

\node[text=drawColor,anchor=base,inner sep=0pt, outer sep=0pt, scale=  0.80] at ( 84.22, 20.39) {-20};

\node[text=drawColor,anchor=base,inner sep=0pt, outer sep=0pt, scale=  0.80] at (131.14, 20.39) {0};

\node[text=drawColor,anchor=base,inner sep=0pt, outer sep=0pt, scale=  0.80] at (178.05, 20.39) {20};

\node[text=drawColor,anchor=base,inner sep=0pt, outer sep=0pt, scale=  0.80] at (224.97, 20.39) {40};
\end{scope}
\begin{scope}
\path[clip] (  0.00,  0.00) rectangle (448.07,260.17);
\definecolor{drawColor}{RGB}{130,106,80}

\path[draw=drawColor,line width= 0.6pt,line join=round] (242.15, 30.40) --
	(443.07, 30.40);
\end{scope}
\begin{scope}
\path[clip] (  0.00,  0.00) rectangle (448.07,260.17);
\definecolor{drawColor}{RGB}{130,106,80}

\path[draw=drawColor,line width= 0.6pt,line join=round] (243.23, 27.90) --
	(243.23, 30.40);

\path[draw=drawColor,line width= 0.6pt,line join=round] (290.15, 27.90) --
	(290.15, 30.40);

\path[draw=drawColor,line width= 0.6pt,line join=round] (337.07, 27.90) --
	(337.07, 30.40);

\path[draw=drawColor,line width= 0.6pt,line join=round] (383.98, 27.90) --
	(383.98, 30.40);

\path[draw=drawColor,line width= 0.6pt,line join=round] (430.90, 27.90) --
	(430.90, 30.40);
\end{scope}
\begin{scope}
\path[clip] (  0.00,  0.00) rectangle (448.07,260.17);
\definecolor{drawColor}{RGB}{85,85,85}

\node[text=drawColor,anchor=base,inner sep=0pt, outer sep=0pt, scale=  0.80] at (243.23, 20.39) {-40};

\node[text=drawColor,anchor=base,inner sep=0pt, outer sep=0pt, scale=  0.80] at (290.15, 20.39) {-20};

\node[text=drawColor,anchor=base,inner sep=0pt, outer sep=0pt, scale=  0.80] at (337.07, 20.39) {0};

\node[text=drawColor,anchor=base,inner sep=0pt, outer sep=0pt, scale=  0.80] at (383.98, 20.39) {20};

\node[text=drawColor,anchor=base,inner sep=0pt, outer sep=0pt, scale=  0.80] at (430.90, 20.39) {40};
\end{scope}
\begin{scope}
\path[clip] (  0.00,  0.00) rectangle (448.07,260.17);
\definecolor{drawColor}{RGB}{130,106,80}

\path[draw=drawColor,line width= 0.6pt,line join=round] ( 36.22, 30.40) --
	( 36.22,238.11);
\end{scope}
\begin{scope}
\path[clip] (  0.00,  0.00) rectangle (448.07,260.17);
\definecolor{drawColor}{RGB}{85,85,85}

\node[text=drawColor,anchor=base east,inner sep=0pt, outer sep=0pt, scale=  0.80] at ( 31.72, 51.05) {-0.2};

\node[text=drawColor,anchor=base east,inner sep=0pt, outer sep=0pt, scale=  0.80] at ( 31.72, 85.60) {-0.1};

\node[text=drawColor,anchor=base east,inner sep=0pt, outer sep=0pt, scale=  0.80] at ( 31.72,120.15) {0.0};

\node[text=drawColor,anchor=base east,inner sep=0pt, outer sep=0pt, scale=  0.80] at ( 31.72,154.71) {0.1};

\node[text=drawColor,anchor=base east,inner sep=0pt, outer sep=0pt, scale=  0.80] at ( 31.72,189.26) {0.2};

\node[text=drawColor,anchor=base east,inner sep=0pt, outer sep=0pt, scale=  0.80] at ( 31.72,223.82) {0.3};
\end{scope}
\begin{scope}
\path[clip] (  0.00,  0.00) rectangle (448.07,260.17);
\definecolor{drawColor}{RGB}{130,106,80}

\path[draw=drawColor,line width= 0.6pt,line join=round] ( 33.72, 53.80) --
	( 36.22, 53.80);

\path[draw=drawColor,line width= 0.6pt,line join=round] ( 33.72, 88.36) --
	( 36.22, 88.36);

\path[draw=drawColor,line width= 0.6pt,line join=round] ( 33.72,122.91) --
	( 36.22,122.91);

\path[draw=drawColor,line width= 0.6pt,line join=round] ( 33.72,157.46) --
	( 36.22,157.46);

\path[draw=drawColor,line width= 0.6pt,line join=round] ( 33.72,192.02) --
	( 36.22,192.02);

\path[draw=drawColor,line width= 0.6pt,line join=round] ( 33.72,226.57) --
	( 36.22,226.57);
\end{scope}
\begin{scope}
\path[clip] (  0.00,  0.00) rectangle (448.07,260.17);
\definecolor{drawColor}{RGB}{85,85,85}

\node[text=drawColor,anchor=base,inner sep=0pt, outer sep=0pt, scale=  1.00] at (239.65,  6.94) {\hspace{0.5em}Aide treatment effect\hspace{9.5em}Small class treatment effect};
\end{scope}
\begin{scope}
\path[clip] (  0.00,  0.00) rectangle (448.07,260.17);
\definecolor{drawColor}{RGB}{85,85,85}

\node[text=drawColor,rotate= 90.00,anchor=base,inner sep=0pt, outer sep=0pt, scale=  1.00] at ( 11.89,134.25) {Contamination weight};
\end{scope}
\end{tikzpicture}